\documentclass[submission, Phys]{SciPost}

\usepackage[utf8]{inputenc} 
\usepackage[T1]{fontenc} 	
\usepackage[english]{babel} 

\usepackage[bitstream-charter]{mathdesign}
\usepackage{geometry} 		
\usepackage{amsmath} 		
\usepackage{mathtools} 		
\usepackage{float} 			
\usepackage{graphicx} 		
\usepackage{tabularx} 		
\usepackage{booktabs} 		
\usepackage{color, xcolor} 	
\usepackage{pdfpages} 		
\usepackage{extarrows} 		
\usepackage{multirow} 		
\usepackage{multicol} 		
\usepackage{enumitem} 		
\usepackage{xspace} 		
\usepackage{stackrel} 		
\usepackage{tikz} 			
\usepackage{braket} 		
\usepackage{bm} 			
\usepackage{tensor} 		
\usepackage{slashed} 		
\usepackage{siunitx} 		
\usepackage{lastpage} 		
\usepackage{cite} 			
\usepackage[normalem]{ulem} 
\usepackage{fontawesome} 	
\usepackage{tocloft} 		
\usepackage{titlesec} 		
\usepackage{doi} 			
\usepackage{hyperref} 		
\usepackage[most]{tcolorbox} 					
\usepackage[nameinlink, capitalize]{cleveref} 	
\usepackage[nottoc, notlot, notlof]{tocbibind} 	
\usepackage[ruled, vlined]{algorithm2e} 		
\usepackage{makecell}
\usepackage[makeroom]{cancel}
\usepackage{feynmf}
\usepackage{pifont}

\usepackage{tabularray}
\UseTblrLibrary{booktabs}

\makeatletter
\def\BState{\State\hskip-\ALG@thistlm}
\makeatother

\makeatletter
\@ifundefined{pdfoutput}{}{\DeclareGraphicsRule{*}{mps}{*}{}}
\makeatother

\makeatletter
\DeclareRobustCommand*{\bfseries}{%
   \not@math@alphabet\bfseries\mathbf
   \fontseries\bfdefault\selectfont
   \boldmath
}
\makeatother

\hypersetup{
	pdftitle={How to Trust Learned Loop Amplitudes},
	pdfauthor={Bahl et al.},
	colorlinks=true, 			
	linkcolor={red!50!black}, 	
	citecolor={blue!50!black}, 	
	urlcolor={blue!80!black} 	
} 

\DeclareSymbolFont{usualmathcal}{OMS}{cmsy}{m}{n}
\DeclareSymbolFontAlphabet{\mathcal}{usualmathcal}

\SetArgSty{textnormal}
\SetKwComment{Comment}{{\small\#}~}{}
\SetCommentSty{mycommfont}

\setitemize{itemsep=0pt, parsep=0pt} 				
\setenumerate{itemsep=0pt, parsep=0pt} 				
\setitemize{itemsep=2pt,topsep=2pt,parsep=0pt,partopsep=0pt,leftmargin=*}
\setenumerate{itemsep=0pt,topsep=2pt,parsep=0pt,partopsep=0pt,labelindent=3pt,leftmargin=*}
\usepackage{amsmath}
 
\definecolor{red_cb}{HTML}{e41a1c}
\definecolor{blue_cb}{HTML}{377eb8}
\definecolor{green_cb}{HTML}{4daf4a}
\definecolor{purple_cb}{HTML}{984ea3}
\definecolor{orange_cb}{HTML}{ff7f00}

\definecolor{EmeraldGreen}{HTML}{1ea78d}
\definecolor{EnglishRed}{HTML}{b02427}
\hypersetup{colorlinks=true,urlcolor=EmeraldGreen,citecolor=EmeraldGreen,linkcolor=EnglishRed}

\newcommand{\ie}{\text{i.e.}\;}

\newcommand{\mwith}{\text{with}}

\newcommand{\qqquad}{\qquad\quad}

\def\d{\mathrm{d}}

\newcommand\one{\leavevmode\hbox{\small1\normalsize\kern-.33em1}}

\newcommand{\mean}[1]{\left\langle#1\right\rangle}

\newcommand{\loss}{\mathcal{L}} 	
\newcommand{\normal}{\mathcal{N}} 	

\newcommand{\gosam}{\textsc{GoSam}\xspace}

\newcommand{\arXiv}[2][]{%
	\ifthenelse{\equal{#1}{}}%
	{\href{http://arxiv.org/abs/#2}{arXiv:#2}}%
	{\href{http://arxiv.org/abs/#2}{arXiv:#2~[#1]}}}

\newcommand{\gev}{\text{GeV}}
\newcommand{\tev}{\text{TeV}}

\def\slashchar#1{\setbox0=\hbox{$#1$}           
   \dimen0=\wd0                                 
   \setbox1=\hbox{/} \dimen1=\wd1               
   \ifdim\dimen0>\dimen1                        
      \rlap{\hbox to \dimen0{\hfil/\hfil}}      
      #1                                        
   \else                                        
      \rlap{\hbox to \dimen1{\hfil$#1$\hfil}}   
      /                                         
   \fi}

\newcommand{\tikznode}[2]{%
\ifmmode%
\tikz[remember picture,baseline=(#1.base),inner sep=0pt] \node (#1) {$#2$};%
\else
\tikz[remember picture,baseline=(#1.base),inner sep=0pt] \node (#1) {#2};%
\fi}

\def\mathswitchr#1{\relax\ifmmode{\mathrm{#1}}\else$\mathrm{#1}$\xspace\fi}
\def\mathswitch#1{\relax\ifmmode#1\else$#1$\xspace\fi}

\newcommand{\cmark}{\ding{51}}%
\graphicspath{{./figs/}}

\begin{document}

\vspace*{-2.5em}
\hfill{\small KA-TP-01-2026, P3H-26-001}
\vspace*{0.5em}

\begin{center}{\Large \textbf{
How to Trust Learned Loop Amplitudes
}}\end{center}

\begin{center}
Henning Bahl\textsuperscript{1},
Jens Braun\textsuperscript{2},
Gudrun Heinrich\textsuperscript{2},
Tilman Plehn\textsuperscript{1,3}, and
Rebecca Revelli\textsuperscript{1,4}
\end{center}

\begin{center}
{\bf 1} Institut für Theoretische Physik, Universität Heidelberg, Germany\\
{\bf 2} Institute for Theoretical Physics, Karlsruhe Institute of Technology (KIT), Germany\\
{\bf 3} Interdisciplinary Center for Scientific Computing (IWR), Universität Heidelberg, Germany\\ 
{\bf 4} Dipartimento di Fisica, Università di Torino, Italy
\end{center}

\begin{center}
\today
\end{center}


\section*{Abstract}
{\bf 
    Higher-order theory predictions are crucial for the precision LHC program, but the time-consuming amplitude evaluation challenges the corresponding Monte-Carlo simulations. Machine-learned amplitude surrogates can resolve this problem, if we can guarantee their precision over the entire phase space. First, we show that our surrogates provide a calibrated learned uncertainty, even for non-Gaussian systematics; second, we describe how less accurate phase space regions can be identified; third, we demonstrate how the precision in these regions can be improved reliably.
}

\vspace{10pt}
\noindent\rule{\textwidth}{1pt}
\tableofcontents\thispagestyle{fancy}
\noindent\rule{\textwidth}{1pt}
\vspace{10pt}

\clearpage
\section{Introduction}

The rapidly increasing amount of LHC data allows analyses and measurements with unprecedented precision, provided the experimental precision is matched by the precision of the theoretical predictions~\cite{Dawson:2018dcd}. This requires simulations including higher-order perturbative corrections, both loop amplitudes and efficient schemes to treat unresolved real radiation. In this regard, we have seen remarkable progress in both QCD and electroweak (EW) corrections. For LHC loop amplitudes, the current frontier are 3-loop box amplitudes with up to two off-shell legs and 2-loop pentagon amplitudes with several mass scales~\cite{Chen:2025utl,Davies:2025otz,Becchetti:2025qlu,Badger:2025ilt,Canko:2025itd}. These kinds of precision predictions have to be numerically available in public codes if we want to benefit from optimal analysis methods, like simulation-based inference~\cite{Brehmer:2019xox,Chatterjee:2021nms,Chatterjee:2022oco,ATLAS:2024jry,Schofbeck:2024zjo,Bahl:2024meb,Benato:2025rgo,Bahl:2025mib}.

However, the availability of analytic or (semi-)numerical loop amplitudes does not guarantee that they can be used for large-scale simulations. Important limitations are the accuracy and speed at which the higher-order amplitudes can be evaluated in practice. Here, we hope that modern machine learning (ML)~\cite{Butter:2022rso,Plehn:2022ftl} will help with three distinct numerical challenges: efficient phase space sampling~\cite{Bendavid:2017zhk,Klimek:2018mza,Gao:2020vdv,Gao:2020zvv,Bothmann:2020ywa,Heimel:2022wyj,Heimel:2023ngj,Deutschmann:2024lml,Heimel:2024wph,Bothmann:2025lwg,Janssen:2025zke}, efficient subtraction schemes for real radiation~\cite{Janssen:2025zke}, and virtual multi-loop amplitudes~\cite{Bishara:2019iwh,Badger:2020uow,Aylett-Bullock:2021hmo,Maitre:2021uaa,Danziger:2021eeg,Winterhalder:2021ngy,Badger:2022hwf,Janssen:2023ahv,Maitre:2023dqz,Brehmer:2024yqw,Bahl:2024gyt,Breso:2024jlt,Herrmann:2025nnz,Favaro:2025pgz,Villadamigo:2025our,Bahl:2025xvx,Beccatini:2025tpk}. 

For virtual loop amplitudes, the numerical complexity rises not only with the number of loops, but also with the number of kinematic scales, related to the number of external legs and their virtuality, and mass scales. At two loops and beyond, their direct evaluation inside the simulation code is often not feasible. 
To construct a surrogate amplitude, the full higher-order amplitude is first used to generate a reference training dataset. This data is then used to construct or learn a surrogate. An interesting aspect of this surrogate is that for training data without significant numerical noise, the network should interpolate rather than fit a function~\cite{Plehn:2022ftl}. For $2\to 2$ amplitudes, we can use two-dimensional interpolation grids~\cite{Heinrich:2017kxx,Czakon:2017dip,Heinrich:2020ckp,Heinrich:2022idm,Agarwal:2024pod,CampilloAveleira:2024rnp}. For $2\to 3$ amplitudes, even with fixed masses, five-dimensional grids are challenging in the required number of data points and in the interpolation techniques. Ultra-fast ML-surrogates have been shown to be more accurate than traditional techniques~\cite{Breso:2024jlt}. A natural application are two-loop amplitudes for $t\bar{t}H$ production, where partial results have been calculated in Refs.~\cite{Catani:2022mfv,FebresCordero:2023pww,Wang:2024pmv,Devoto:2024nhl,Agarwal:2024jyq}.

For precision simulations, it is crucial that we control surrogate uncertainties. In addition to globally defined numerical noise, a given phase space region might be poorly described, leading to a biased prediction of the observables in this region. A comparably expensive solution is to use the surrogate amplitude just to increase the unweighting efficiency while still using the true amplitude during event generation~\cite{Janssen:2023ahv,Bothmann:2025lwg}. An alternative approach avoiding the evaluation of the true amplitude during event generation --- particularly useful for multi-loop amplitudes --- is to equip the surrogate with a calibrated uncertainty estimate~\cite{Badger:2022hwf,ATLAS:2024rpl,Bahl:2024gyt,Bahl:2025xvx, Beccatini:2025tpk}, and then ensure that the largest uncertainties do not correspond to a localized failure mode in phase space. Depending on the failure mode, a correctly learned uncertainty can then be used to improve the training data and training procedure~\cite{Badger:2022hwf,Beccatini:2025tpk}.

The paper is structured as follows. In Sec.~\ref{sec:surr} we introduce a set of conceptual improvements to probabilistic (amplitude) regression: in Sec.~\ref{sec:surr_prob} we review the existing concepts and introduce a Student's $t$-likelihood approach for non-Gaussian problems; in Sec.~\ref{sec:surr_syst} we show how Gaussian systematics only appear for phase space dimensions beyond $2\to 2 $ scattering; in Sec.~\ref{sec:surr_local} we introduce a way to analyze if low-accuracy amplitudes cluster in phase space; and in Sec.~\ref{sec:surr_sample} we propose an adaptive two-step sampling procedure to optimize the network training locally.  We demonstrate these new techniques for di-Higgs production at leading order (LO) and next-to-leading order (NLO) in QCD in Sec.~\ref{sec:2to2}. We then illustrate the performance gain for top-associated Higgs production at NLO as an example for a more challenging $2\to 3$ process in Sec.~\ref{sec:2to3}. In Sec.~\ref{sec:conclusions} we conclude that this phase space-controlled training will allow us to use ML-surrogate amplitudes for perturbative precision simulations.

\section{Efficient amplitude surrogates}
\label{sec:surr}

The ML-task to provide ultrafast loop amplitude surrogates is to learn an amplitude $A_\text{NN}$ approximating the amplitude $A_\text{true}$ for every phase space point $x$,
\begin{align}
    A_\text{NN}(x) \approx A_\text{true}(x)\;.
\label{eq:def_a}
\end{align}
Here, $A_\text{true}$ is calculated from first principles at a given order in perturbation theory. Ultraviolet (UV) and infrared (IR) singularities need to be removed before an amplitude can be used in a Monte Carlo simulation. We typically learn UV-renormalized and IR-subtracted amplitudes~\cite{Heinrich:2017kxx, Agarwal:2024jyq}. The finite amplitude is exact up to negligible numerical noise, originating for example from the evaluation of special loop functions or from numerical integration over Feynman parameters. Our goal is to replace the amplitudes in the Monte Carlo generator entirely by the ML-surrogate.

\subsection{Non-Gaussian probabilistic regression}
\label{sec:surr_prob}

A statistically sound surrogate describes the probability $p(A|x)$. The target probability depends implicitly on the training data $D_\text{train} = \left\{ (A_\text{train}, x_\text{train}) \right\}$. We can encode this probability in a set of network parameters $\theta$, generalizing beyond the training data, as 
\begin{align}
    p(A\,|\,x) = \int \d\theta\; p(A\,|\,x,\theta) \; p(\theta\,|\,D_\text{train})\;
    \approx
    \int \d\theta\;p(A\,|\,x,\theta) \; q(\theta) \;,
    \label{eq:predictive_dist}
\end{align}
where we replaced the true but normally intractable posterior with an approximate distribution $q(\theta)$. Here, we omit the explicit conditioning of $q$ on $D_\text{train}$ for notation simplicity.  We compute the mean and variance as 
\begin{align}
A_\text{NN}(x)
    &= \int \d A\;A\;p(A\,|\,x)\notag\\
    &= \int \d\theta\; q(\theta)\;\overline{A}(x,\theta)
        \qquad\mwith\quad
        \overline{A}(x,\theta) = \int \d A\; A \;p(A\,|\,x,\theta) \notag \\
    \sigma_\text{tot}^2(x)
    &= \int \d A \,\left[A - A_\text{NN}(x) \right]^2\,p(A\,|\,x) \notag\\
    &= \int \d\theta\; q(\theta)
    \left[
    \overline{A^2}(x,\theta) - \overline{A}(x,\theta)^2 + \left( \overline{A}(x,\theta) -A_\text{NN}(x) \right)^2
    \right] \notag \\
    &\equiv \sigma_\text{syst}^2(x) + \sigma_\text{stat}^2(x)
    \;.
    \label{eq:sigma-tot}
\end{align}
In the last step, we split the variance into a systematic and a statistical contribution,
\begin{align}
    \sigma_\text{syst}^2(x) 
    &= \int \d\theta\; q(\theta)\; \left[
    \overline{A^2}(x,\theta) - \overline{A}(x,\theta)^2 \right] \notag \\
    \sigma_\text{stat}^2(x) 
    &= \int \d\theta \; q(\theta) \left[\overline{A}(x,\theta) - A_\text{NN}(x)\right]^2\;,
    \label{eq:unc_types}
\end{align}
where $\overline{A^2}(x,\theta) = \int \d A\; A^2 \;p(A\,|\,x,\theta)$.

The systematic uncertainty encodes the data-intrinsic uncertainty as well as the uncertainty induced by a lack of model expressivity. The data inherent uncertainty vanishes for noise-free data --- \ie for $p(A\,|\,x,\theta) \to \delta(A(x)-A_0(x))$ with $A_0$ being the noiseless value of the amplitude --- if the model is fully expressive and perfectly trained. It also captures the uncertainty induced by a lack of model expressivity. The statistical uncertainty represents the uncertainty due to the limited size of the training dataset and vanishes for an infinitely large training dataset and perfect training, \ie if $q(\theta)\to \delta(\theta - \theta_0)$. It is also referred to as reducible uncertainty.

The statistical uncertainty can be inferred using Bayesian neural networks (BNNs)~\cite{bnn_early,bnn_early2,bnn_early3,deep_errors}, repulsive ensembles~\cite{repulsive_ensembles_ml,ATLAS:2024rpl,Rover:2024pvr,Bahl:2024meb,Bahl:2024gyt,Bahl:2025xvx}, or evidential regression~\cite{DBLP:journals/corr/abs-1910-02600,2021arXiv210406135M,Kriesten:2024ist,Khot:2025kqg,Bahl:2025xvx}. We have benchmarked these methods for amplitude regression~\cite{Bahl:2024gyt,Bahl:2025xvx} and found the statistical uncertainties to be small compared to the systematics. Consequently, we will focus on the systematic uncertainties.

\subsubsection*{Gaussian systematics}

To train our surrogate we minimize the negative log-likelihood over the training dataset,
\begin{align}
    \loss = - \log p(A\,|\, x, \theta) \; .
    \label{eq:nll}
\end{align}
To capture the systematic uncertainty reflected in the variability of the training data, the simplest ansatz is a Gaussian likelihood,
\begin{align}
    p\left(A\;\middle|\;x,\theta\right) = \normal\left(A\;\middle|\;\overline{A}(x, \theta),\, \sigma^2(x, \theta)\right) \;,
    \label{eq:gaussian_likelihood}
\end{align}
where $\overline{A}(x, \theta)$ and $\sigma^2(x, \theta)$ are learned and $\normal(\mu,\sigma^2)$ denotes a Gaussian with mean $\mu$ and variance $\sigma^2$. The Gaussian ansatz works well for amplitude regression~\cite{Bahl:2024gyt,Bahl:2025xvx} and can be validated via pull distributions. The Gaussian likelihood leads us to the heteroskedastic loss
\begin{align}
    \loss_\text{het} = 
        \frac{[\overline{A}(x, \theta) - A_\text{true}(x)]^2}{2\sigma_\text{syst}^2(x, \theta)} + \log \sigma_\text{syst}(x, \theta) \; .
    \label{eq:het_loss}
\end{align}
For each phase space point, the training minimizes the numerator by learning the true amplitude. The second logarithmic contribution encourages the network to adjusts $\sigma_\text{syst}(x)$ according to the observed deviation. Consequently, large values of $\sigma_\text{syst}(x)$ are predicted if the network struggles to learn the amplitude accurately.

\subsubsection*{Non-Gaussian systematics}

If the simple Gaussian likelihood does not describe the variability of the training data, it can be replaced by a Gaussian mixture model (GMM)~\cite{ATLAS:2024rpl,Bahl:2025xvx},
\begin{align}
    p_\text{GMM}\left(A\;\middle|\;x,\theta\right) &= \sum_k \omega_k(x,\theta)\;\normal\!\left(A \;\middle|\; \overline A_k(x,\theta),\,\sigma_k^2(x,\theta)\right)
    \qquad \text{with} \qquad \sum_k \omega_k(x,\theta) = 1 \; .
    \label{eq:gmm_likelihood}
\end{align}
The mean and standard deviation become
\begin{align}
    \overline{A}_\text{GMM}(x,\theta) &= \sum_k \omega_k(x,\theta)\; \overline A_k(x,\theta) \notag\\
    \sigma_\text{GMM}^2(x,\theta) &= \sum_k \omega_k(x,\theta)\;
    \left[ \sigma_k^2(x,\theta) + \overline A_k^2(x,\theta)\right]
    - \overline{A}_\text{GMM}^2(x,\theta) \; .
    \label{eq:gmm_mean_and_variance}
\end{align}
The GMM loss does not have a simple analytic form and has to be computed numerically,
\begin{align}
    \loss_\text{GMM} 
    &= - \log \left[ \sum_k
        \frac{\omega_k(x,\theta)}{\sqrt{2\pi\sigma_k^2(x,\theta)}} 
        \exp\!\left[-\frac{[\overline A_k(x,\theta) - A_\text{true}(x)]^2}{2\sigma_k^2(x,\theta)}\right]
        \right]
          \;.
    \label{eq:gmm_loss}
\end{align}
If the task is to just describe non-Gaussian tails, the GMM can be simplified by identifying the means,
\begin{align}
    \overline A_k(x,\theta) = \overline A(x,\theta) \; .
\end{align}
This significantly simplifies the calculation of confidence intervals.

An alternative approach to non-Gaussian likelihoods is a Student's $t$-distribution 
\begin{align}
  p_\text{St}\left(A\;\middle|\;x,\theta\right)
  &= \text{St}\left(A\;\middle|\;\overline A(x,\theta),\sigma^2(x,\theta),\nu(x,\theta)\right) \notag \\
  &= \frac{\Gamma\left(\dfrac{\nu+1}{2}\right)}{\Gamma\left(\dfrac{\nu}{2}\right)}
    \; \frac{1}{\sqrt{\pi\nu}\sigma(x,\theta)} \; 
  \left[1 + \frac{\left[ \overline A(x,\theta) - A_\text{true}(x) \right]^2}{\nu\sigma^2(x,\theta)}\right]^{-\frac{\nu+1}{2}}\;,
\end{align}
where $\Gamma$ is the Gamma function. Two limits of the Student's $t$-distribution are
\begin{align}
  p_\text{St}\left(A\;\middle|\;\overline A,\sigma^2,\nu\right)
  = \begin{cases}
     \dfrac{\sigma}{\pi}\dfrac{1}{(A - \overline A)^2 + \sigma^2}
    \qqquad &\nu \to 1 \\[4mm]
    \normal\left(A\;\middle|\;\overline A,\sigma^2\right)
    \qqquad &\nu \to \infty \; ,
  \end{cases}
\label{eq:stt_limits}
\end{align}
where the Breit-Wigner or Cauchy distribution allows the corresponding likelihood to describe much larger tails. For $\nu >2$, the variance of the Student's $t$-distribution is
\begin{align}
    \sigma_\text{St}^2(x, \theta) = \frac{\nu(x,\theta)}{\nu(x,\theta) - 2} \; \sigma^2(x,\theta) \; ,
\end{align}
and the corresponding negative log-likelihood loss reads
\begin{align}
  \loss_\text{St}
  =&\frac{1+\nu(x,\theta)}{2} \log\left[1 + \frac{\left[ \overline A(x,\theta) - A_\text{true}(x)\right]^2}{\nu(x,\theta) \sigma^2(x,\theta)}\right] + \log \sigma(x,\theta)  \notag \\
  +& \frac{1}{2}\log\left(\pi \nu(x,\theta)\right) + \log \Gamma\left(\frac{\nu(x,\theta)}{2}\right) - \log \Gamma\left(\frac{\nu(x,\theta)+1}{2}\right)
  \; .
\label{eq:stt_loss}
\end{align}
Compared to a Gaussian likelihood, only one additional quantity, $\nu(x,\theta)$, needs to be learned.

\subsubsection*{Architecture and preprocessing}

For the surrogate architecture we use a simple MLP, consisting of 5 hidden layers with 512 hidden channels and GELU activation. The surrogates learns the logarithm of the amplitude, scaled to zero mean and unit standard deviation~\cite{Bahl:2025xvx}. The inputs are the 4-momenta $p_i$ of the external particles, complemented by the Lorentz invariants
\begin{align}
    z_{ij} = \log\left(p_i\cdot p_j\right)\; .
\end{align}
All network inputs are standardized, and the hyperparameters are summarized in App.~\ref{app:hyperparameters}. For this preprocessing we need to guarantee that all $p_i\cdot p_j$ are positive. We use the convention that the sum of the initial-state 4-momenta is equal to the sum of the final-state 4-momenta, $\sum_i p_i = \sum_f p_j$. The energy components of all physical 4-vectors are positive, as well as $E_i \ge |\vec p_i|$, with the equality fulfilled for massless particles. This gives us 
\begin{align}
    p_i\cdot p_j = E_i E_j - |\vec p_i||\vec p_j|\cos\theta_{ij} \ge 0 \; .
\end{align}
We also learn the systematic uncertainty using the heteroskedastic loss of Eq.\eqref{eq:het_loss} or its GMM variant as $\log\sigma_\text{syst}$, to cover large variations. For the same reason, we learn $\log\nu$ when using the Student's $t$-likelihood ansatz.

Advanced architectures like equivariant transformers have advantages for processes with a substantial number of identical final state and/or initial state particles, since they allow to implement permutation invariance into the NN architecture~\cite{Brehmer:2024yqw,Favaro:2025pgz}. For the processes studied in this paper, permutation invariance does not yield a substantial performance improvement~\cite{Breso:2024jlt}.

\subsection{Systematic uncertainty shape}
\label{sec:surr_syst}

For precision physics applications it is important that we understand the shape of the systematic uncertainties and then check the calibration of the learned uncertainties. This ensures that amplitude values with low accuracy are indeed those with large learned uncertainties, an aspect we will need below.

\subsubsection*{Accuracy and calibration}

To measure the accuracy of the amplitude surrogates, we define 
\begin{align}
    \Delta (x) = \frac{A_\text{NN}(x) - A_\text{true}(x)}{A_\text{true}(x)} \;. \label{eq:relative-precision}
\end{align}
To test the calibration of the uncertainties, we define the systematic pull
\begin{align}
t_\text{syst}(x) = \frac{A_\text{NN}(x) - A_\text{true}(x)}{\sigma_\text{syst}(x)} \;.
\end{align}
If the shape of this pull distribution is not Gaussian, the Gaussian assumption used for deriving the heteroskedastic loss might not be justified.

When a non-Gaussian likelihood is appropriate, we cannot rely on the pull distribution to test the uncertainty calibration. Instead, we evaluate the empirical coverage. For a test dataset we evaluate for what fraction of the true amplitude the surrogate lies within a given $\gamma$ confidence region. This defines the empirical coverage
\begin{align}
    c_\gamma = \left\langle \mathbb{1}\left(p_\text{Gauss}(A_\text{true}(x)|A_\text{NN}(x),\sigma^2_\text{syst}(x)) > 1-\gamma\right) \right\rangle_{D_\text{test}}\;,\label{eq:def_coverage}
\end{align}
which is here written down for the Gaussian case. The indicator function $\mathbb{1}$ is one if the associated condition is met and zero otherwise. $p_\text{Gauss}$ is the $p$-value for the true amplitude given the learned phase space dependent Gaussian probability distribution,
\begin{align}
    p_\text{Gauss}(A_\text{true}(x)|A_\text{NN}(x),\sigma^2_\text{syst}(x)) 
    = 2\left[ 1 - \Phi\left(\frac{|A_\text{true}(x) - A_\text{NN}(x)|}{\sigma_\text{syst}(x)}\right)\right] \; , \label{eq:pval_gauss}
\end{align}
where $\Phi$ is the cumulative Gaussian distribution function. If the empirical coverage is larger than the nominal confidence level, the learned uncertainty is conservative or underconfident. If it is smaller, the learned uncertainty is overconfident.

If instead of the Gaussian likelihood we use a GMM or a Student's $t$-distribution, the $p$-value and cumulative distribution functions in Eq.~\eqref{eq:def_coverage} and~\eqref{eq:pval_gauss} have to be replaced by their respective equivalents for the GMM or Student's $t$-distribution.

\subsubsection*{Central limit theorem}

\begin{figure}[t]
  \includegraphics[width=0.495\textwidth]{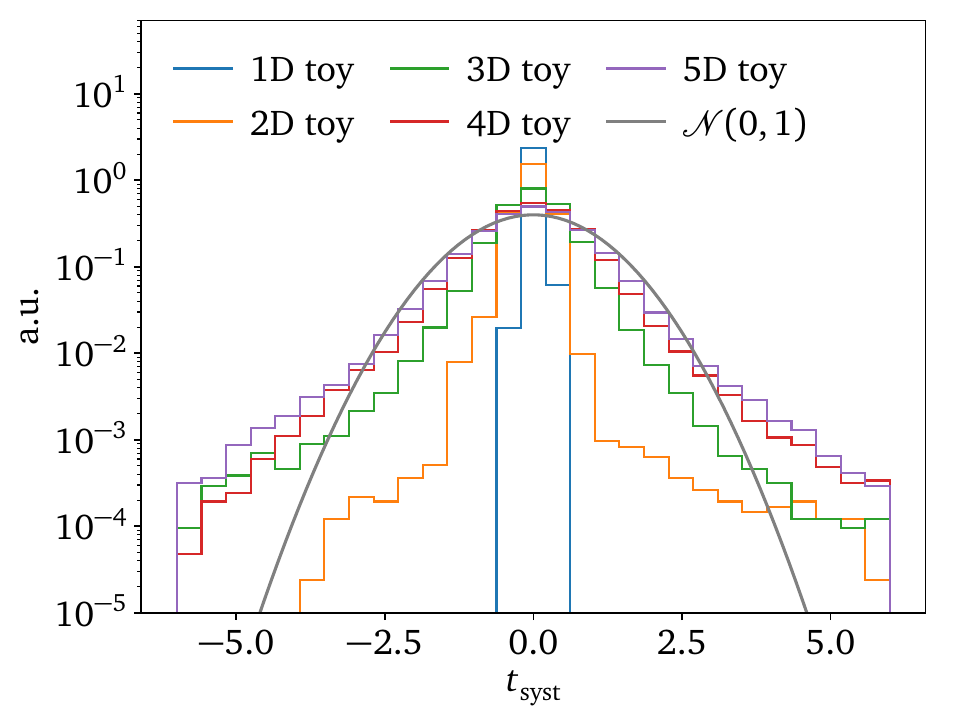}
  \includegraphics[width=0.495\textwidth]{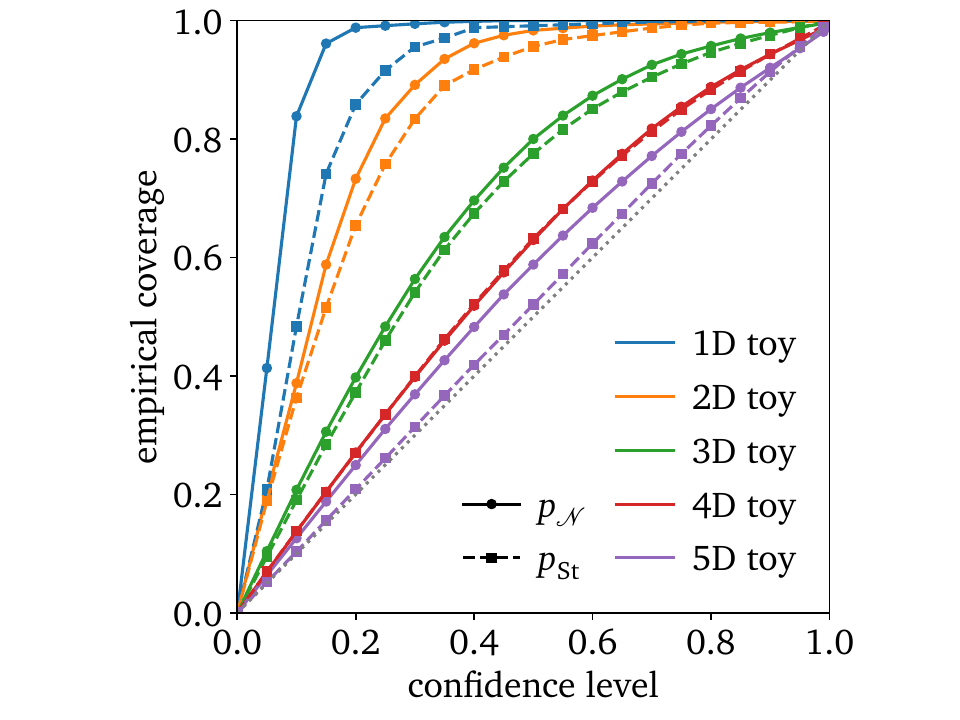}
  \caption{Left: systematic pull of toy models. Right: empirical coverages, comparing Gaussian and Student's $t$-likelihoods.
  }
  \label{fig:toy}
\end{figure}

Above, we have seen that statistical regression is greatly simplified when the variability of the training data is captured by a Gaussian likelihood. If needed, we can capture non-Gaussian behavior using a GMM or a Student's $t$-likelihood. 
To study the impact of the dimensionality, we start with a simple $n$-dimensional toy model where we regress the sum of the input coordinates drawn from the $n$-dimensional hypercube
\begin{align}
  f(x) = \sum_{i = 1}^n x_i
  \qquad\text{with}\qquad
  x_i \sim \mathcal{U}(0,1) \; .
  \label{eq:toy_f_true}
\end{align}
The one-dimensional output of the surrogate $f_\text{NN}$ will effectively approximate the contribution of each dimension with a given uncertainty,
\begin{align}
    f_\text{NN}(x) = \sum_{i=1}^n \left(x_i + \epsilon_i(x)\right)\;,
\end{align}
where we assume independent $\epsilon_i$. In that case, the deviation or residual $r(x)$ is
\begin{align}
    r(x) \equiv f_\text{NN}(x) - f(x)  = \sum_{i=1}^n \epsilon_i(x)\; ,
\end{align}
and the central limit theorem applies. The distribution of the residuals $r(x)$ converges towards a Gaussian,
\begin{align}
    r(x) \sim \normal \left( \mean{r}, \sigma_r^2 \right)
    \qquad \text{for} \qquad
    n\rightarrow\infty \; .
\end{align}
We confirm this behavior by training simple MLP surrogates with a heteroskedastic loss to reproduce $f(x)$ for $n \le 5$ using $10^5$ training events. We list the used NN and training setting in App.~\ref{app:hyperparameters}. We show the systematic pulls in the left panel of Fig.~\ref{fig:toy}. Although the systematic uncertainty is overestimated for small $n$, the calibration of the learned Gaussian uncertainty improves for higher dimensions, just leaving the tails being too large.

The convergence towards a Gaussian likelihood is also visible for the empirical coverage in the right panel of Fig.~\ref{fig:toy}. For few dimensions the uncertainties are clearly overestimated, while for larger dimensionality the empirical coverage shifts towards the diagonal. We also show the results for a Student's $t$-likelihood, given in Eq.\eqref{eq:stt_loss}. Within numerical uncertainties, relaxing the Gaussian assumption can lead to an improved uncertainty calibration in particular for low dimensions.

A similar argument can be made for general target functions. First, if the true function is a product of input components we look at $\log f_\text{NN}(x) - \log f(x)$, so the differences or residuals follow a log-normal distribution. For small residuals, it approaches a normal distribution. More generally, the central limit theorem applies when the residuals can be represented as a sum of an increasing number of independent errors. This is true for a wide range of regression problems, for a strict mathematical discussion for linear regression we refer to Ref.~\cite{0f8094d1-2ebf-3360-90f1-98e50c172f31}.

\subsubsection*{Amplitudes for \texorpdfstring{$Z$}{Z} plus gluons}

\begin{figure}[t]
  \includegraphics[width=0.495\textwidth]{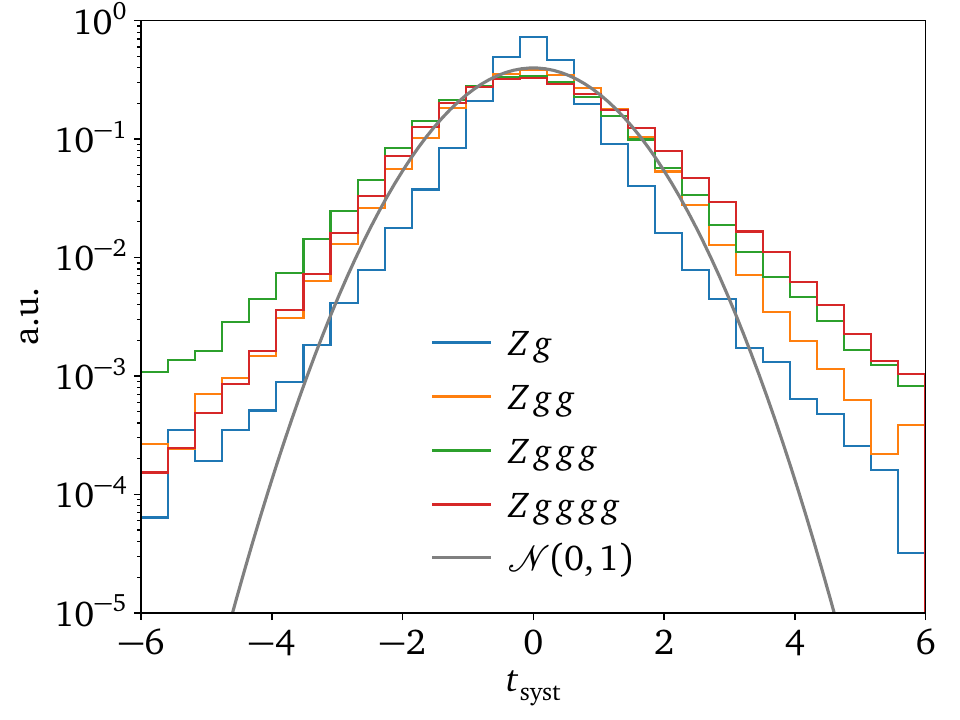} 
  \includegraphics[width=0.495\textwidth]{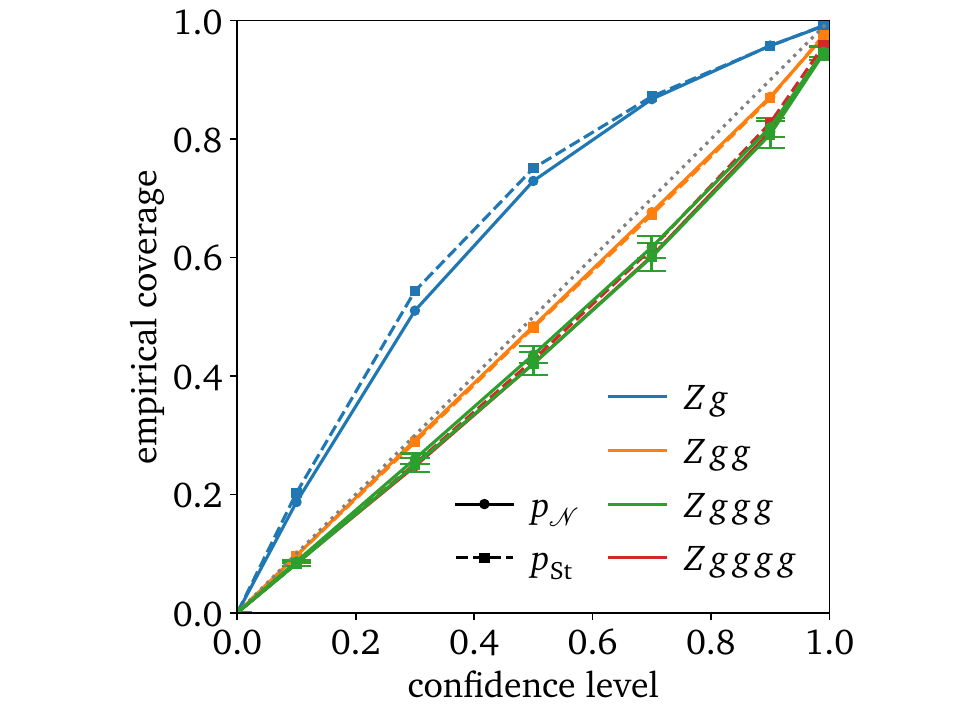} 
  \caption{Left: systematic pulls. Right: coverage of $Z + n_gg$ surrogates. The uncertainty bars for the $Zggg$ surrogates indicate the mean and standard deviation obtained by averaging over five independent runs.}
  \label{fig:Zgluons}
\end{figure}

Next, we test the applicability of the central limit theorem for amplitudes for the production of a $Z$-boson in association with gluons~\cite{Brehmer:2024yqw,Favaro:2025pgz},
\begin{align}
  q\bar q \to Z+ \{ 1,...,4 \} g \; .
\end{align}
Cuts are applied such that all final-state objects have a transverse momentum of at least 20~GeV and such that the angular separation $\Delta R$ is larger than 0.4 for all pairs of final-state objects. 

We use 70k training points for each multiplicity. The phase space dimensionalities for $n_g$ final-state gluons are $3 (n_g +1) -4$. As usual, the relative accuracies degrade for an increasing number of gluons. The corresponding pull distributions in the left panel of Fig.~\ref{fig:Zgluons} become increasingly Gaussian for an increasing number of gluons and hence phase space dimensions. Like for the toy model, the remaining challenge consists in increased tails for higher-dimensional regression. Finally, we show the empirical coverage. The uncertainty is overestimated for a small number of gluons, again as previously observed in the toy model. For $n_g > 1$ only a slight underestimate of the uncertainties remains. We do not observe any benefit from using a Student's $t$-likelihood. The error bars for the $Zggg$ surrogate indicate that the empirical coverage curves are stable.

\subsection{Low-accuracy regions}
\label{sec:surr_local}

If low accuracy or large uncertainties of the surrogate amplitude represent distinct failure modes, the main question is if this failure is of statistical or systematic origin. In the latter case, the network typically fails to learn a feature and the points with low accuracy and large uncertainties cluster in phase space. Identifying such clusters will allow us to improve the network training, for instance by adding more training data, see Sec.~\ref{sec:surr_sample}.

\subsubsection*{Phase space metrics}

To define low-accuracy clusters or regions, we need a distance metric on our phase space. We define two different distance metrics, one based directly on the 4-vectors of the external particles and one based on the kinematic degrees of freedom.
\begin{itemize}
\item The first metric between two phase space points with $N_f$ final-state particles is defined as
\begin{align}
  d_\text{kin}
  = \frac{
\sqrt{\displaystyle\sum_{i=1}^{N_f} \left| \vec{p}_i^{\,(1)} - \vec{p}_i^{\,(2)} \right|^2}
}{
\displaystyle\frac{1}{2N_f} \sum_{i=1}^{N_f} \left( E_i^{(1)} + E_i^{(2)} \right)
}\;,
\end{align}
The initial-state particles are excluded from the sum because they are always aligned with the beam axis. We divide by the sum of the energies because otherwise high-energy events would always be more separated from each other than low-energy ones. The advantage of this metric is that it can be defined straightforwardly. An obvious disadvantage is that it is not Lorentz-invariant, even though the underlying amplitude is. Moreover, since this metric does not account for the symmetries of the system, it can distinguish physically equivalent configurations, leading to redundant clusters. Therefore $d_\text{kin}$ was only considered as a first exploratory approach and is not used in the final study.

\item The second metric is defined in terms of given kinematic directions $x_i$,
\begin{align}
d_x = \left[ \sum_i \left(x_i^{(1)} - x_i^{(2)}\right)^2 \right]^{1/2}\;,
\end{align}
We define these directions explicitly for the $q\bar q\to t\bar t H$ and $gg\to t\bar t H$ processes in Sec.~\ref{sec:2to3}. The number of $x_i$ directions does not have to correspond to the phase space dimension, as some directions add more numerical noise than benefit. If we choose the kinematic directions as Lorentz invariants, so is the metric.
\end{itemize}
For our study, $d_x$ provides more stable and physically meaningful results, so we use it throughout Sec.~\ref{sec:2to3}. 

\subsubsection*{Clustering algorithm}
\label{sec:cluster}

To identify where amplitude results with low accuracy or large uncertainties are clustered, we use the HDBSCAN algorithm~\cite{2015arXiv150606422E,HDBSCAN}. It is a density-based clustering method that does not require a pre-defined number of clusters and that can identify outliers rather than forcing all points into clusters. 

As weights for the edges of a graph made of phase space points, the algorithm uses the reachability distance
\begin{align}
  d_{\text{mreach-}k}(a , b) =
  \text{max}\left[ \text{core}_k(a),\text{core}_k(b),d(a,b)\right] \;,
\end{align}
where $\text{core}_k(a)$ is the distance of $a$ to its $k$th nearest neighbor. Dense points with low core distances keep their original distance $d(a,b)$. For sparse points the reachability distance is at least the maximum core distance. This way, the algorithm is robust against single points forming bridges between separate clusters:
\begin{enumerate}
\item First, HDBSCAN constructs the minimum spanning tree (MST) as a unique graph representation of a set of phase space points. It involves the minimal number of edges, such that dropping any further edge disconnects the graph. Moreover, it ensures that there is no edge with a lower reachability distance that could provide an alternative connection of the phase space points.

\item We then convert the MST into an ordered list of connected clusters. For this, we sort the MST edges by their reachability distance. 
We then go through the distance values in decreasing order and cut all connections above this distance. Each edge removal splits the MST into two clusters. If one of the resulting new clusters has fewer phase space points than a predefined minimum size, it is labeled an outlier and removed from the tree. If both new clusters have more events than the minimum size, they are kept as a valid cluster split.

\item The actual clusters are selected from the above cluster candidates as those that survive the longest during this edge-removal procedure. The notion of time is the inverse distance $1/d_{\text{mreach-}k}$. If a cluster is created at inverse $d^\text{birth}_{\text{mreach-}k}$, its stability is defined as 
\begin{align}
    \text{stability} = 
    \sum_{e\in\text{cluster}} \left(
    \frac{1}{d^{(e)}_{\text{mreach-}k}} - \frac{1}{d^\text{birth}_{\text{mreach-}k}}
    \right)\;,
\end{align}
where the sum runs over all points $e$ in the cluster. Based on this stability, we select the most stable clusters. 

\item The one additional constraint we apply is that a stable cluster cannot be a descendant of an already selected cluster. This uniquely defines a set of clusters and outlier events.
\end{enumerate}

\subsection{Adaptive sampling of training data}
\label{sec:surr_sample}

Even though the training data distribution is a key ingredient to precision regression, we usually just use some kind of unweighted phase space points. Known modifications are boosted training~\cite{Badger:2022hwf} and uniform sampling~\cite{Breso:2024jlt}. In both cases, the full training dataset is still generated before we begin with the network training. Whereas boosted training reweights an existing training dataset to optimize the surrogate accuracy, expensive loop amplitudes require a maximally efficient training. Therefore, we split the generation of the training data into two sets and only start with the usual unweighted events. We then identify phase space regions where the surrogate lacks precision, using an independent test dataset or based on the learned uncertainty. For these regions, we generate additional training data from the amplitudes. 

The advantage of this adaptive procedure can be understood by considering for example a flat distribution with a dip. A flat sample will have the same density of points in the flat region as in the dip region. However, the surrogate will likely need far fewer points in the flat region to describe it to the same accuracy as the dip region. Only generating training data in the critical dip region will therefore increase the overall accuracy more than placing some of the additional points in the flat region.

\subsubsection*{Adaptive phase space sampling with kernel density estimation (KDE)}

When we generate the dataset for this second step, we need to take into account the accuracy of the surrogate in different phase space regions. The second sample should favor low-accuracy regions through an appropriate probability density function (PDF). We construct it using a weighted kernel density estimation, with weights assigned according the surrogate's accuracy. A general multivariate kernel estimator reads~\cite{scott_multivariate_2014}
\begin{align}
    f(x) = \frac{1}{\det H \cdot \sum_{i} w_i}\sum_{i}w_i\; K\left(H^{-1}(x - x_i)\right).
\end{align}
The $x_i$ are the $n$ points of the original sample, $K$ a kernel function, and $H$ the bandwidth matrix. We assign a different weight $w_i$ to each point. The kernel function and bandwidth matrix can be chosen freely, and the resulting PDF can significantly depend on their choice. However, as long as the PDF favors the correct phase space regions, we are not interested in its precise shape. This allows us to work with a simple $d$-dimensional Gaussian~\cite{gramacki_nonparametric_2018},
\begin{align}
    K(x) = \frac{1}{\sqrt{(2\pi)^d \det V}} \exp \left(- \frac{1}{2} x^T V^{-1} x\right) \; ,
\end{align}
where $V$ is the covariance matrix of the initial sample. With the data covariance being included in the kernel, we use a scalar bandwidth following Scott's rule,
\begin{align}
  h = n_\text{eff}^{-1/(d+4)}
  \qquad \text{with} \qquad
  n_\text{eff} = \frac{\left(\sum_i w_i\right)^2}{\sum_i w_i^2} \; ,
\end{align}
the effective dataset size.

Finally, we choose the absolute relative accuracy $|\Delta_i| = |\Delta(x_i)|$ of Eq.\eqref{eq:relative-precision} as weights. The kernel density estimation PDF then reads
\begin{align}
    f(x) = \frac{1}{h^d\sum_{i}|\Delta_i|} \frac{1}{\sqrt{(2\pi)^d \det V}} \sum_{i}|\Delta_i| \; \exp\left[- \frac{1}{2h^2} (x - x_i)^T V^{-1} (x - x_i)\right] 
    \; .
\end{align}
The second sample, drawn from this PDF, will cluster in regions with small surrogate accuracy and neglect regions with large accuracy. Because the Gaussian kernel falls off exponentially, it will not explore unknown regions, but only revisit regions already seen in the initial sample. Therefore, this method can only be used when the initial sample is sufficiently large.

\clearpage
\section{\texorpdfstring{$HH$}{HH} production}
\label{sec:2to2}

As a first instructive and phenomenologically relevant example, we apply our advanced amplitude learning to di-Higgs production via gluon fusion. As a $2\to 2$ process, it only depends on two independent kinematic invariants,
\begin{align}
    \hat{s} = (p_1 + p_2)^2 \qquad\text{and}\qquad \hat{t} = (p_1 - p_a)^2\;,    
\end{align}
where $p_{1,2}$ are the 4-momenta of the incoming gluons and $p_{a,b}$ are the 4-momenta of the outgoing Higgs bosons. 

The differential cross-section is given by
\begin{align}
    d\sigma = \frac{1}{2\hat s}|\mathcal{M}|^2\, d\Phi \, d\rho_{a,b}(\hat s, s)\;,    
\end{align}
where $\hat s$ is the partonic center-of-mass energy, $d\Phi$ is the element of the Lorentz-invariant phase space and $d\rho_{a,b}$ is the probability of finding partons $a$ and $b$ with total energy $\sqrt{\hat s}$ in the colliding protons. The surrogate amplitude  is defined in Eq.\eqref{eq:def_a}. Depending on the optimal pre-processing it is derived from $|\mathcal{M}|^2$ or from $d\sigma$, for our di-Higgs production setup it is 
\begin{align}
 A(x) = |\mathcal{M}(x)|^2  \; .
\end{align}    
We parameterize the ($2 \to 2$)-phase space $x$ in terms of the dimensionless quantities~\cite{Heinrich:2017kxx}
\begin{align}
    \beta_H &= \sqrt{1 - \frac{4 m_H^2}{\hat s}} \in [0, 1]
    \qquad\text{and}\qquad
    \cos\theta = \frac{\hat t - \hat u}{\hat s \beta_H}\;,
\end{align}
where $\hat u = (p_1 - p_b)^2$.

A characteristic feature, stemming from the top-quark loop mediating this process at leading order, is the (virtual) top quark pair production threshold at $\hat s = 4m_t^2$. Due to the onset of an imaginary part of the amplitude, the differential cross section as a function of $\hat s$ reaches a maximum just above this threshold. 

We work with four different datasets. The first three are generated at one-loop (LO) level using \gosam-3~\cite{Braun:2025afl}. We use $m_H = 125\,\gev$ and a fixed value of $\alpha_S = 0.1184$. The Higgs mass is measured to per-mille level, and its uncertainty translates into a negligible uncertainty on the di-Higgs amplitude.
For the first two (LO) datasets, we use $8 \cdot 10^4$ events for training and $10^5$ events for testing.
\begin{enumerate}
\item For the first dataset, we sample the center-of-mass energy uniformly in the range 
\begin{align}
    \hat s \in \left[(2m_H)^2, (10\,\tev)^2\right]\; ,
\end{align}
and fix the top-quark mass to $m_t = 173\,\gev$. 

\item A second dataset allows us to study the effects of the phase space sampling. Instead of sampling uniformly in $\hat s$, we take into account the flux factor $1/(2\hat s)$, effectively sampling from a reciprocal distribution.

\item 
Different renormalization schemes result in a range of top-quark mass values for which one potentially wants to evaluate the amplitude. While the on-shell value is about $172.5\,\gev$, the $\overline{\text{MS}}$ value depends on the renormalization scale, and the NLO amplitude shows a variation of about 20\% if the envelope between predictions in different top mass renormalization schemes is taken as an uncertainty~\cite{Baglio:2018lrj,Bagnaschi:2023rbx,Jaskiewicz:2024xkd,Davies:2025ghl}.
For our third dataset, we learn the di-Higgs amplitude without fixing the top-quark mass value, i.e.\ for uniformly sampled
\begin{align}
 m_t \in [130\,\gev, 190\,\gev] \; ,
\end{align}
corresponding roughly to $\overline{\text{MS}}$ mass values $m_t(\mu_t)$ with $\mu_t\in [250\,\gev,2\,\tev]$.
\item The fourth dataset is generated at NLO in QCD including the full top-quark mass dependence with $m_t = 173\,\gev$~\cite{Borowka:2016ehy,Borowka:2016ypz}. This dataset has previously been used as input for the interpolation grid in the corresponding \texttt{POWHEG-Box} implementation~\cite{Heinrich:2017kxx,Heinrich:2020ckp,Heinrich:2022idm}. It contains only $6320$ events.
\end{enumerate}
%

\subsection{One-loop with fixed top mass}

\begin{figure}[b!]
  \includegraphics[width=0.495\textwidth]{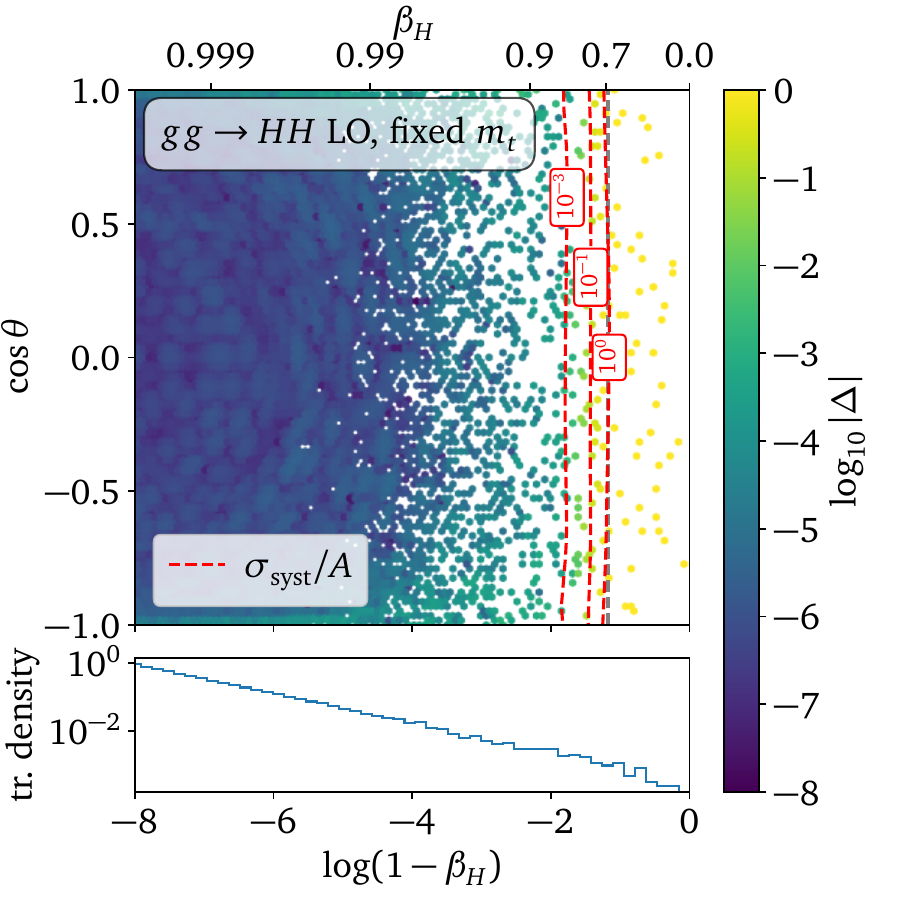}
  \includegraphics[width=0.495\textwidth]{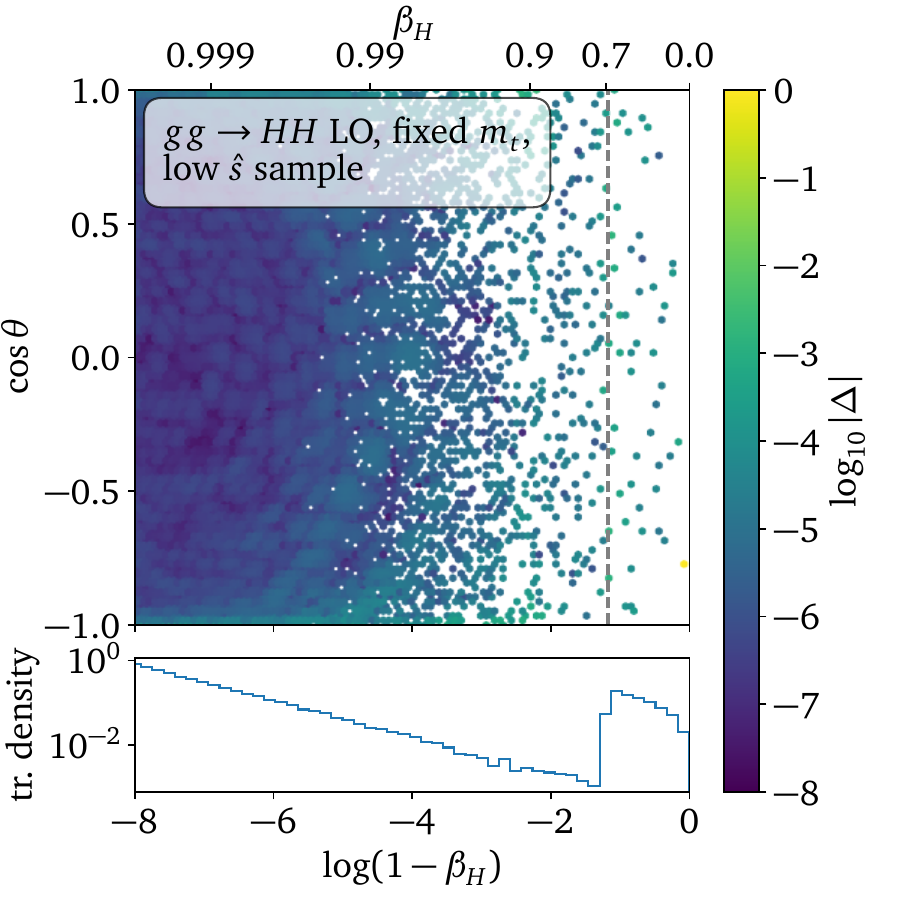}\\
  \includegraphics[width=0.495\textwidth]{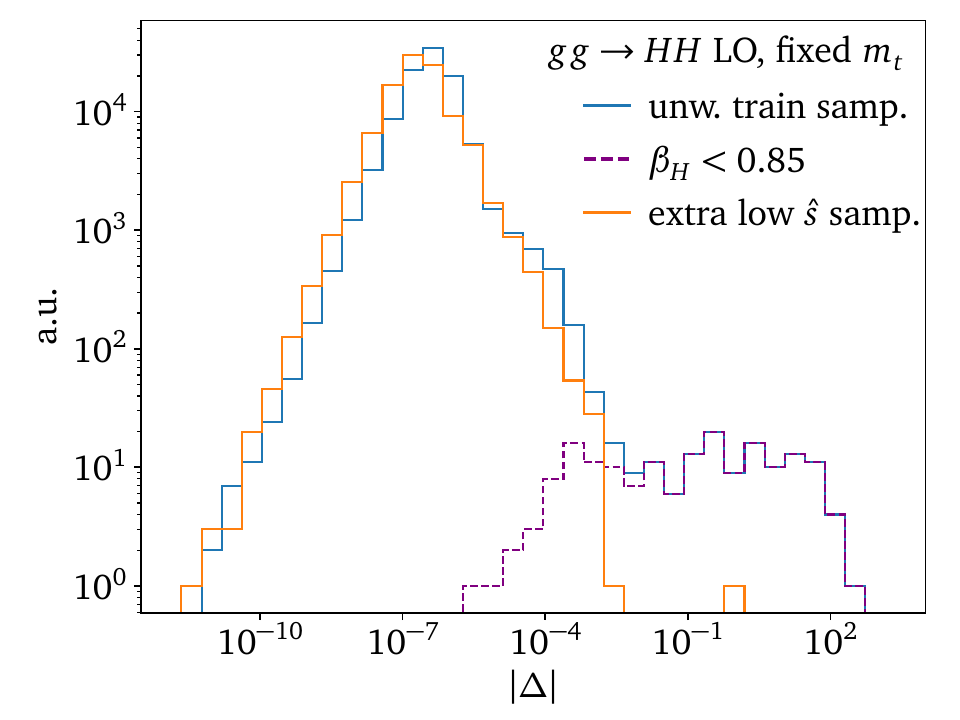}
  \includegraphics[width=0.495\textwidth]{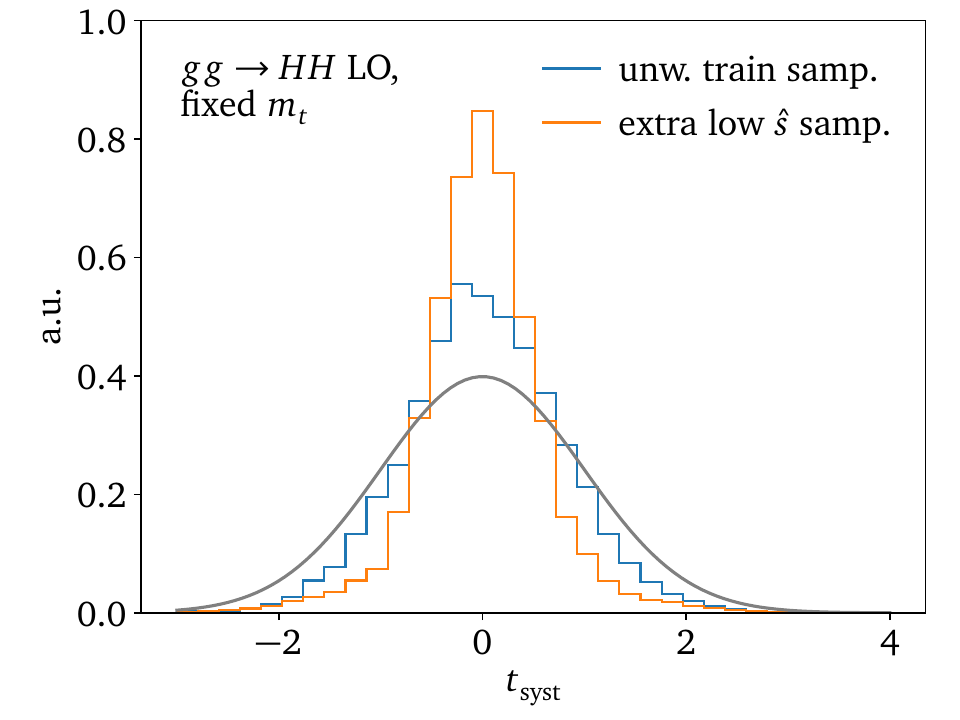}
  \caption{Upper: mean relative accuracy of the LO $HH$ surrogate with fixed $m_t$, trained on an unweighted sample (left) and including an additional low-$\hat s$ training sample (right). The sub-panels show the density of the training dataset. Lower: relative accuracy, we also show the amplitudes below the top threshold (left), and corresponding pull distributions (right).}
  \label{fig:ggHH_LO_phase_space_fixed_mt}
\end{figure}

We first investigate the mean relative accuracy over the phase space, $\Delta(x)$, after training on the first unweighted event sample in the upper left panel of Fig.~\ref{fig:ggHH_LO_phase_space_fixed_mt}. We observe no effect from the scattering angle $\theta$, but as $\beta_H$ approaches zero close to the Higgs pair production threshold, the accuracy drops. This starts for values right above the top threshold at $\beta_H\simeq 0.69$. Below this threshold, the amplitude drops significantly, the uniformly sampled training dataset becomes more sparse, and the surrogate accuracy decreases. This reduced accuracy is captured correctly by the learned uncertainty.

We can improve the training with an additional small dataset with uniform sampling in
\begin{align}
 \hat s \in [(2m_H)^2,2\cdot (2m_t)^2] \; .
\end{align} 
This low-$\hat s$ dataset includes $10^4$ events. In order to allow for a fair comparison, we moreover reduce the size of the normal training dataset to $7\cdot 10^4$ events. The performance of the surrogate trained on the low-$\hat s$-enhanced dataset is shown in the upper right panel of Fig.~\ref{fig:ggHH_LO_phase_space_fixed_mt}.  The accuracy in the low-energy region improves by almost two orders of magnitude. Indeed, a clever choice of training data significantly boosts performance. The learned uncertainty normalized by the amplitude is essentially flat across phase space, so we do not show it. 

The effect of the additional training sample can also be seen in the accuracy distributions in the lower left panel of Fig.~\ref{fig:ggHH_LO_phase_space_fixed_mt}, with a maximum around a relative accuracy of $10^{-6}$. The lower boundary around $\Delta \sim 10^{-11}$ corresponds to the numerical precision of \gosam. For the unweighted training dataset the accuracy stretches to $\Delta \sim 10^2$, where the tail includes almost exclusively phase space points below the top threshold. This maximum of the accuracy distribution is reduced by almost three orders of magnitude when we add the low-$\hat s$ training data.

\begin{figure}[b!]
  \includegraphics[width=0.495\textwidth]{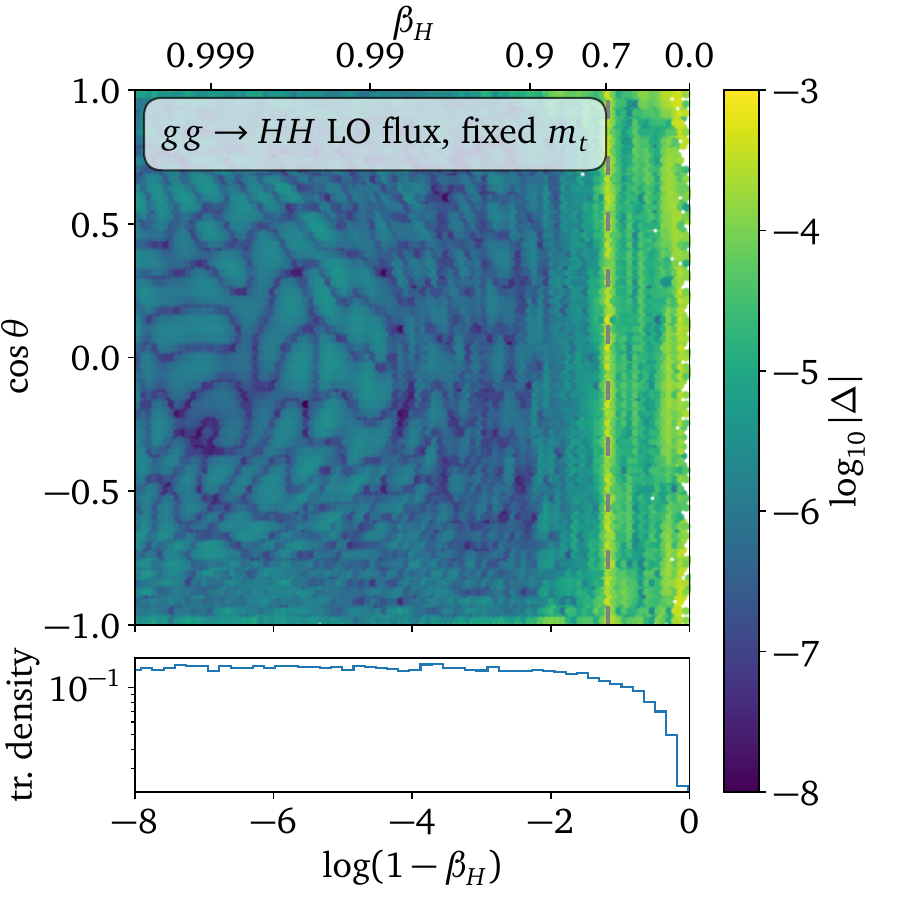}
  \includegraphics[width=0.495\textwidth]{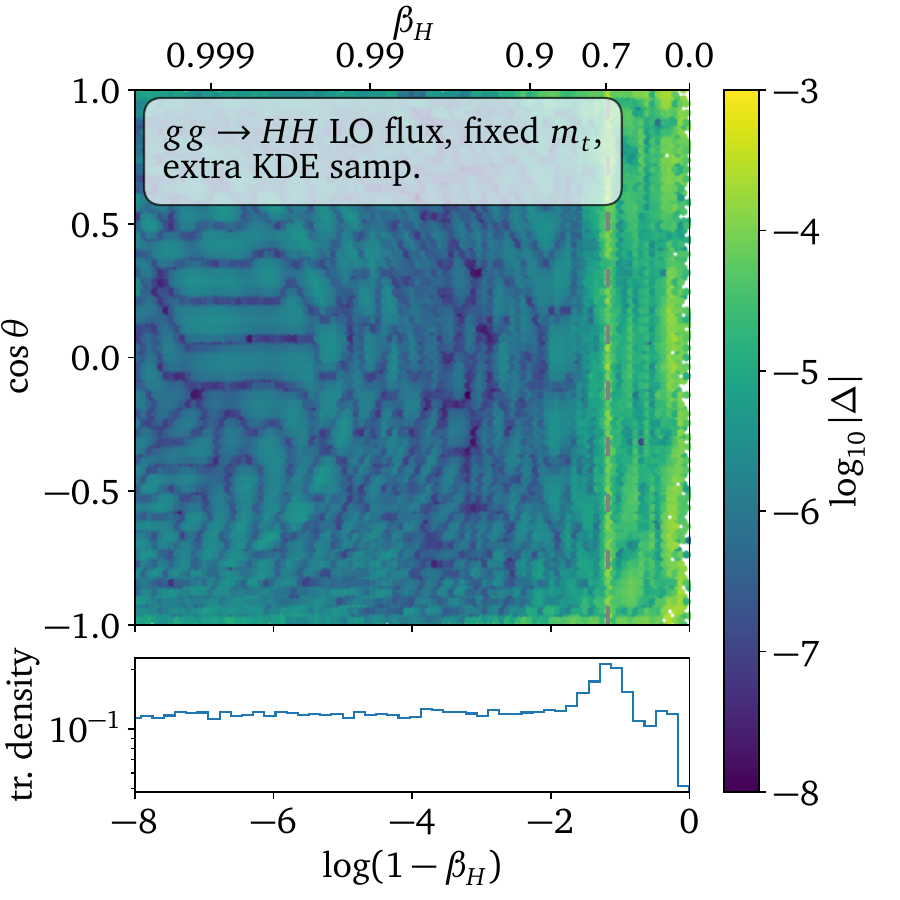} \\
  \includegraphics[width=0.495\textwidth]{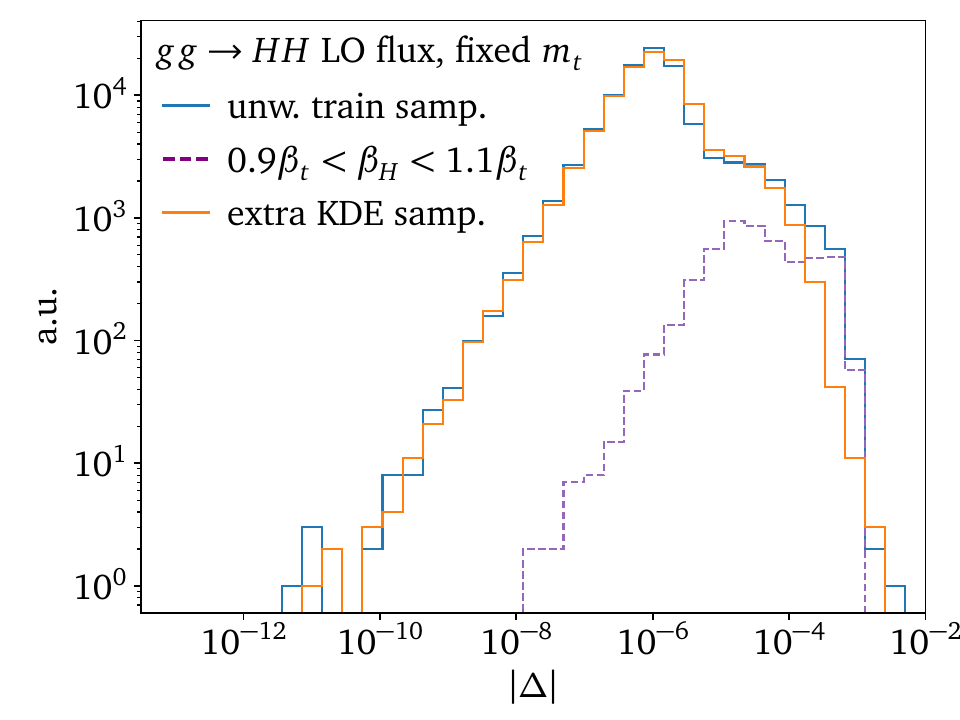}
  \includegraphics[width=0.495\textwidth]{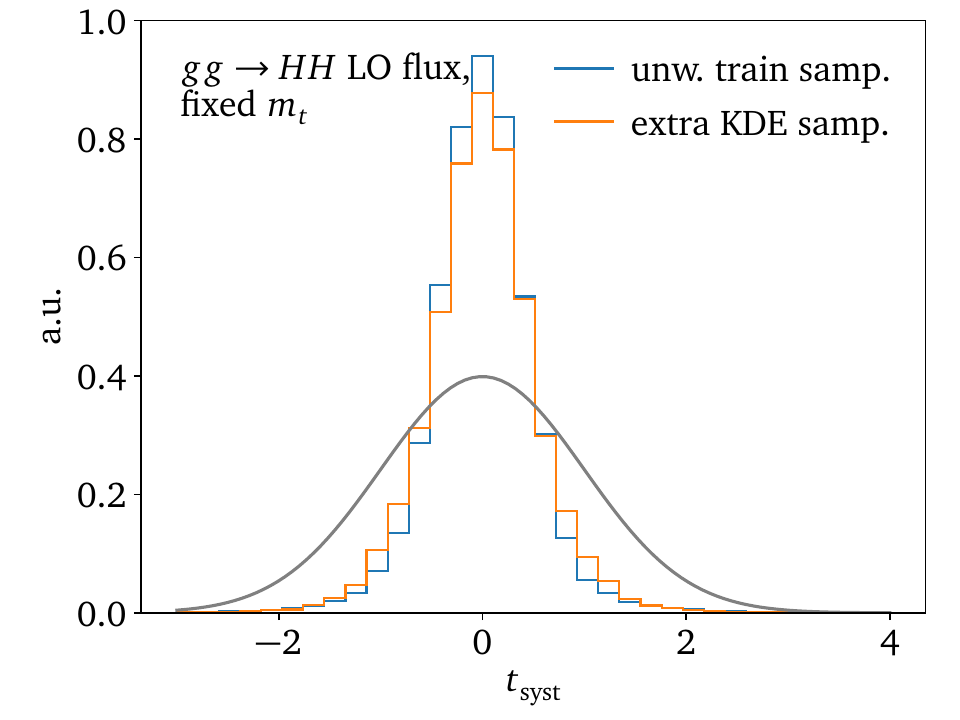}
  \caption{Upper: mean relative accuracy of the LO $HH$ surrogate with fixed $m_t$, trained on a flux-reweighted sample (left), and including an additional KDE sample for training (right). The dashed gray vertical lines indicate the top threshold. The sub-panels show the density of the training dataset. Lower: relative accuracy, we also show the amplitudes around the top threshold (left), and corresponding pull distributions (right).}
  \label{fig:ggHH_LO_flux_phase_space}
\end{figure}

For the calibration of the learned uncertainty, we look at the corresponding pull in the lower right panel of Fig.~\ref{fig:ggHH_LO_phase_space_fixed_mt}. Both distributions, without and with the additional low-$\hat s$ training data, are approximately Gaussian, validating our Gaussian ansatz for the likelihood. However, the width of the pull distribution is narrower than expected. This slightly underconfident uncertainty estimate is not improved when using the Student's $t$-likelihood.

\subsubsection*{Sampling with flux factor}

Instead of flat sampling in $\hat s$, we can account for the flux factor. This changes the density of the training dataset, in particular, at low energies. The results for this second training dataset are shown in the upper left panel of Fig.~\ref{fig:ggHH_LO_flux_phase_space}. The density of the training data is  much more evenly distributed, leading to an accuracy $\Delta \sim 10^{-5}$ for most of the phase space. Only close to the di-top threshold, indicated by the vertical dashed line, the surrogate loses accuracy. This is a consequence of the behavior of the amplitude close to the threshold, which we will study in more detail below. Below the top threshold, the accuracy improves again but does not reach the level above the threshold, due to the limited amount of training data in this region.

To improve accuracy, we generate $10^4$ additional training data points using the KDE sampling strategy outlined in Sec.~\ref{sec:surr_sample}. As always, we reduce the original training dataset, such that the surrogates with and without additional KDE sample always use $8\cdot 10^4$ training points. In the upper right panel of Fig.~\ref{fig:ggHH_LO_flux_phase_space} we see that the KDE sampling improves the accuracy at and below the di-top threshold. 

\begin{figure}[b!]
  \includegraphics[width=0.495\textwidth]{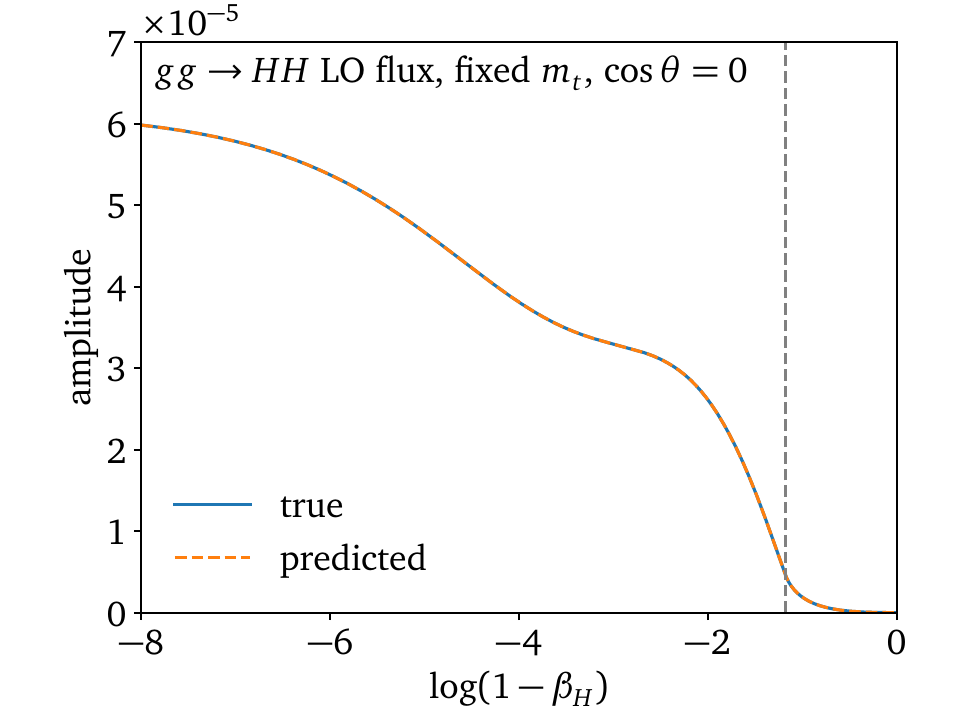}
  \includegraphics[width=0.495\textwidth]{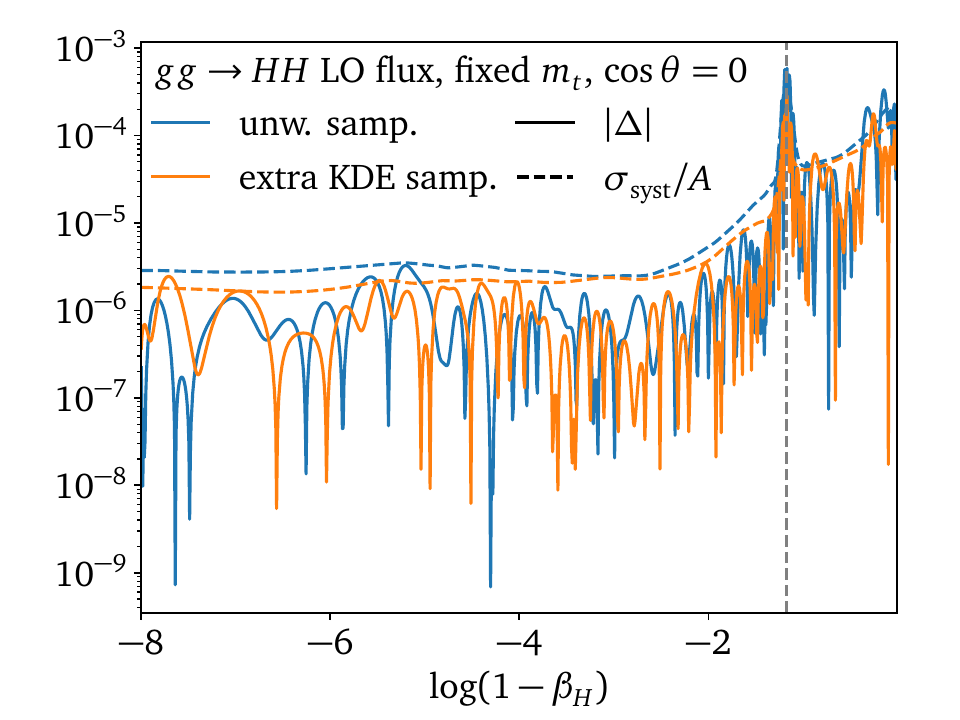}
  \includegraphics[width=0.495\textwidth]{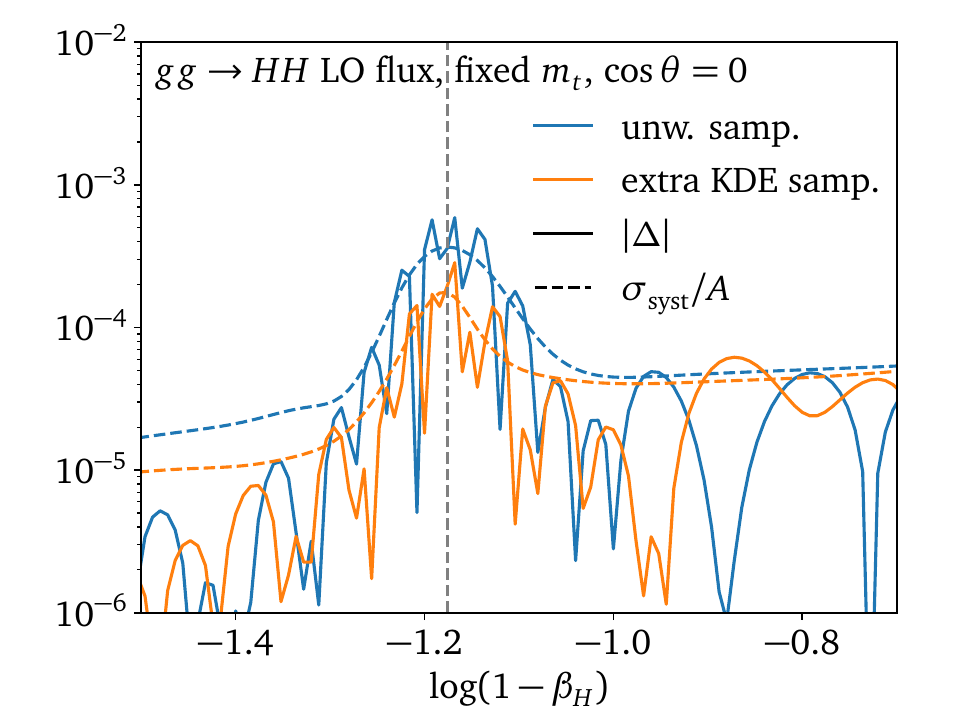}
  \includegraphics[width=0.495\textwidth]{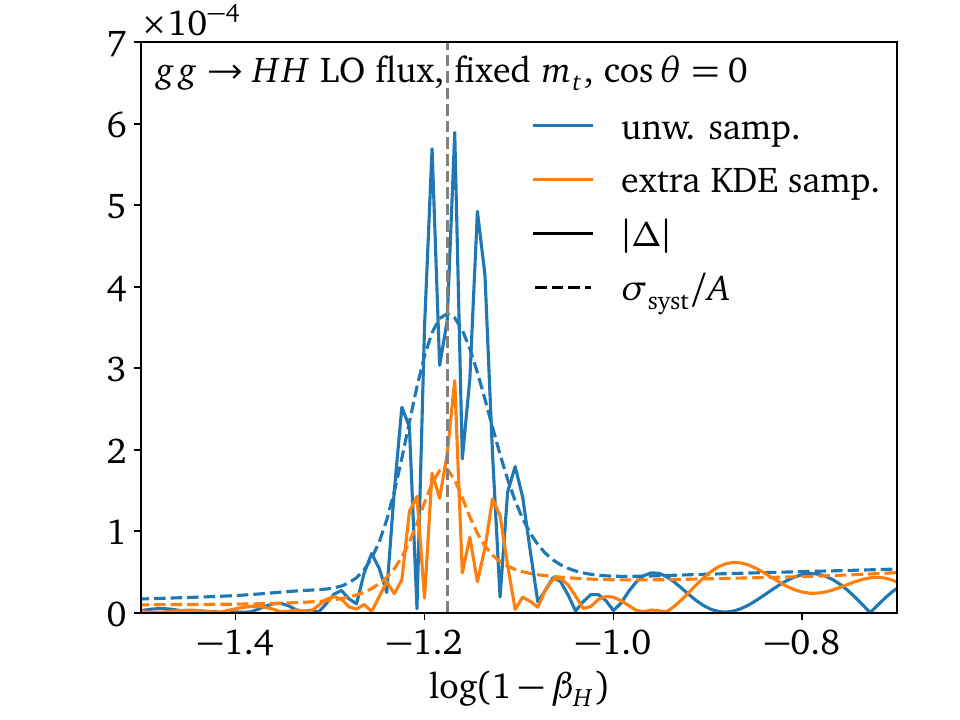}
  \caption{From upper left to lower right: kinematic distributions of the LO $HH$ amplitude and the learned surrogate with fixed $m_t$; effect of the additional KDE sampling on the relative accuracy and the relative systematic uncertainty; the same distribution, shown only around the top threshold; same as before, but with a linear $y$-axis.}
  \label{fig:ggHH_LO_flux_threshold_scan}
\end{figure}

We confirm this pattern in the $|\Delta|$ distributions in the lower left panel of Fig.~\ref{fig:ggHH_LO_flux_phase_space}. The additional KDE sampling visibly reduces the large-$\Delta$ tail, which is mainly caused by the top threshold region. 
The systematic pull shown in the lower right panel of Fig.~\ref{fig:ggHH_LO_flux_phase_space} show again that the learned uncertainties significantly overestimate the actual deviations, a behavior that does not improve when we switch to a Student's $t$-likelihood. 

Finally, we investigate the behavior of the surrogate close to the di-top threshold in Fig.~\ref{fig:ggHH_LO_flux_threshold_scan}. The upper left panel shows the true and learned amplitudes as a function of $\beta_H$ for fixed $\cos\theta = 0$. At this scale, we cannot see any deviation of the surrogate. For increasing $\log(1-\beta_H)$, the amplitude decreases, with the steepest descent below the top threshold at $\beta_H \simeq 0.69$. At the top threshold, the amplitude has a turning point before it plateaus for large $\log(1-\beta_H)$.

The steep decline is generally hard to learn by a network with a fixed typical resolution. The effect is shown in the upper right panel depicting the relative accuracy $|\Delta|$ and the relative estimated uncertainty $\sigma_\text{syst}/A$ as a function of $\log(1-\beta_H)$ comparing the surrogates trained with and without the additional KDE sample.. $|\Delta|$ fluctuates at the level of $\lesssim 10^{-6}$ for $\log(1-\beta_H) \lesssim -2$. Above this point, the steep decline lowers the accuracy, and close to the top threshold, the accuracy drops to $\sim 3\cdot 10^{-4}$. This behavior is well captured by the uncertainty estimate. In the lower two panels, we show the same curves as in the upper right panel focusing on the top threshold region. The lower left panel uses a logarithmic scaling for the $y$ axis; the right one, a linear scaling. We can see how the additional KDE sampling does not remove the accuracy drop at the top threshold, but alleviates it significantly.

\subsection{One-loop with varying top-quark mass}

\begin{figure}[b!]
  \includegraphics[width=0.495\textwidth]{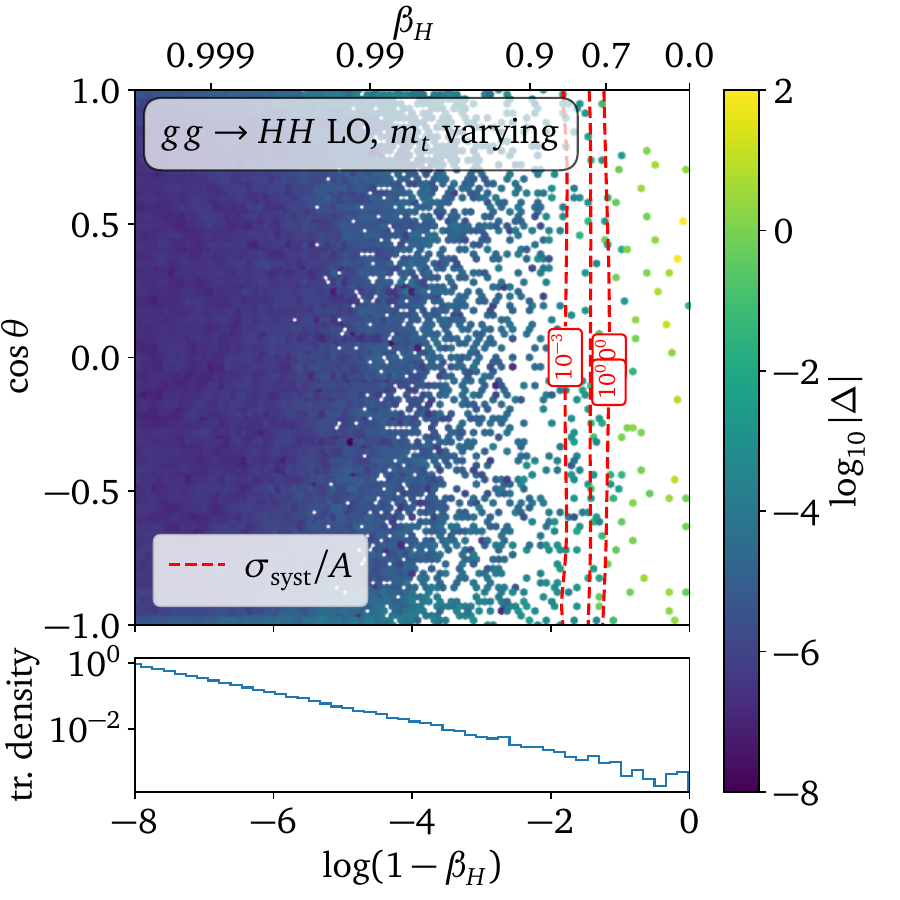}
  \includegraphics[width=0.495\textwidth]{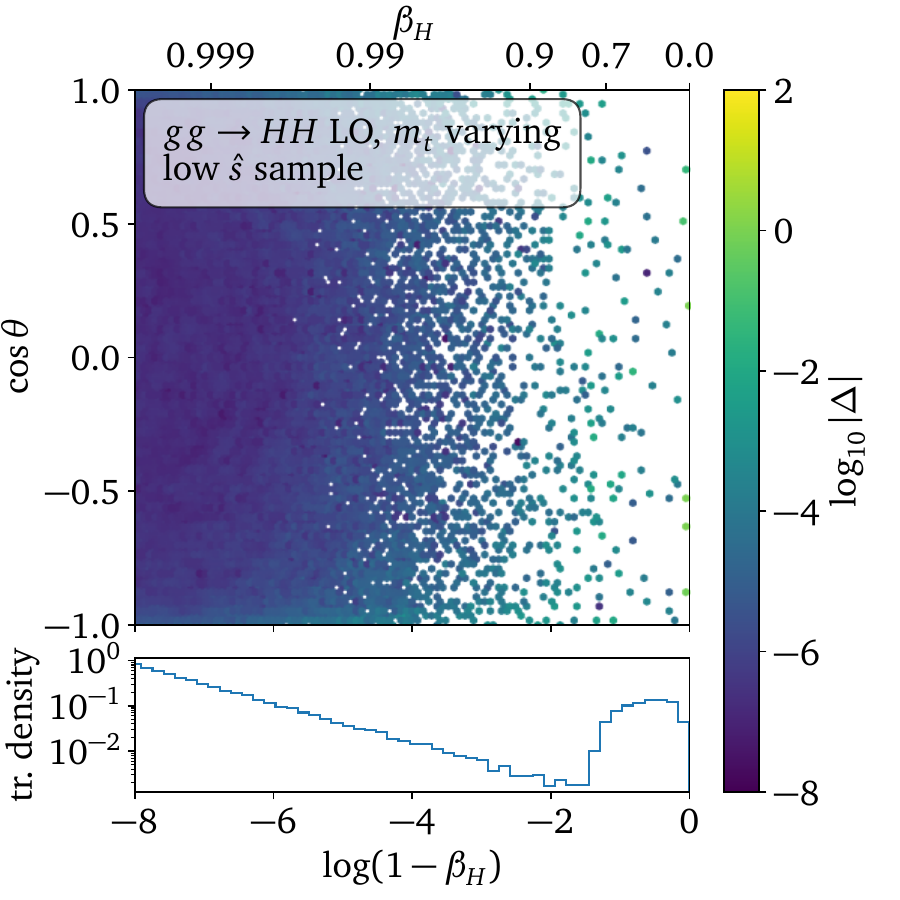} \\
  \includegraphics[width=0.495\textwidth]{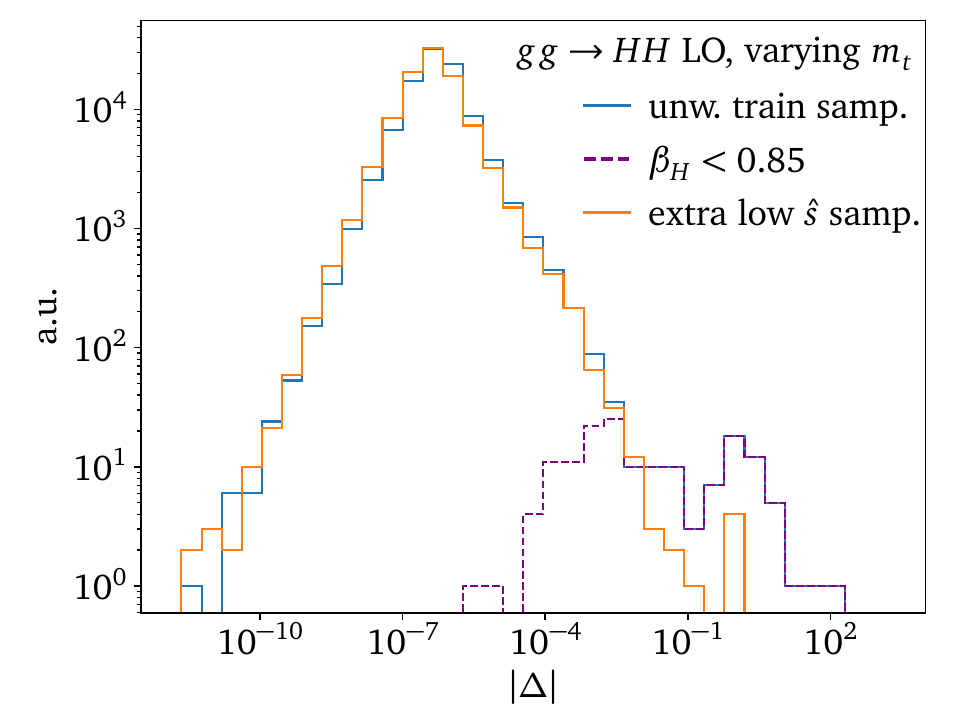}
  \includegraphics[width=0.495\textwidth]{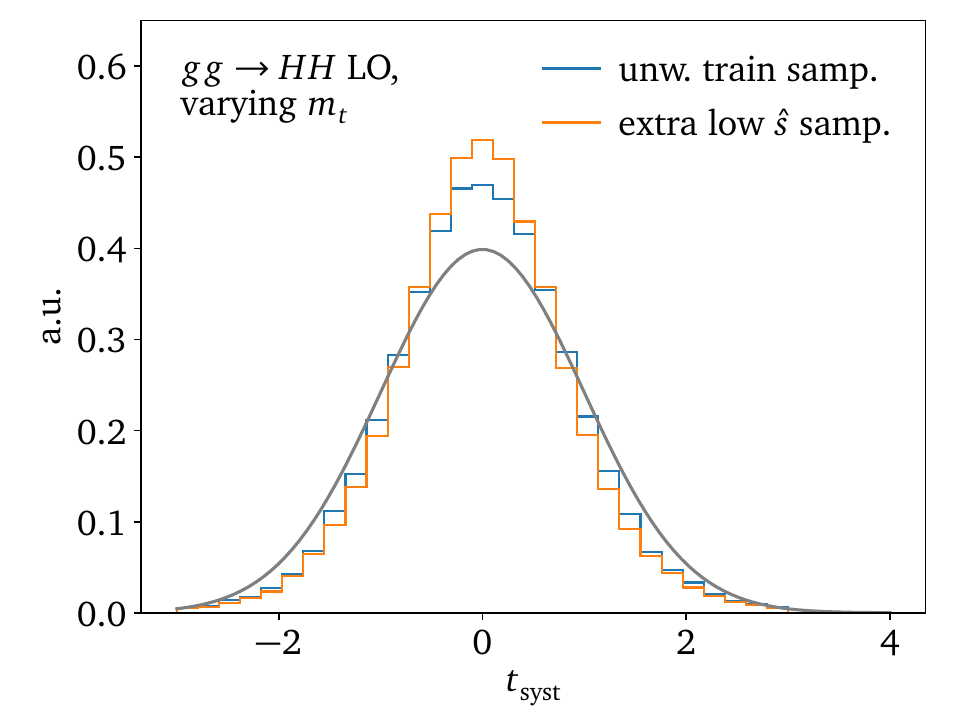}
  \caption{Upper: mean relative accuracy of the LO $HH$ surrogate with variable $m_t$, trained on an unweighted sample (left) and including an additional low-$\hat s$ sample for training (right). The sub-panels show the density of the training dataset. Lower: relative accuracy, we also show the amplitudes below the top threshold (left), and corresponding pull distributions (right).} 
  \label{fig:ggHH_LO_phase_space_varying_mt}
\end{figure}

Given the accuracy the surrogate reaches for the fixed top-quark mass, we can complicate the regression task by varying the top-quark mass and build on the network interpolation over phase space and over the (conditional) parameter input. The accuracy as a function of phase space is shown in the upper left panel of Fig.~\ref{fig:ggHH_LO_phase_space_varying_mt}. 
Compared to Fig.~\ref{fig:ggHH_LO_phase_space_fixed_mt}, we see a significantly lower accuracy in the low-energy region. An additional low-energy training sample enhances the accuracy in the low-energy region, as visible in the upper right panel of Fig.~\ref{fig:ggHH_LO_phase_space_varying_mt}. The ultimate performance of the amplitude surrogate is comparable to the one based on the fixed-$m_t$ dataset.

This is confirmed in the lower left panel of Fig.~\ref{fig:ggHH_LO_phase_space_varying_mt}. Including the low-energy sample shifts the end point of the accuracy distribution from around $10^3$ to order-one, albeit not as low as for the fixed-$m_t$ case. The peak of the relative accuracy remains around $10^{-6}$, on par with the fixed-$m_t$ case. For training without an extra low-energy sample, the tail of the distribution is again almost exclusively constituted of below top-quark threshold events. We analyze the calibration of the learned uncertainties in the lower right panel of Fig.~\ref{fig:ggHH_LO_phase_space_varying_mt}. The pull distributions for the more complex regression problem are clearly Gaussian and much better calibrated. 

\begin{figure}[b!]
  \includegraphics[width=0.495\textwidth]{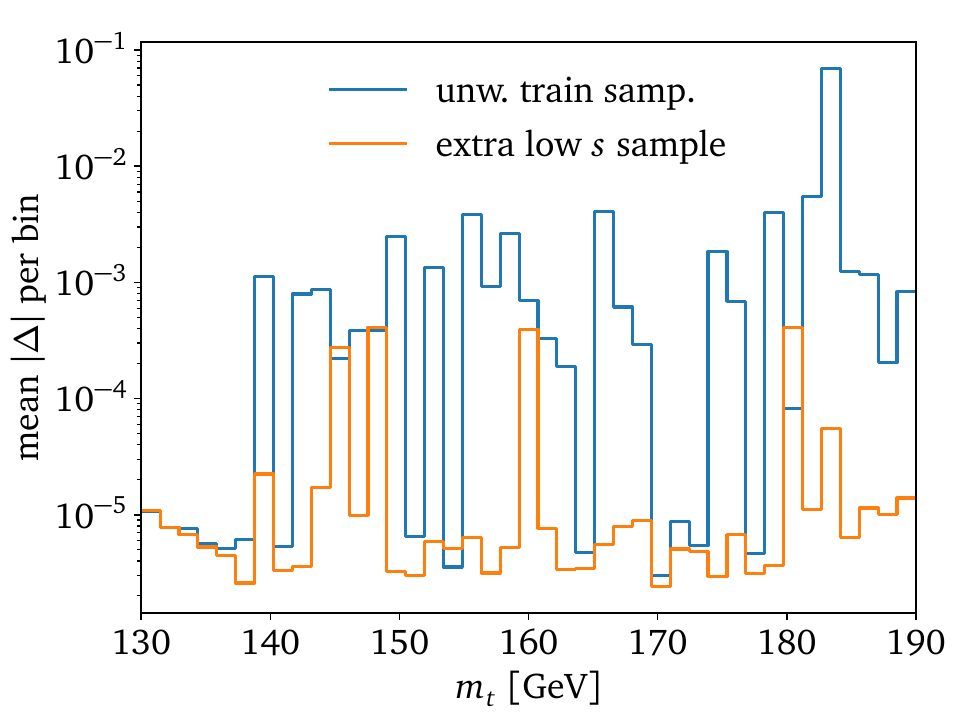}
  \includegraphics[width=0.495\textwidth]{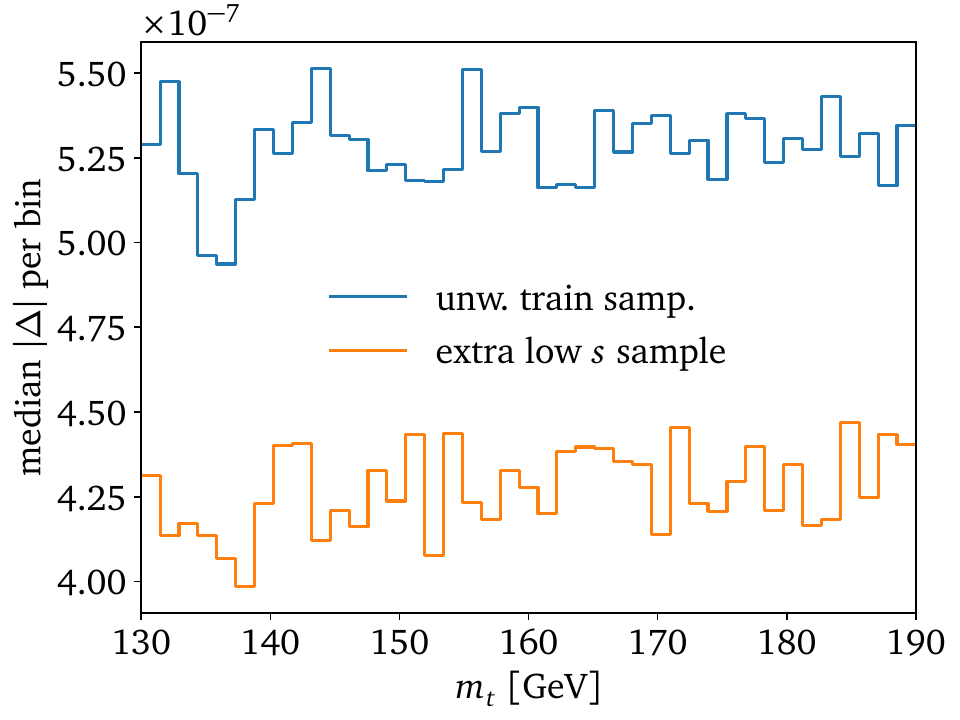}
  \caption{Left: mean relative accuracy of the LO $HH$ surrogate as a function of $m_t$, comparing the training datasets without and with an additional low-$\hat s$ training sample. Right: corresponding median relative accuracy.}
  \label{fig:ggHH_LO_mt_dep}
\end{figure}

We investigate the dependence of the accuracy on the top-quark mass in Fig.~\ref{fig:ggHH_LO_mt_dep}. In the left panel, showing the mean accuracy as a function of $m_t$, we see how the low-energy training sample improves the accuracy. It fluctuates strongly as a function of $m_t$, reflecting the random sampling. A more regular sampling strategy could help flatten the distribution. However, if an additional low-energy training sample is included, the accuracy is never worse than $\sim 5\cdot 10^{-4}$.

Comparing the mean to the median accuracy, shown in the right panel of Fig.~\ref{fig:ggHH_LO_mt_dep}, indicates that the upwards fluctuations in the mean accuracy are mainly caused by outliers. For the median, the surrogate trained using an additional low-energy sample has an almost flat accuracy, indicating that for the bulk of the test data the amplitude is predicted at the relative $10^{-6}$ level. Without the additional low-energy sample, the surrogate struggles to learn the $m_t$-dependence and the median relative accuracy decreases towards the edges of the considered $m_t$-range.

\begin{table}[t]
    \centering
    \begin{booktabs}{colspec={cc|cc|r|r}, colsep=4pt}
        \toprule
         varying $m_t$ & flux samp. &low-$\hat s$ sample & KDE sample &  \SetCell{c} mean $|\Delta|$ & \SetCell{c} median $|\Delta|$ \\
         \midrule
                &         &         &        &  $(1.3 \pm 0.3)\cdot 10^{-2}$ & $(3.8 \pm 0.5) \cdot 10^{-7}$\\
                & \cmark  &         &        &  $(9.3 \pm 1.1)\cdot 10^{-6}$ & $(1.1 \pm 0.1) \cdot 10^{-6}$\\
         \cmark &         &         &        &  $(0.9 \pm 1.0)\cdot 10^{-2}$ & $(4.7 \pm 0.3) \cdot 10^{-7}$\\ \midrule
                  &        & \cmark &         &  $(1.2\pm 0.02)\cdot 10^{-5}$ & $(2.5 \pm 0.2) \cdot 10^{-7}$ \\
                 & \cmark &       &\cmark   &   $(6.4 \pm 0.4)\cdot 10^{-6}$ & $(1.1 \pm 0.1) \cdot 10^{-6}$\\
         \cmark  &        & \cmark &         &  $(4.4\pm 0.1)\cdot 10^{-5}$ & $(4.2 \pm 0.2) \cdot 10^{-7}$\\
         \bottomrule
    \end{booktabs}
    \caption{Mean and median test accuracies of the LO $HH$ surrogate for fixed and varying $m_t$, including training with additional low-$\hat s$ and KDE samples. The average and standard deviations of five independent runs are shown.}
    \label{tab:ggHH_LO_deltaabs_comparison}
\end{table}

Finally, we compare the mean and median relative accuracies of the different LO fixed-$m_t$ and $m_t$-dependent surrogates based in Tab.~\ref{tab:ggHH_LO_deltaabs_comparison}. All values are based on five independent runs with the given uncertainty indicating the standard deviation across runs. All medians are somewhat comparable, reflecting the surrogate accuracy and precision for the bulk of phase space points. Additional sampling only has a minor effect on the performance. However, including an additional low-energy training sample or a KDE sample does significantly boost the mean accuracy, which is sensitive to outliers. This effect, which also shifts the upper endpoints in the $|\Delta|$ distributions, is more pronounced for varying $m_t$ and brings the datasets with fixed and varying $m_t$ to roughly comparable accuracy, without increasing the size of the training dataset.

\subsection{Two-loop virtual amplitude}

\begin{figure}[b!]
  \includegraphics[width=0.495\textwidth]{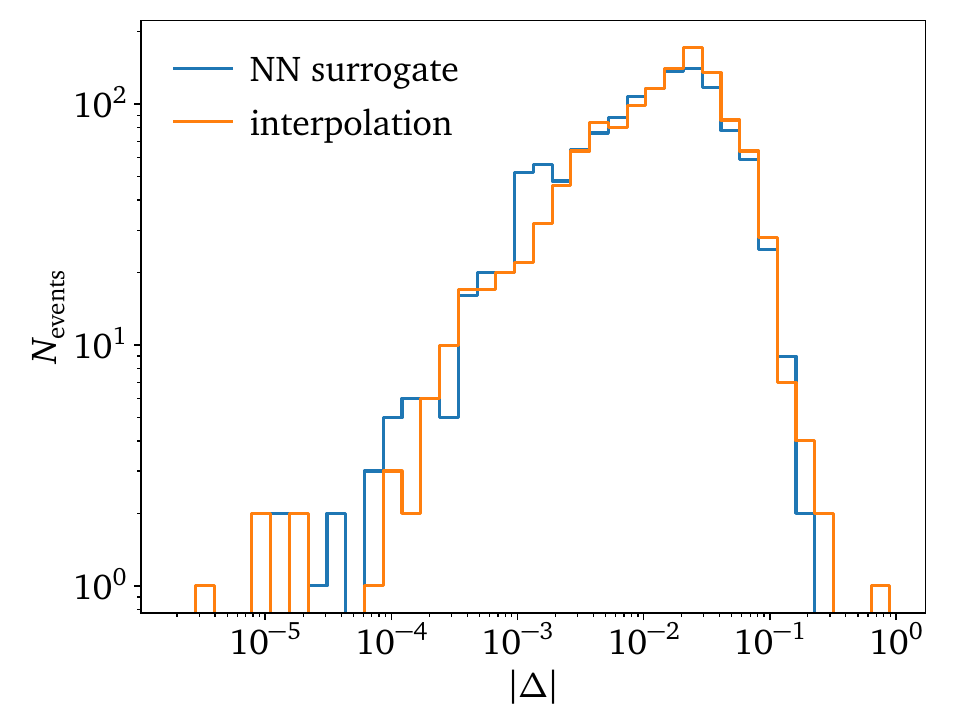}
  \includegraphics[width=0.495\textwidth]{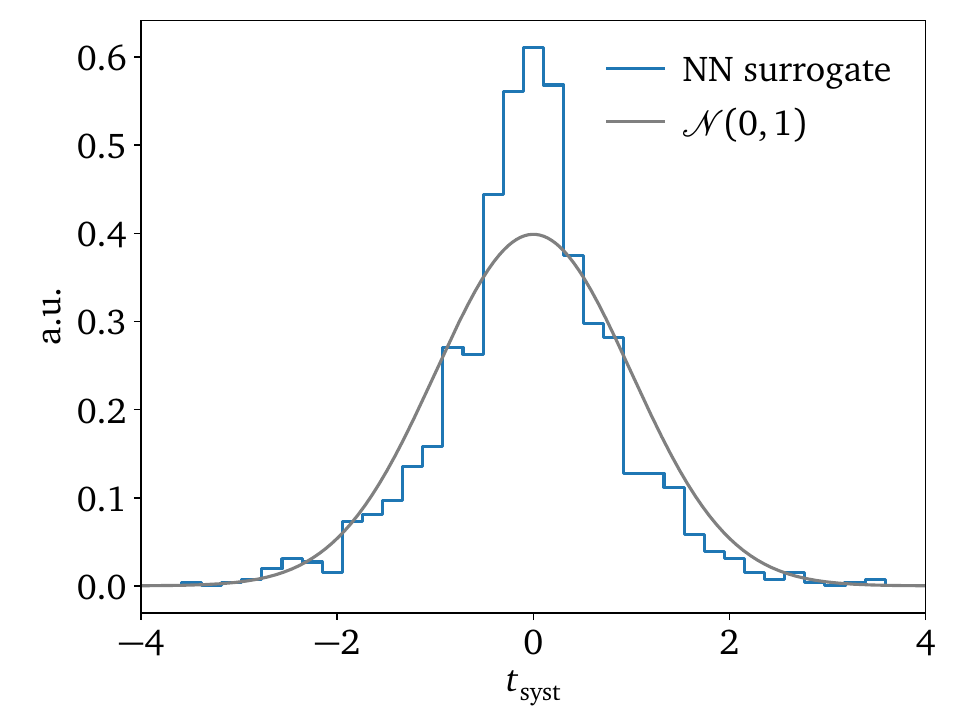}
  \caption{Left: comparison of the accuracy of the NN surrogate and a grid interpolation. Right: pull distribution of the total uncertainty predicted by the surrogate.}
  \label{fig:ggHH_NLO_deltaabs}
\end{figure}

Finally, we investigate a surrogate trained on the much smaller dataset for the two-loop virtual amplitude, interfered with the Born amplitude, UV-renormalized and after subtracting the IR singularities~\cite{hhgrid,Borowka:2016ehy,Borowka:2016ypz,Heinrich:2017kxx,Davies:2019dfy}. To fully exploit the limited training and validation datasets, we use K-fold cross-validation. The folds are obtained by holding out a fixed test set and randomly partitioning the remaining training and validation events into five equally-sized, mutually exclusive subsets ($K = 5$).

The accuracy is shown in the left panel of Fig.~\ref{fig:ggHH_NLO_deltaabs}, compared to a grid interpolation~\cite{Heinrich:2017kxx} based on convenient data pre-processing and the Clough-Tocher routine from {\tt SciPy}~\cite{scipy}. We observe a very similar performance, with the NN surrogate performing slightly better, in particular towards large relative deviations. The similar performance of the grid interpolation and the NN surrogate reflects the low phase space dimensionality and the limited amount of training data~\cite{Breso:2024jlt}. An investigation of the surrogate accuracy across phase space did not reveal any significant patterns. This is due to the relatively flat distribution of the training data.

\begin{figure}[t]
  \includegraphics[width=0.495\textwidth]{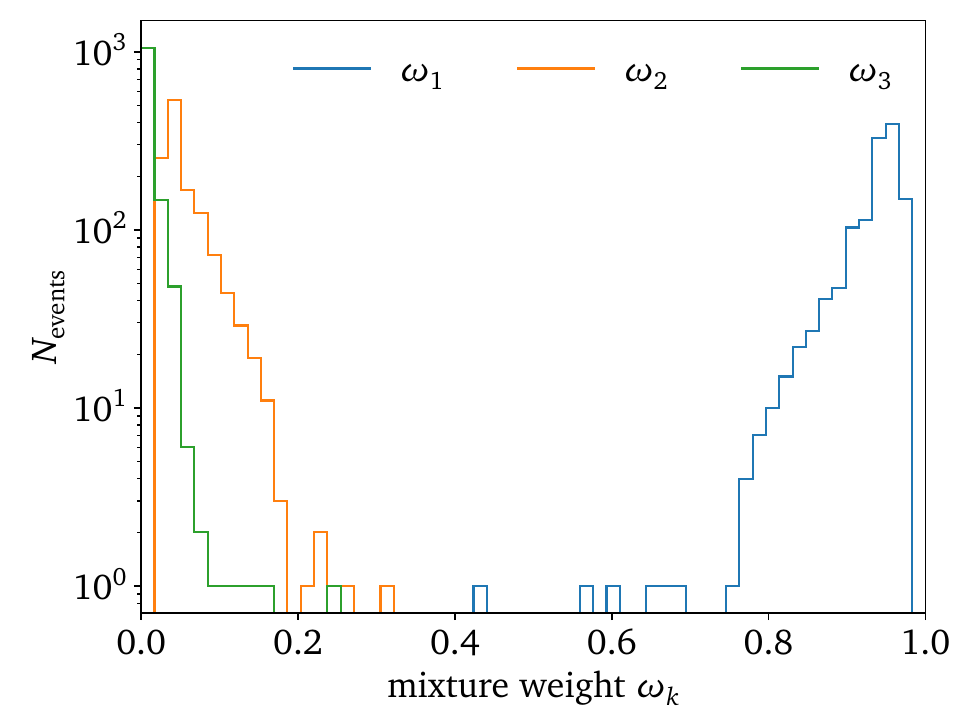}
  \includegraphics[width=0.495\textwidth]{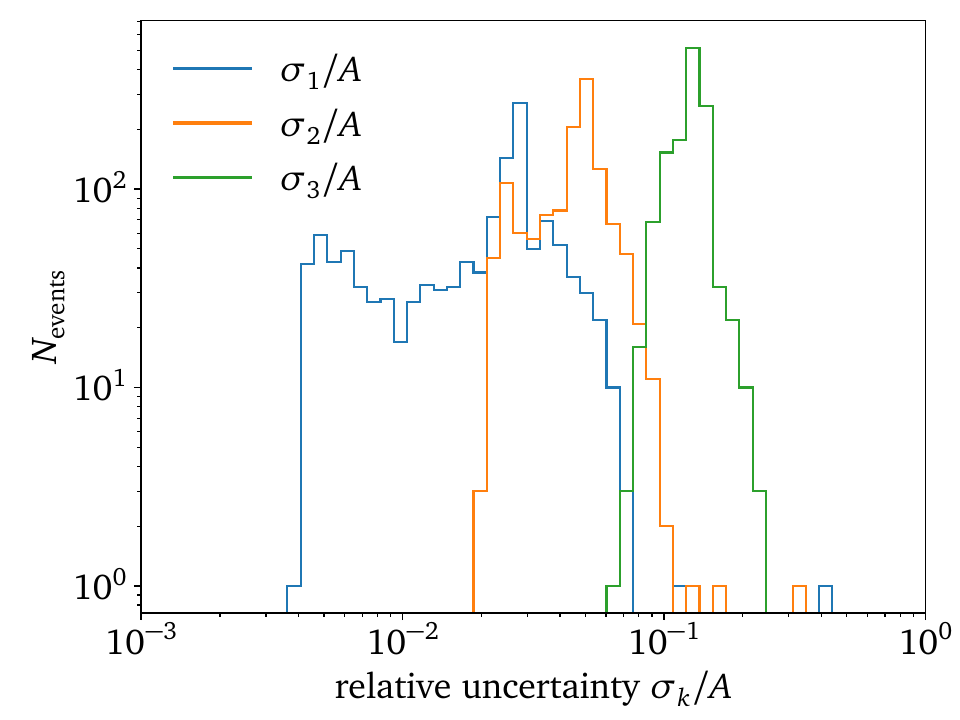} \\
  \includegraphics[width=0.495\textwidth]{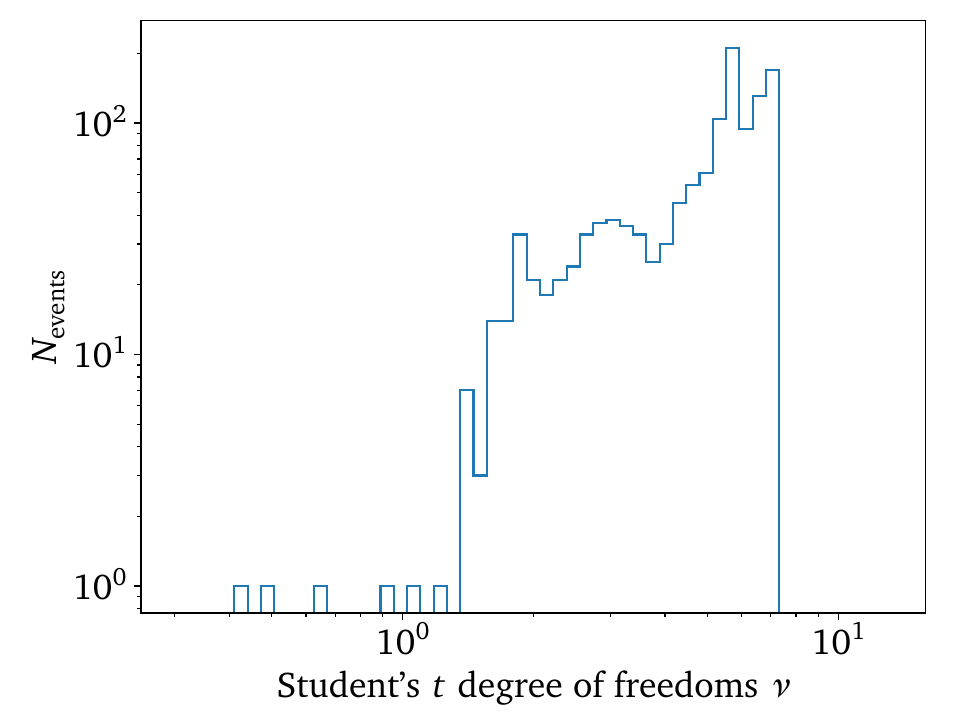}
  \includegraphics[width=0.495\textwidth]{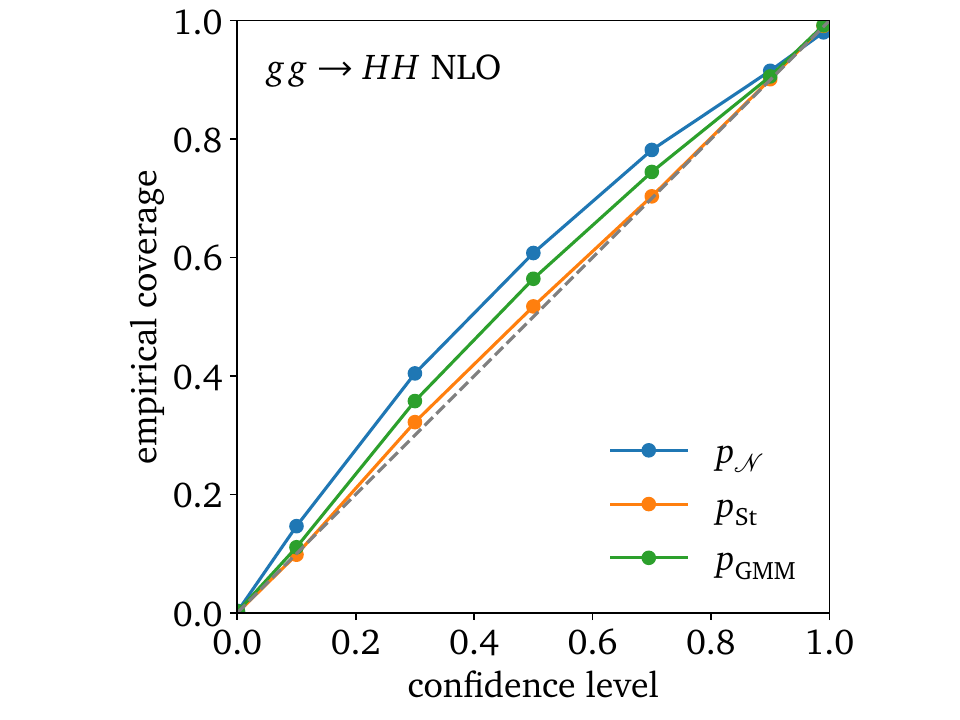}
  \caption{Non-Gaussian effect for the NLO $HH$ surrogate. We employ a GMM (upper panels) and a Student's $t$-likelihood (lower left panel). The empirical coverage is shown in the lower right panel.}
  \label{fig:ggHH_NLO_St_GMM}
\end{figure}

Unlike the grid interpolation, the NN surrogate also provides an uncertainty estimate, for which we show the pull distribution in the right panel of Fig.~\ref{fig:ggHH_NLO_deltaabs}. The shape resembles a Gaussian, but with enhanced tails. As discussed in Sec.~\ref{sec:surr_syst}, the non-Gaussianity is a consequence of the low dimensionality of the phase space. 

Consequently, we investigate a potential improvement from a GMM, with three modes having the same mean, and a Student's $t$-likelihood. The mixtures weights are shown in the upper left panel of Fig.~\ref{fig:ggHH_NLO_St_GMM}. One mode dominates, but there exists a significant contribution from one additional mode. We show the distributions of the associated learned widths in the upper right panel. The second and third mode have significantly larger widths than the dominating first mode. This shows that the second and to a lesser extent the third mode are covering the low tail of the residual distribution. 

We see a similar effect for the Student's $t$-likelihood. In the lower left panel of Fig.~\ref{fig:ggHH_NLO_St_GMM}, we see the distribution of the learned degree of freedom $\nu(x,\theta)$. Almost all values lie between 1 and 10, for which the Student's $t$-likelihood has significantly enhanced tails, as compared to a Gaussian likelihood. Finally, we show all empirical coverages in the lower right panel of Fig.~\ref{fig:ggHH_NLO_St_GMM}. The GMM slightly outperforms the poor Gaussian likelihood ansatz. However, the Student's $t$-likelihood provides almost perfect coverage, indicating an excellent calibration of the learned uncertainty all over phase space.

\clearpage
\section{\texorpdfstring{$t\bar t H$}{ttH} production}
\label{sec:2to3}

As a more challenging application, we target top-pair-associated Higgs production. In the main body of the paper we show the gluon fusion results to LO and NLO, whereas the related quark-antiquark scattering results are given in App.~\ref{app:qqttH}. For this process it has been demonstrated that NN-amplitude surrogates can outperform classic interpolation techniques~\cite{Breso:2024jlt}. Following Refs.~\cite{Breso:2024jlt,Agarwal:2024jyq}, we parametrize the $2\to 3$ phase space in terms of five dimensionless order-one variables $x_i$, as required in Sec.~\ref{sec:surr_local}. They are
\begin{itemize}
    \item distance from the production threshold $s_0 = (2m_t + m_H)^2$
    \begin{align}
        \beta^2 
        = 1 - \frac{s_0}{\hat{s}} 
        \equiv 1 - \frac{(2 m_t + m_H)^2}{(p_t+p_{\bar{t}}+p_H)^2}\;, 
        \qquad \text{where we use} \quad
        x_1 = \frac{100}{86}\left(\beta^2 - \frac{1}{10}\right)\;,
    \end{align} 
    corresponding to $\beta^2\in [0.1,0.96]$, or  $\hat{s}\in [480\,\gev, 2.4\,\tev]$, which is similar to the range accessible by experiment, see also Fig.~2 of Ref.~\cite{Agarwal:2024jyq}. 
    \item energy fraction carried by the $t\bar{t}$ system 
    \begin{align}
        x_2 
        = \frac{s_{t\bar{t}} - 4 m_t^2}{(\sqrt{\hat{s}} - m_H)^2 - 4 m_t^2}
        \qquad \text{with} \qquad  
        s_{t\bar{t}} = (p_t + p_{\bar{t}})^2 \; .
    \end{align}
    \item polar angle of the Higgs boson relative to the beam axis 
    \begin{align}
        \cos\theta_H = \frac{p_{H,3}}{\sqrt{E_H^2 - m_H^2}}
        \qquad \Rightarrow \qquad         
        x_3 = \frac{\theta_{H}}{\pi}  \; .
    \end{align}
    \item polar angle of the top relative to the $H$-$t\bar{t}$ axis in the $t\bar{t}$ system 
    \begin{align}
        \cos\theta_t = \frac{\sqrt{s_{t\bar{t}}}\; \left(E_H - \frac{1}{2}\sqrt{E_H^2 - m_H^2 + s_{t\bar{t}}}\right)}{2\;\sqrt{s_{t\bar{t}} - 4m_t^2}\;\sqrt{E_H^2 - m_H^2}}
        \qquad \Rightarrow \qquad         
        x_4 = \frac{\theta_{t}}{\pi}  \; .
    \end{align}
    \item azimuthal angle of the top relative to the $H$-$t\bar{t}$ axis in the $t\bar{t}$ system 
    \begin{align}
        \tan\varphi_t = \frac{p_{H,2}}{p_{H,3} \sin\theta_H + p_{H,1} \cos\theta_H} 
        \qquad \Rightarrow \qquad         
        x_5 = \frac{\varphi_t}{2\pi} \; .
    \end{align}
\end{itemize} 
For the partonic sub-process 
\begin{align}
 gg \to  t \bar{t} H
\end{align}
we train the amplitude surrogate on two different gluon-fusion datasets, the tree-level amplitude $f_3 = |\mathcal{M}_0|^2$  and the finite part of the one-loop amplitude $f_4 = 2\,\text{Re}(\mathcal{M}_0^\dag\mathcal{M}_1)$.
All amplitudes are evaluated using \gosam. No additional preprocessing is applied to 
$f_3$, while the NLO amplitudes are normalized to the corresponding tree-level amplitude, so we actually learn
\begin{align}
  A_\text{LO} = f_3
  \qqquad \text{and} \qqquad 
  A_\text{NLO} = \frac{f_4}{f_3}\,.
\end{align}
As for di-Higgs production, we use $8\cdot 10^4$ phase space points to train each amplitude and $10^5$ phase space points for testing. The training data corresponds to an unweighted event sample generated with densities according to pseudo parton distribution functions~\cite{Breso:2024jlt}. They mimic the behavior of proper parton distribution function sets in the suppression or enhancement of certain phase space regions. 
This preprocessing allows the surrogate training to focus on relevant phase space regions.

\subsection{Singularity structure}
\label{sec:ttH_singularities}

The physical singularities of the $t\bar{t}H$ amplitudes provide a challenge to the network training, which can be alleviated through an appropriate preprocessing. 
Even though ultraviolet and infrared singularities are removed from our target amplitudes, 
there can still remain integrable singularities, which manifest themselves by steep localized amplitude patterns.

\subsubsection*{Forward and backward regions}

If one of the top quarks is parallel to the beam axis, the amplitude approaches a singularity which is protected by the top quark mass. In the high-energy limit, the top quark mass becomes negligible, $m_t^2 \ll \hat s$, and the amplitude rises steeply.
Using the above parametrization, the transverse momentum of the top quark is 
\begin{align}
    p^2_{t,T} ={}& \left[-|\vec p_t| \sin\theta_t \cos \phi_t \cos \theta_H - \left(\frac{1}{2}\beta_B \sqrt{s_{t \bar{t}}} + \gamma_B |\vec p_t| \cos \theta_t\right) \sin \theta_H \right]^2 \nonumber\\
    & + \left[|\vec p_t| \sin\theta_t \sin \phi_t\right]^2 \;,
\end{align}
where $\beta_B = |\vec p_{H,t\bar t}|/\sqrt{s_{t \bar{t}}}$, $\gamma_B = \sqrt{1 + \beta_B^2}$, and $p_{H,t\bar t}$ is the Higgs 3-momentum in the $t\bar t$ rest frame.

This forward or backward singularity is most severe in the high-energy limit, $\hat s\gg m_t^2$, and if most of the energy is carried by the $t\bar t$ system, $s_{t\bar t} \simeq \hat s$. In this regime, $\beta_B \simeq 0$ and $\gamma_B \simeq 1$, and thereby,
\begin{align}
    p^2_{t,T} \simeq |\vec p_t|^2\left[\left(\sin\theta_t \cos \phi_t \cos \theta_H + \cos \theta_t \sin \theta_H \right)^2 + \left( \sin\theta_t \sin \phi_t\right)^2\right]\;.
\end{align}
The transverse momentum is minimized either for $\sin\theta_t\simeq 0$ or for $\sin\phi_t\simeq 0$. In the first case, also $\sin\theta_H\simeq 0$ has to fulfilled, implying that $\theta_t,\,\theta_H \simeq 0,\,\pi$. In the second case, $\phi_t \simeq 0,\,\pi$ means that we need to solve either
\begin{align}
    \sin \theta_t \cos \theta_H + \cos \theta_t \sin \theta_H &= 0  \qquad\text{for}\qquad \phi_t\simeq 0
    \qquad \text{or} \notag \\
    \sin \theta_t \cos \theta_H - \cos \theta_t \sin \theta_H &= 0  \qquad\text{for}\qquad \phi_t\simeq \pi  \; .
\end{align}
This implies 
\begin{align}
 \theta_t =
 \begin{cases} \pi - \theta_H 
 \qquad &\text{for} \qquad \phi_t\simeq 0 \notag \\
 \theta_H
 \qquad &\text{for} \qquad \phi_t\simeq\pi \; .
\end{cases}
\end{align}  
These two solutions also incorporate the first case, for which $\sin\theta_t \simeq 0$ to begin with. In terms of the $x_i$, the limits correspond to
\begin{align}
 x_3 = x_4
 \qquad \text{or} \qquad 
 x_3 = 1 - x_4  \; .
\label{eq:define_fb}
\end{align}
In the first case, the top quark is moving in negative $\hat z$ direction; in the second case, in positive $\hat z$ direction. This derivation is valid analogously for the anti-top quark.

\subsubsection*{Coulomb singularities}

A second key feature, starting at NLO, is a Coulomb-type singularity (Sommerfeld enhancement~\cite{Sommerfeld}) in the limit
$x_2 \to 0$.
By default, we subtract the Coulomb singularity from the NLO amplitudes~\cite{Breso:2024jlt}, based on the results of Ref.~\cite{Beenakker:2002nc}. For the $q\bar q \to t\bar t H$ NLO amplitude, this reads explicitly
\begin{align}
    2 \text{Re}\left[\braket{\mathcal{M}_0^{q\bar q t\bar t H}|\mathcal{M}_1^{q\bar q t\bar t H}} + \frac{\pi^2}{\beta_{t\bar t}}\braket{\mathcal{M}_0^{q\bar q t\bar t H}|\mathbf{T}_{t\bar t}|\mathcal{M}_0^{q\bar q t\bar t H}}\right]\;,
\end{align}
where $\mathbf{T}_{t\bar t}$ is the product of the $t$ and $\bar{t}$ color operators and $\beta_{t\bar t}=\sqrt{1-4m_t^2/s_{t\bar t}}$~\cite{Beenakker:2002nc}. In analogy, the NLO amplitude for $gg \to t\bar t H$ gives us 
\begin{align}
    2 \text{Re}\left[\braket{\mathcal{M}_0^{gg t\bar t H}|\mathcal{M}_1^{gg t\bar t H}} + \frac{\pi^2}{\beta_{t\bar t}}\braket{\mathcal{M}_0^{gg t\bar t H}|\mathbf{T}_{t\bar t}|\mathcal{M}_0^{gg t\bar t H}}\right]\;.
\end{align}
Due to the color structure the Coulomb singularity is more severe for gluon-induced production than for quark-induced production. In particular, Ref.~\cite{Beenakker:2002nc} finds the following contributions to the partonic cross-section, singular in the limit $\beta_{t\bar t}\to 0$,
\begin{align}
    \sigma_\text{Coul} = \frac{8\alpha_s}{3} & \sqrt{\frac{2m_t}{2m_t + m_H}}\frac{C_\text{Coul}}{\beta_{t\bar t}}\;, \notag \\
    \text{where} \qquad
    C_\text{Coul}^{q\bar q} &= -\frac{1}{6} \;,\notag \\
    C_\text{Coul}^{gg} &= \frac{11}{42}\frac{[4 m_t^2 - m_H^2]^2 
    - \dfrac{9}{11}m_H^4}{[4 m_t^2 - m_H^2]^2  + \dfrac{9}{7} m_H^4} \simeq 0.25\;.
    \label{eq:coulomb_sing_strength}
\end{align}
If not mentioned otherwise, the Coulomb singularity is subtracted from the NLO amplitudes, and for $pp\to t\bar{t} H$ production, this singularity is also suppressed by the phase space. However, we will also consider learning amplitudes without Coulomb subtraction, to show that such features can be learned reliably and independently of phase space factors.

\subsection{Leading order \texorpdfstring{$gg\xrightarrow{} t \bar{t} H $}{gg->ttH}}

\begin{figure}[t]
  \includegraphics[width=0.495\textwidth]{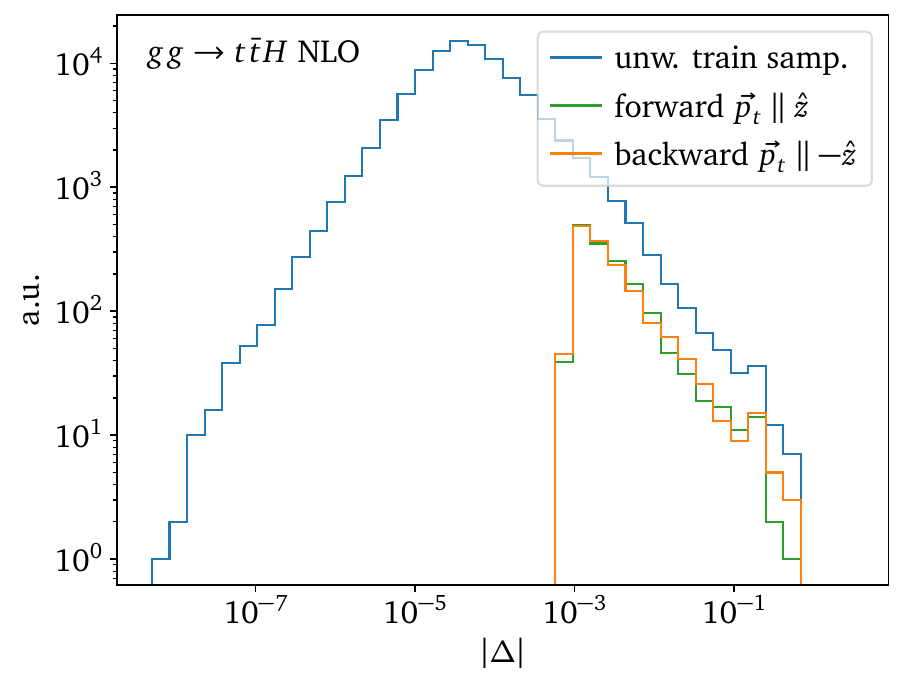}
  \includegraphics[width=0.495\textwidth]{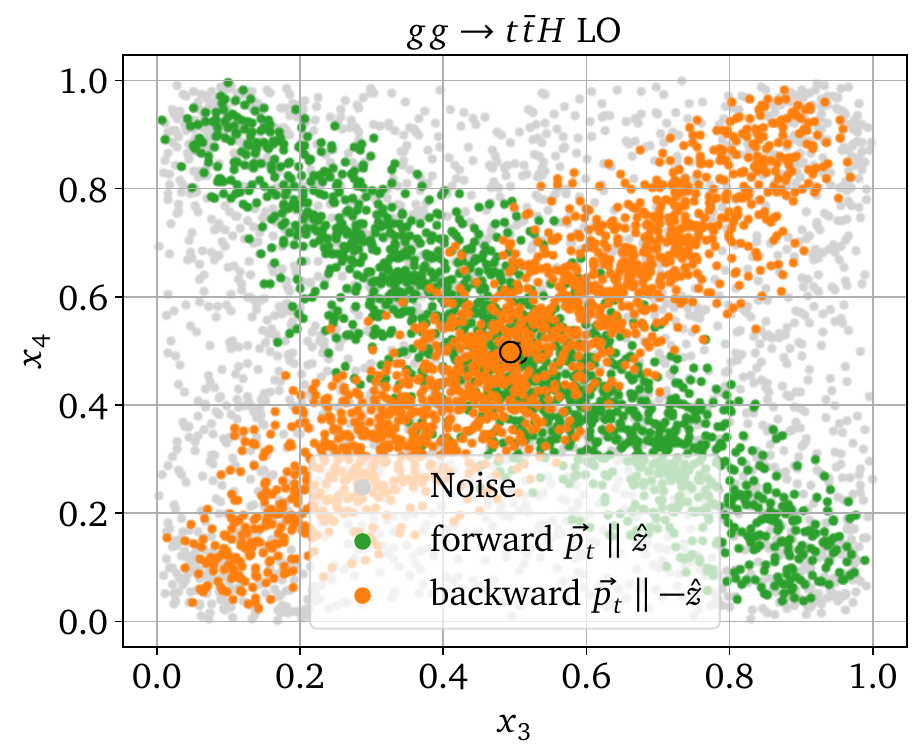} \\
  \includegraphics[width=0.495\textwidth]{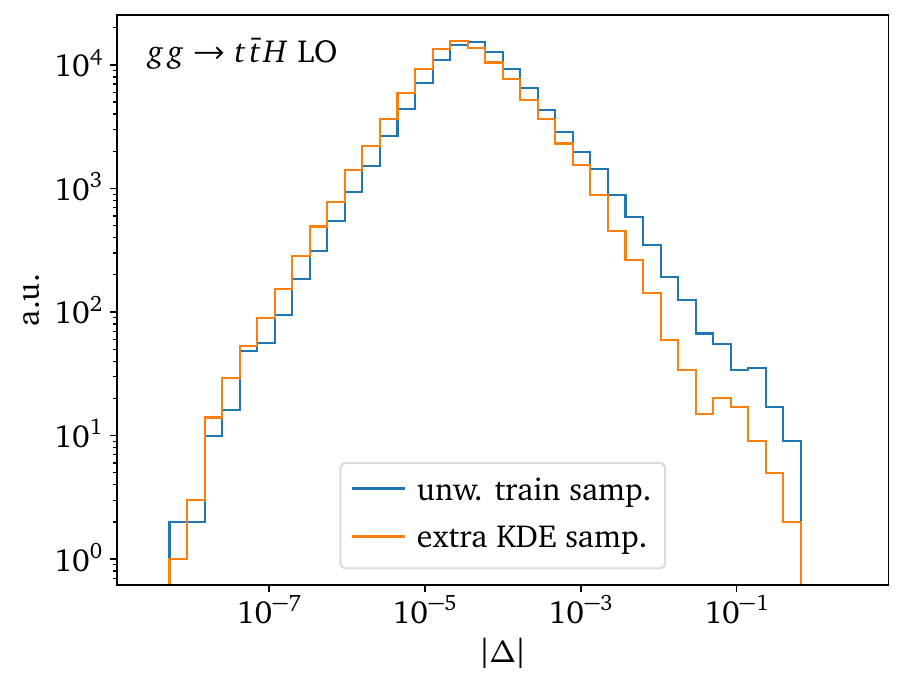}
  \includegraphics[width=0.495\textwidth]{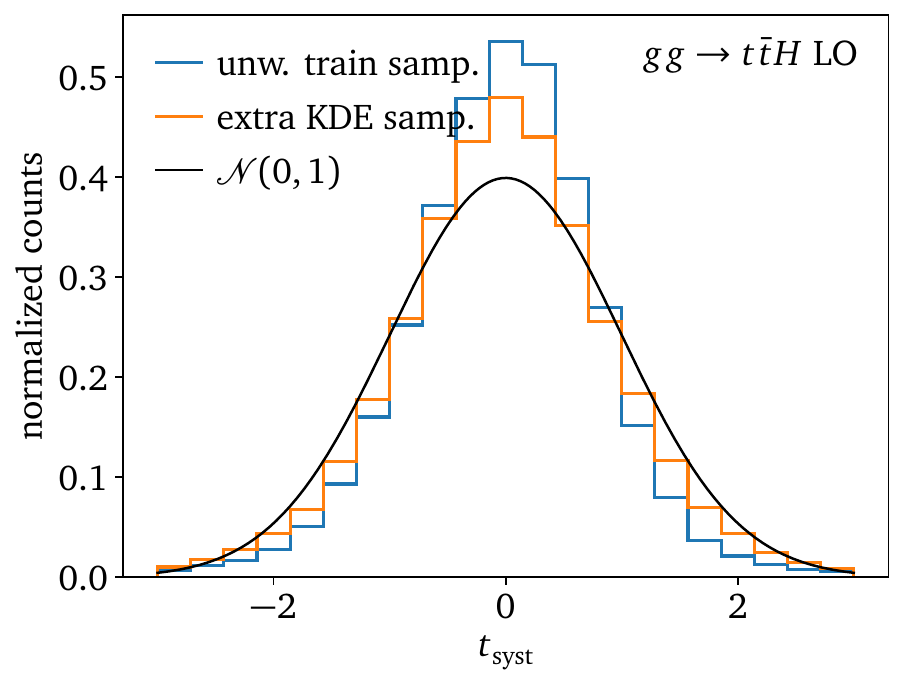}
  \caption{Upper: relative accuracy of the $g g \to t \bar{t}H$ LO surrogate, highlighting the contribution of the identified low-accuracy clusters (left) and clustering representing the forward and backward singularities (right). Lower: relative accuracy (left) and pull distributions (right) without and with additional KDE training. }
  \label{fig:ggttH_LO}
\end{figure}

We show the relative accuracy for the $gg \xrightarrow{} t \bar{t} H $ amplitude in the upper left panel of Fig.~\ref{fig:ggttH_LO}, where ``a.u.'' is used as an abbreviation for ``arbitrary units''. The distribution peaks around $|\Delta|\sim 10^{-4}$ and the upper tail stretches to $|\Delta|\sim 1$. As described in Sec.~\ref{sec:cluster}, we search for clusters in the upper 5\% percentile of the $|\Delta|$ distribution. The clustering algorithm identifies two clusters, which are most easily understood as diagonals in the $(x_3, x_4)$ plane, shown in the upper right panel of Fig.~\ref{fig:ggttH_LO}. These kinematic configurations correspond to the forward/backward scattering pseudo-singular regions, where the top-quark moves into positive  or negative beam direction. Each cluster contains approximately $30\%$ of the $10\%$ of phase space points with the highest $|\Delta|$. The remaining $40\%$ low-accuracy points are not clustered in phase space.

Next, we cover low-accuracy regions with an additional training sample using the KDE algorithm described in Sec.~\ref{sec:surr_sample}. This additional dataset contains $1.5 \cdot 10^4$ events. To provide a fair comparison, we reduce the normal unweighted training dataset to $6.5\cdot 10^4$, such that the training dataset has the same size as without the KDE sample. The resulting accuracy distributions are shown in the lower left panel of Fig.~\ref{fig:ggttH_LO}, with a visible improvement in the large-$\Delta$ tail. We confirmed that the probability density constructed by the KDE also features the cross-like structure in the $(x_3,x_4)$-plane, see App.~\ref{app:kde_densities}.

Subsequently, we test the calibration of the learned uncertainties. As visible in the lower right panel of Fig.~\ref{fig:ggttH_LO}, the pull distributions of the surrogates with and without the additional KDE sample are reasonably well compatible, and the approximately unit Gaussian distribution pull confirms the Gaussian ansatz. We also checked that the Student's $t$-likelihood does not improve the empirical coverage.

\subsection{Next-to-leading order \texorpdfstring{$gg\xrightarrow{} t \bar{t} H $}{gg->ttH}}

\begin{figure}[t]
  \includegraphics[width=0.495\textwidth]{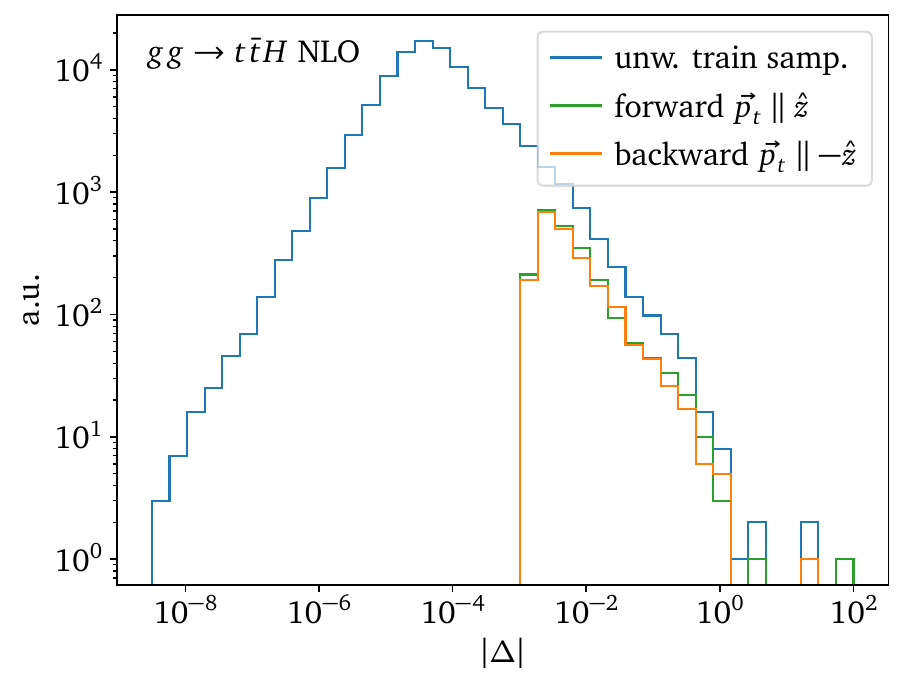}
  \includegraphics[width=0.495\textwidth]{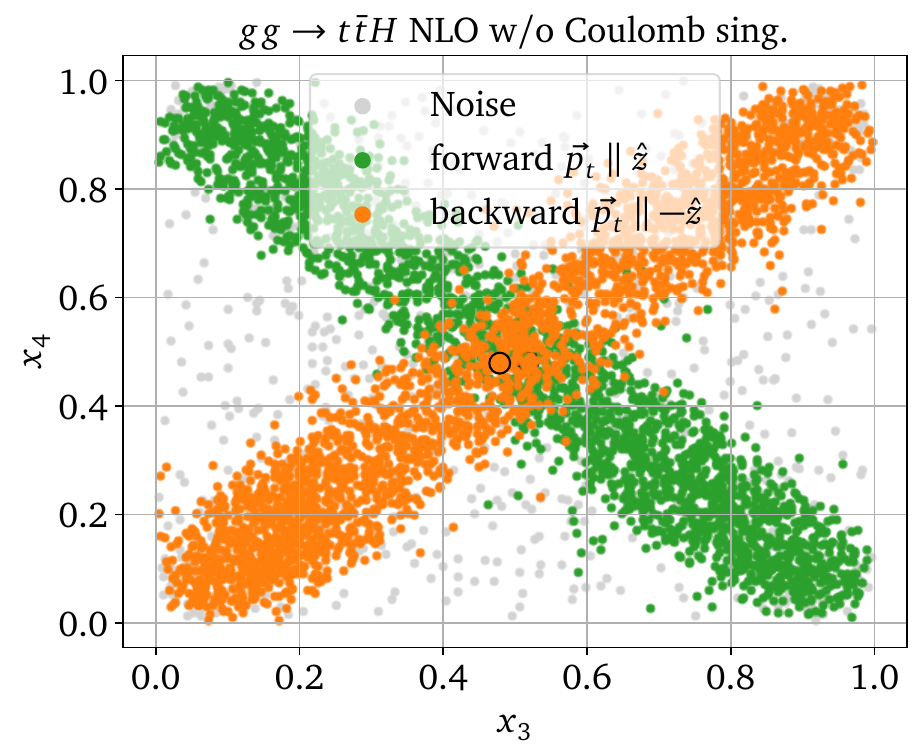} \\
  \includegraphics[width=0.495\textwidth]{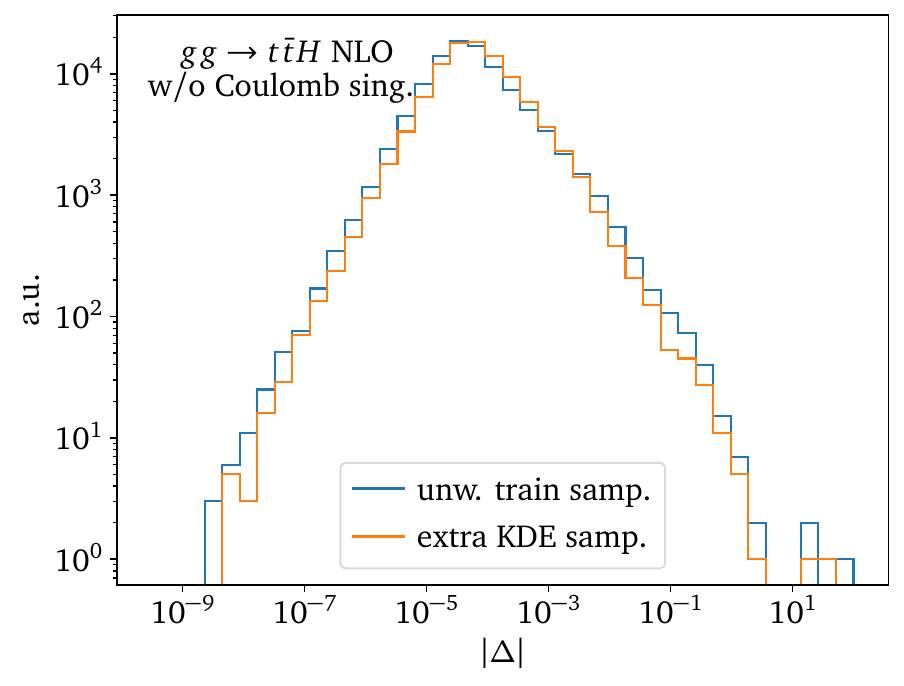}
  \includegraphics[width=0.415\textwidth]{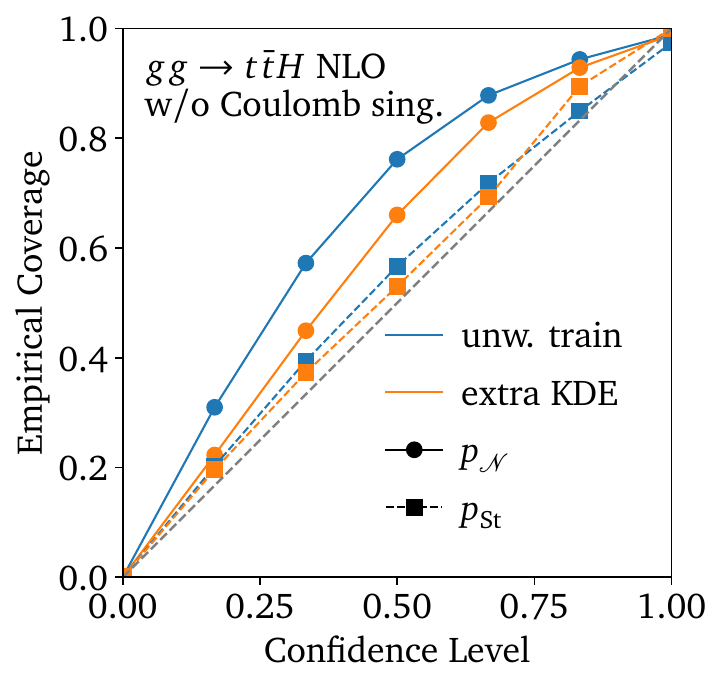}
  \caption{Upper: relative accuracy of the $g g \to t \bar{t}H$ NLO surrogate, highlighting the contribution of the identified low-accuracy clusters (left) and clustering representing the forward and backward singularities (right). Lower: relative accuracy  (left) and empirical coverage (right) without and with additional KDE training.}
  \label{fig:ggttH_NLO}
\end{figure}

Now we move to the NLO $gg \xrightarrow{} t \bar{t} H $ surrogate with subtracted Coulomb singularity. In the upper left panel of Fig.~\ref{fig:ggttH_NLO}, we show the $|\Delta|$ distributions including the contribution from the identified low-accuracy cluster. As for the LO amplitude, two clusters correspond to two forward/backward scattering regions. As mentioned above, we do not directly learn the NLO amplitude but the ratio between NLO and LO amplitudes. In this ratio, the forward/backward beam singularity cancels almost completely. Even though the amplitude ratio is flatter than the LO amplitude, a significant feature sill remains, see App.~\ref{app:NLO_collinear}. The remaining singularity structure is still identified by the clustering algorithm, as shown in the upper right panel of Fig.~\ref{fig:ggttH_NLO}. The two clusters for top quarks along the beam pipe constitute around $88\%$ of the large-$|\Delta|$ points, with very little noise left in the $(x_3, x_4)$-plane.

Generating additional training data in the low-accuracy regions, using KDE sampling, slightly improves the tail of the $|\Delta|$ distribution, as shown in the lower left panel of Fig.~\ref{fig:ggHH_NLO_deltaabs}. For the NLO case, we show the calibration of the learned uncertainties without a Gaussian assumption and in terms of the empirical coverage in the lower right panel of Fig.~\ref{fig:ggHH_NLO_deltaabs}. The surrogates with the additional KDE sample improves the calibration of the uncertainties, leading to a reliable calibration when used together with a Student's $t$-likelihood rather than the standard Gaussian likelihood. 

\subsubsection*{NLO with Coulomb singularity}

\begin{figure}[t]
  \includegraphics[width=0.495\textwidth]{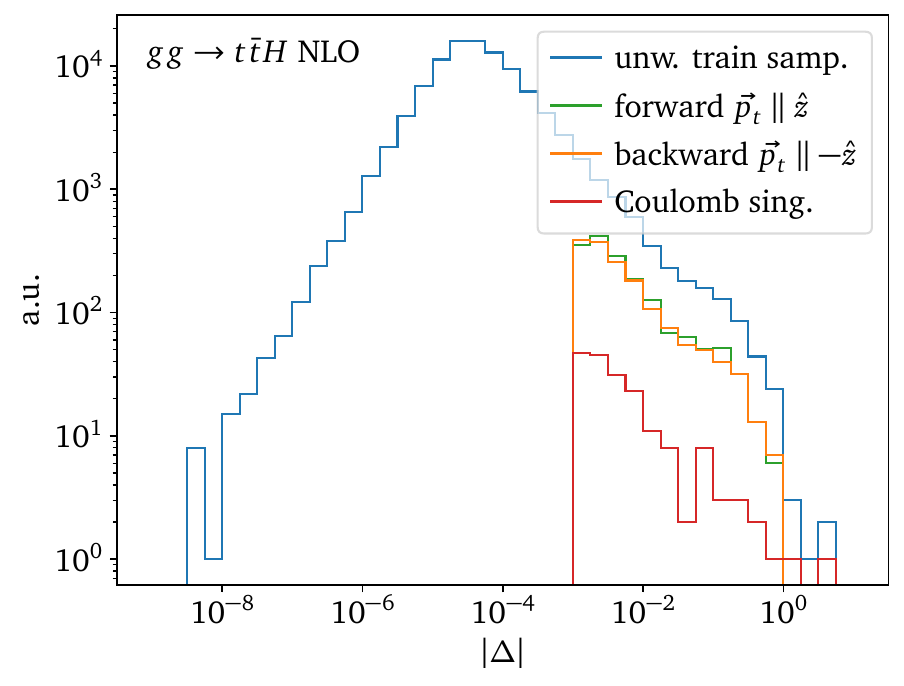}
  \includegraphics[width=0.495\textwidth]{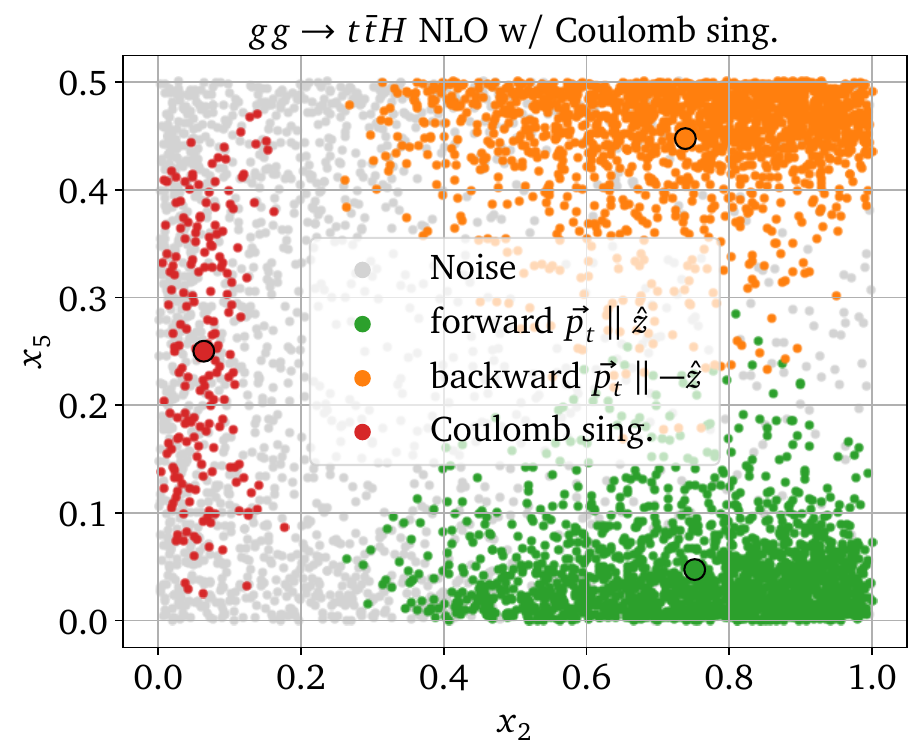} \\
  \includegraphics[width=0.495\textwidth]{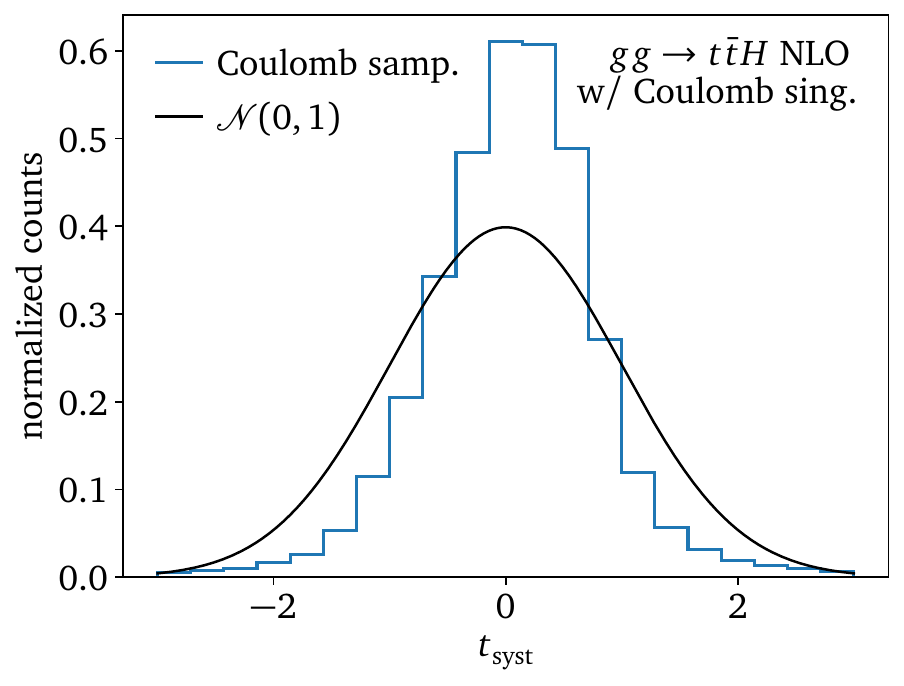}\hfill
  \includegraphics[width=0.405\textwidth]{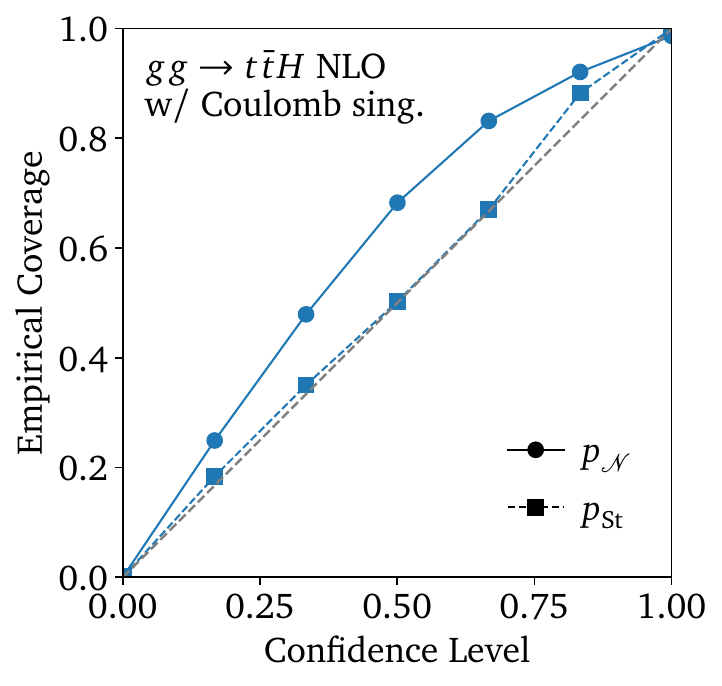}
  \caption{Upper: relative accuracy of the $g g \to t \bar{t}H$ NLO surrogate with  Coulomb singularity, highlighting the contribution of the identified low-accuracy clusters (left) and clustering representing the forward, backward, and Coulomb singularities (right). Lower: pull distibution (left) and empirical coverage (right).}
  \label{fig:ggttH_NLO_C}
\end{figure}

As a final test of our phase space cluster finding, we investigate the $gg\to t\bar t H$ NLO amplitude without subtracting the Coulomb singularity. Indeed, the clustering algorithm now picks up three clusters, including the Coulomb singularity, as shown in the upper right panel of Fig.~\ref{fig:ggttH_NLO_C}. The Coulomb cluster appears for small $x_2$ values, corresponding to a small energy of the $t\bar t$ system. However, in the corresponding $|\Delta|$ distribution, we observe that the Coulomb singularity cluster only contributes around $4\%$ to the poor-accuracy tail, whereas the two collinear clusters are responsible for $65\%$ of this tail. This is because the radiation of the Higgs boson from a top-quark protects the Coulomb singularity, as the top propagators only can go both on-shell in the soft Higgs limit. As an additional test presented in App.~\ref{app:sigma_clustering}, we also performed the clustering based on the learned uncertainty finding similar results.

The pull distribution in the lower left panel of Fig.~\ref{fig:ggttH_NLO_C} shows that, as for the subtracted NLO amplitude, the learned uncertainty using a Gaussian likelihood is underconfident. Replacing the Gaussian with a Student's $t$-likelihood significantly  improves the empirical coverage, as shown in the lower right panel.

\subsubsection*{Accuracy}

\begin{table}[t]
    \centering
    \begin{small}
    \begin{booktabs}{colspec={llc|r|r|r}, colsep=3.5pt}
        \toprule
          \SetCell[c=2]{c} dataset && KDE & \SetCell{c} mean $|\Delta|$ & \SetCell{c} median $|\Delta|$ & \SetCell{c} $\epsilon$\\
         \midrule
          \SetCell[r=6]{c} $gg \to t \bar{t} H$
          & LO && $(5.5 \pm 0.4)\cdot 10^{-4}$&$(7.3 \pm 2.8)\cdot 10^{-5}$& $(5.5 \pm 0.4)\cdot 10^{-4}$\\
          & LO & \cmark & $(2.3 \pm 0.04)\cdot 10^{-4}$&$(3.2 \pm 0.1)\cdot 10^{-5}$& $(2.34 \pm 0.04)\cdot 10^{-4}$\\
          & NLO && $(2.1 \pm 0.8)\cdot 10^{-3}$&$(6.9 \pm 3.2)\cdot 10^{-5}$& $(1.3 \pm 0.1)\cdot 10^{-3}$\\
          & NLO & \cmark & $(1.5 \pm 0.3)\cdot 10^{-3}$&$(6.4 \pm 0.8)\cdot 10^{-5}$& $(9.6 \pm 0.7)\cdot 10^{-4}$\\
          & NLO with Coul.\ sing. && $(1.5 \pm 0.1)\cdot 10^{-3}$&$(5.6 \pm 1.6)\cdot 10^{-5}$& $(2.2 \pm 0.1)\cdot 10^{-3}$\\
          & NLO with Coul.\ sing. & \cmark & $(4.8 \pm 0.1)\cdot 10^{-4}$&$(1.8 \pm 0.6)\cdot 10^{-5}$& $(1.6 \pm 0.01)\cdot 10^{-3}$\\
          \midrule
          \SetCell[r=6]{c} $q \bar{q} \to t \bar{t} H$
          & LO &&  $(9.6 \pm 0.7)\cdot 10^{-5}$& $(2.0 \pm 0.1)\cdot 10^{-5}$ & $(9.6 \pm 0.7)\cdot 10^{-5}$\\
          & LO & \cmark &  $(5.5 \pm 0.1)\cdot 10^{-5}$& $(2.0 \pm 0.1)\cdot 10^{-5}$& $(5.5 \pm 0.1)\cdot 10^{-5}$\\
          & NLO &&  $(1.1 \pm 0.2)\cdot 10^{-3}$& $(4.9 \pm 1.0)\cdot 10^{-5}$& $(2.2 \pm 0.1)\cdot 10^{-4}$\\
          & NLO & \cmark &  $(8.0 \pm 1.4)\cdot 10^{-4}$& $(4.5 \pm 0.4)\cdot 10^{-5}$& $(1.8 \pm 0.1)\cdot 10^{-4}$\\
          & NLO with Coul.\ sing. &&  $(1.4 \pm 0.04)\cdot 10^{-3}$& $(3.3 \pm 0.5)\cdot 10^{-5}$& $(2.8 \pm 0.2)\cdot 10^{-3}$\\
          & NLO with Coul.\ sing. & \cmark &  $(3.1 \pm 0.04)\cdot 10^{-4}$& $(3.9 \pm 3.4)\cdot 10^{-6}$& $(1.8 \pm 0.2)\cdot 10^{-3}$\\
          \bottomrule
    \end{booktabs}
    \end{small}
    \caption{Comparison of mean and median accuracies of the various $t\bar tH$ surrogates. For comparison, also values for the $\epsilon$ metric introduced in Ref.~\cite{Breso:2024jlt} are shown. The metrics are averaged over five independent runs with the quoted uncertainty corresponding to the standard deviation of the five runs.
    }
    \label{tab:ttH_deltaabs_comparison}
\end{table}

Finally, in Tab.~\ref{tab:ttH_deltaabs_comparison}, we compare the mean and median accuracies reached for the various $t\bar t H$ datasets. For comparison, we also show the $\epsilon$ metric introduced in Ref.~\cite{Breso:2024jlt}, which is given by
\begin{align}
    \epsilon = \frac{\sum_i|A_\text{NN}(x_i) -A_\text{true}(x_i)|}{\sum_i A_\text{true}(x_i)}  
\end{align}
for an unweighted phase space sample and independent of the way the amplitude is actually learned.

For the $gg\to t\bar tH$ amplitudes, the mean relative accuracy of the LO amplitude is roughly an order of magnitude better than the one of the NLO amplitude. However, the median relative accuracy is roughly the same for the LO and NLO amplitudes. This indicates that the loss of mean accuracy at NLO is caused by a small number of low-accuracy phase space points. Including an additional training dataset covering low-accuracy regions described by KDE clustering improves, in particular, the mean accuracies by almost a factor of two. Training a surrogate for the NLO amplitude without subtracting the Coulomb singularity does not lead to a visible degradation and can, within uncertainties, even improve the performance.

For the NLO gluon-fusion dataset, the shown $\epsilon$ value is slightly worse than the MLP value found in Ref.~\cite{Breso:2024jlt} for the same size of the training dataset. This is due to the used L1 loss in Ref.~\cite{Breso:2024jlt} which more directly optimizes the $\epsilon$ metric.

Finally, we also show results for the $q\bar q\to t\bar tH$ amplitude. The same kind of study as for the gluon-fusion process is given in App.~\ref{app:qqttH}. As a consequence of the missing $t$-channel diagram, the forward/backward scattering singularities are weaker. As shown in Eq.~\eqref{eq:coulomb_sing_strength} and in App.~\ref{app:coulomb}, also the Coulomb singularity is weaker. Consequently, the low-accuracy clusters are smaller and the upper tail of the $|\Delta|$ distribution is dominated by noise.

\section{Outlook}
\label{sec:conclusions}

Accurate and fast higher-order predictions of kinematic distributions are at the heart of the precision-LHC program. The challenges of the upcoming HL-LHC runs force us to accelerate the corresponding simulation tools and to include even higher loop orders. For both challenges, NN-surrogates of transition amplitudes provide a promising direction, especially if we can use these surrogates without further reweighting. Trustworthy amplitude surrogates have to cover two aspects: they have to accurately reproduce the truth and they have to include a reliable and calibrated uncertainty. 

We have first established a way to learn uncertainties in a non-Gaussian regime, where the standard pull benchmark to assess the uncertainties has to be replaced by a coverage test. Just like for linear regression, surrogates with a larger phase space dimensionality behave more Gaussian, thanks to the central limit theorem. To identify potential failure modes, we have developed a novel method to search for localized patterns with low accuracy or large uncertainties. Finally, we have proposed a way to alleviate such localized problems through an adaptive sampling of the training dataset.

We have established these novel techniques for the finite part of the virtual amplitudes for di-Higgs production in gluon fusion up to two loops, and for Higgs production associated with a top quark pair up to one loop. In both cases, challenging phase space regions, such as the virtual top quark pair production threshold in the di-Higgs case or a Coulomb-type singularity in the $t\bar{t}H$ case, are reproduced adequately by the adaptively trained surrogate.

Using simple network architectures, surrogate amplitudes can be evaluated at least as fast as low-multiplicity leading-order amplitudes, \ie faster than analytically available (multi-scale) two-loop amplitudes and much faster than numerical calculations. An efficient way to generate optimal training data is a critical step. Our new, adaptive training provides us with accurate surrogates including a reliable and conservative uncertainty estimate, even for non-Gaussian likelihoods and in the presence of challenging phase space features. Our methods will be especially important for pushing the amplitude surrogates to higher loop orders where  the singularity structures are richer. Another complication at higher loop orders is given by the smaller amount of available training data. As the performance of the NN-based methods scales favorably with the number of phase space dimensions, surrogate amplitudes can pave the way to fast Monte Carlo programs for multi-scale processes at NNLO and beyond, provided that similar progress is made on the unresolved real radiation side.

\section*{Acknowledgments}

We would like to thank Victor Breso, Vitaly Magerya and Anton Olsson for useful discussions. GH would like to thank Stephen Jones and Matthias Kerner for their work on \texttt{hhgrid}.
This work is supported by the Deutsche Forschungsgemeinschaft (DFG, German Research Foundation) under grant
396021762 – TRR 257 Particle Physics Phenomenology after the Higgs Discovery. The work of HB, TP, and RR is supported by Deutsche Forschungsgemeinschaft (DFG) under Germany’s Excellence Strategy EXC-2181/1 - 390900948 (the Heidelberg STRUCTURES Excellence Cluster). JB is supported in parts by the Federal Ministry of Technology and Space (BMFTR) under grant number 05H24VKB. The authors acknowledge support by the state of
Baden-Württemberg through bwHPC and the German Research Foundation (DFG) through
the grants INST 35/1597-1 FUGG and INST 39/1232-1 FUGG.

\newpage
\appendix

\section{Hyperparameters}
\label{app:hyperparameters}

\begin{table}[H]
    \centering
    \begin{booktabs}{l|ccccc} 
    \toprule
        Parameter & toy & $Z+\text{gluons}$ & $HH$ LO & $HH$ NLO & $t\bar tH$ \\ 
        \midrule
        Activation function & GELU & GELU & GELU & GELU & GELU \\
        Number of hidden layers & 3 & 5 & 5 & 6 & 5 \\
        Hidden nodes        & 64 & 128 & 512 & 256 & 512\\
        Batch size          & 256 & 256 & 256 & 1024  & 256\\
        Scheduler           & Cosine & Cosine & Cosine & Cosine  & Cosine\\
        Max learning rate   & $2 \cdot 10^{-4}$ & $2 \cdot 10^{-4}$ & $2 \cdot 10^{-4}$ & $10^{-4}$ & $2 \cdot 10^{-4}$\\
        Number of epochs    & 1000 & 4000 & 4000 & 4000 & 2000\\
        \bottomrule
    \end{booktabs}
    \caption{Network and training parameters.}
    \label{tab:ae}
\end{table} 

We provide an overview of the hyperparameter settings used for the various processes in Tab.~\ref{tab:ae}. For the $gg\to t\bar t H$ and $q\bar q\to t\bar t H$ surrogates, we used the same hyperparameters.

\section{Forward/backward regions: NLO to LO ratio}
\label{app:NLO_collinear}

\begin{figure}[h!]
    \centering
    \includegraphics[width=0.495\textwidth]{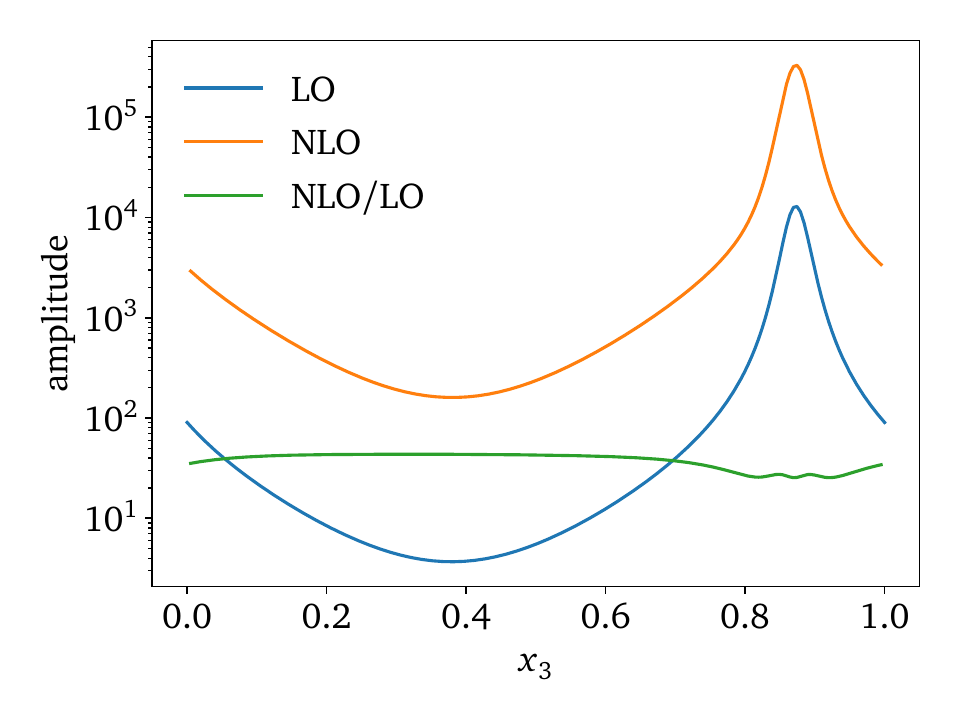}
    \includegraphics[width=0.495\textwidth]{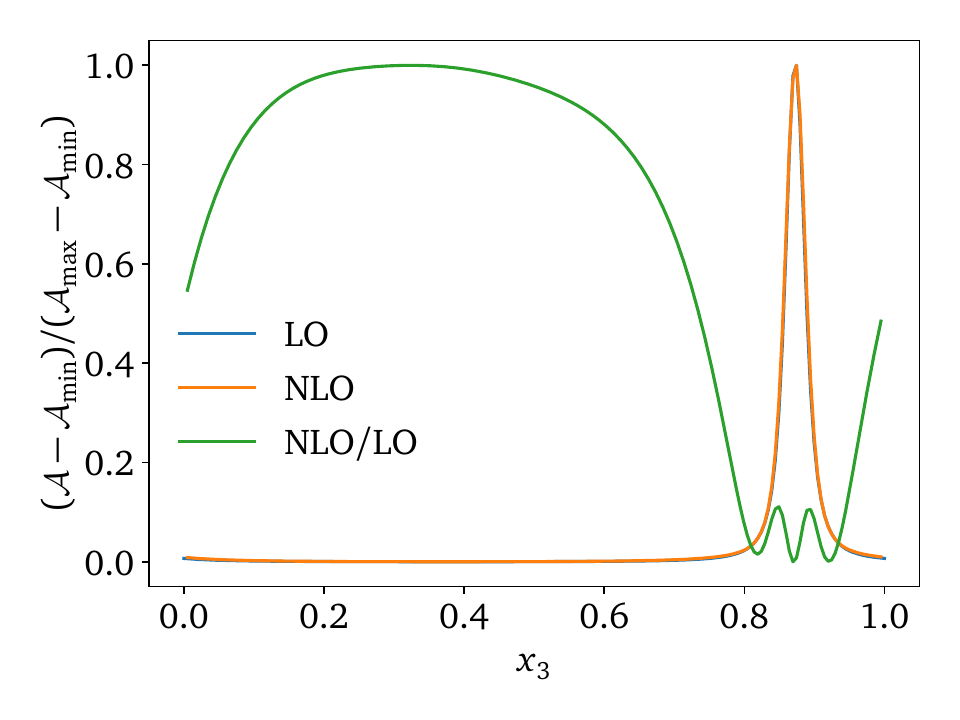}
    \caption{Left: One-dimensional slice of the $gg\rightarrow ttH$ amplitude for $x_1 = 0.99$, $x_2 = 0.95$, $x_4 = 0.40$, $x_5 = 0.01$. Right: same as left, but each function is individually normalized to $[0, 1]$.}
    \label{fig:gg_tth_amp_slice}
\end{figure}

Fig. \ref{fig:gg_tth_amp_slice} shows a one-dimensional slice in $x_3$ of the $gg \rightarrow ttH$ amplitude for fixed values of $x_1 = 0.99$, $x_2 = 0.95$, $x_4 = 0.40$ and $x_5 = 0.01$. The forward/backward region described in section \ref{sec:ttH_singularities} is clearly visible at $x_3 \sim 0.87$ in the LO and NLO amplitudes, whereas the absolute value of the ratio is nearly flat compared to the amplitudes themselves. When each function is individually normalized to $[0, 1]$ however, the ratio also shows steep changes in the forward/backward region. This results in a challenging region for the surrogate persisting even when learning only the ratio.

\section{Accuracy close to Coulomb singularity}
\label{app:coulomb}

\begin{figure}[H]
  \centering
  \includegraphics[width=0.495\textwidth]{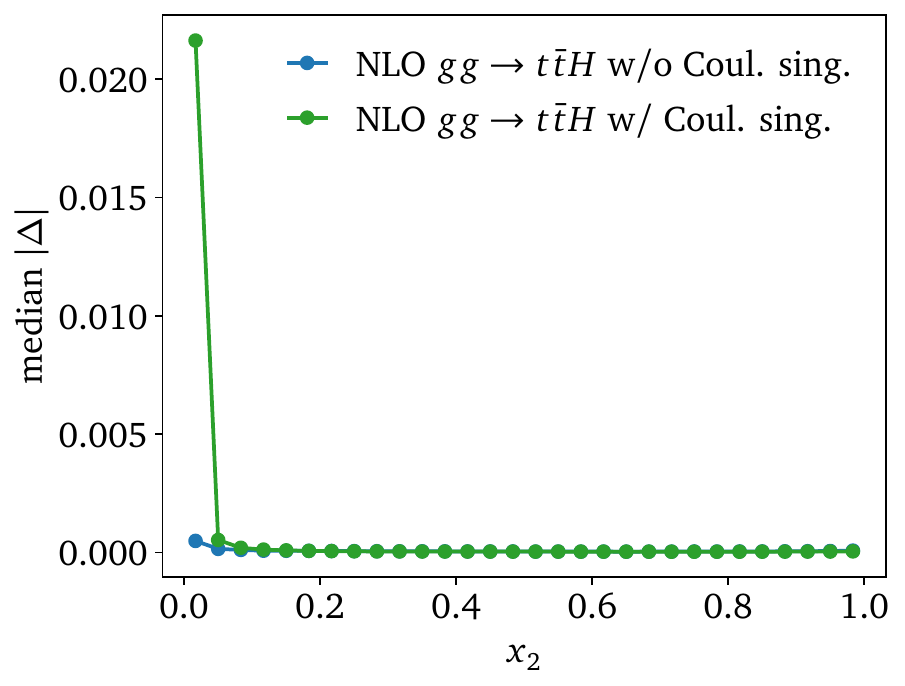}
  \includegraphics[width=0.495\textwidth]{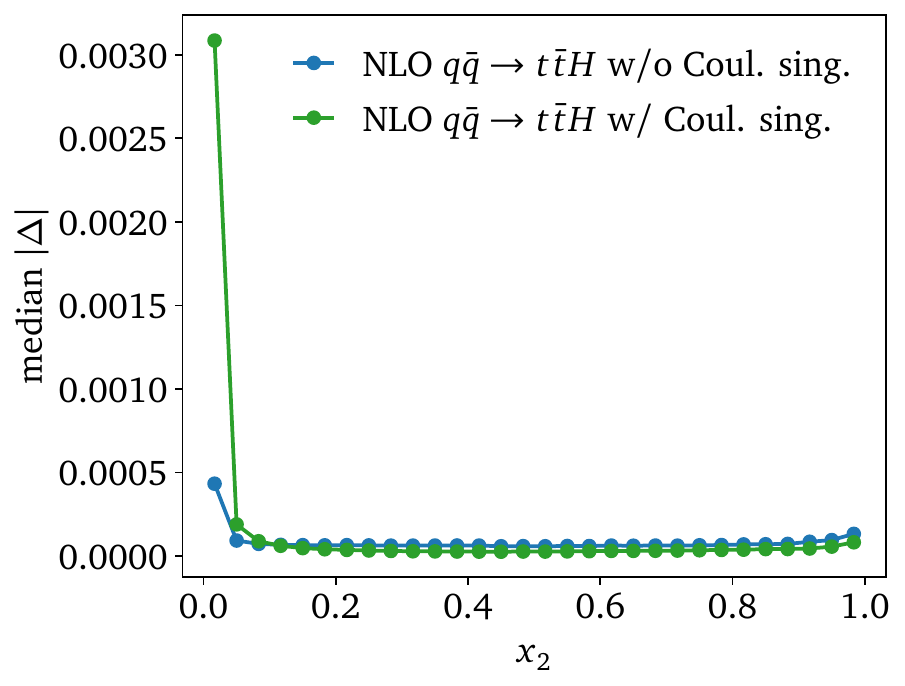}
  \caption{Left: median accuracy of the $gg\to t\bar tH$ NLO surrogate as a function of $x_2$ with and without subtraction of the Coulomb singularity. Right: Same as left, but the median accuracy of the $q\bar q \to t\bar t H$ NLO surrogate is shown.}
  \label{fig:ttH_NLO_coulomb}
\end{figure}

In Fig.~\ref{fig:ttH_NLO_coulomb}, we show the median relative accuracy of the $t\bar t H$ NLO surrogates as a function of $x_2$ with and without subtraction of the Coulomb singularity. For the $gg \to t\bar tH$ surrogate shown in the left panel, the loss in relative accuracy for $x_2 \to 0$ without subtraction of the Coulomb singularity is significantly larger than for the $q\bar q\to t\bar tH$ amplitude shown in the right panel. This is in agreement with the theoretical expectation outlined in Sec.~\ref{sec:ttH_singularities}.

\section{Clustering based on estimated uncertainties}
\label{app:sigma_clustering}

\begin{figure}[H]
  \centering
  \includegraphics[width=0.495\textwidth]{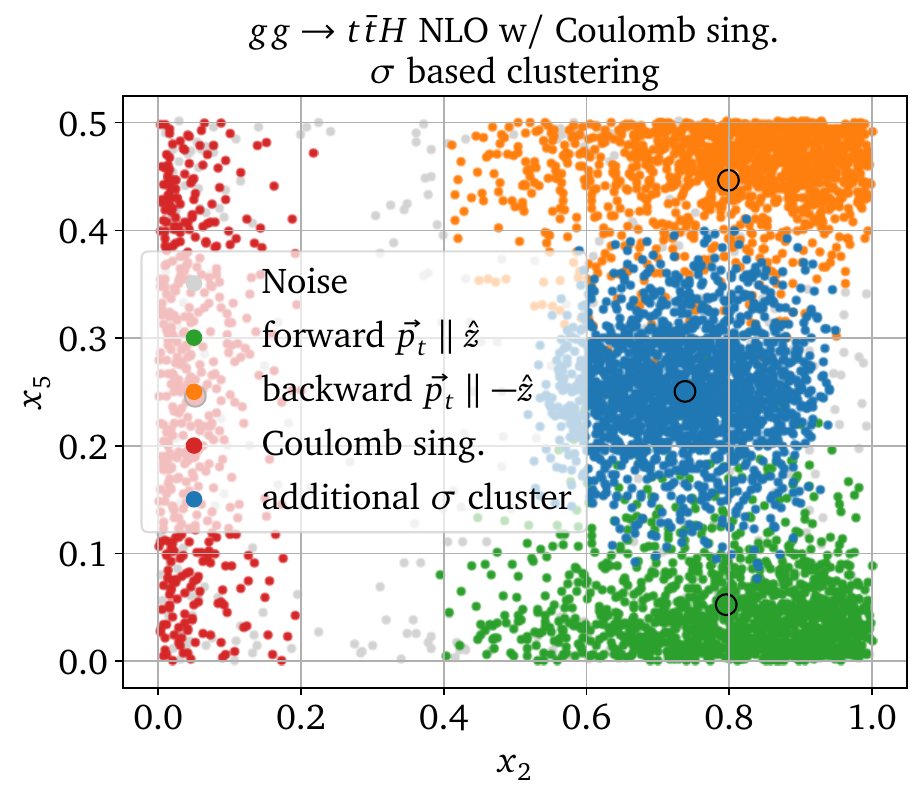}
  \caption{Identified clusters based on the estimated uncertainty for the NLO $gg\rightarrow t\bar tH$ surrogate without subtraction of the Coulomb singularity in the $x_2$-$x_5$ plane}
  \label{fig:ttH_NLO_coulomb_clustering_unc}
\end{figure}

In Fig.~\ref{fig:ttH_NLO_coulomb_clustering_unc}, we show the result of the clustering algorithm for the NLO $gg\rightarrow t\bar tH$ surrogate without subtraction of the Coulomb singularity based on the estimated uncertainties. In particular, we use $\sigma/A$ instead of $|\Delta|$ as measure to selecting the least accurate phase-space points. As before, the forward/backward regions and the Coulomb singularity are identified. In addition, the algorithm identifies an additional cluster, which we could not identify with any specific features of the amplitude. We attribute this to the fact that the clustering is performed based on the 5\% points with the largest $\sigma/A$ and not the largest $|\Delta|$. Since the uncertainty is not perfectly calibrated, a different set of points is selected.

\section{Extracted KDE estimates}
\label{app:kde_densities}

\begin{figure}[H]
  \centering
  \includegraphics[width=0.495\textwidth]{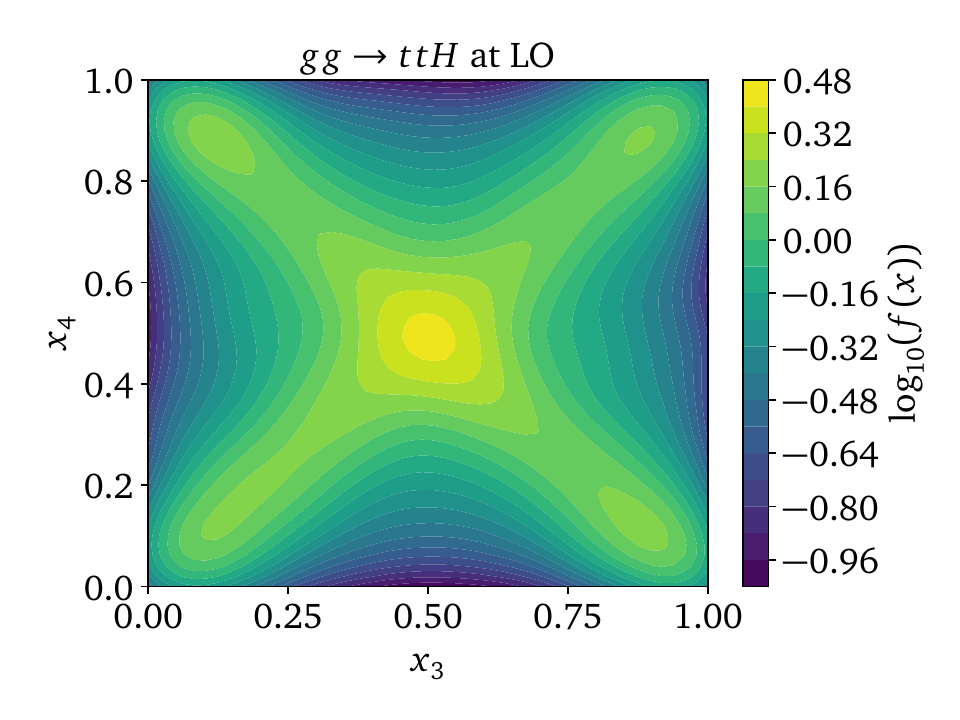}
  \includegraphics[width=0.495\textwidth]{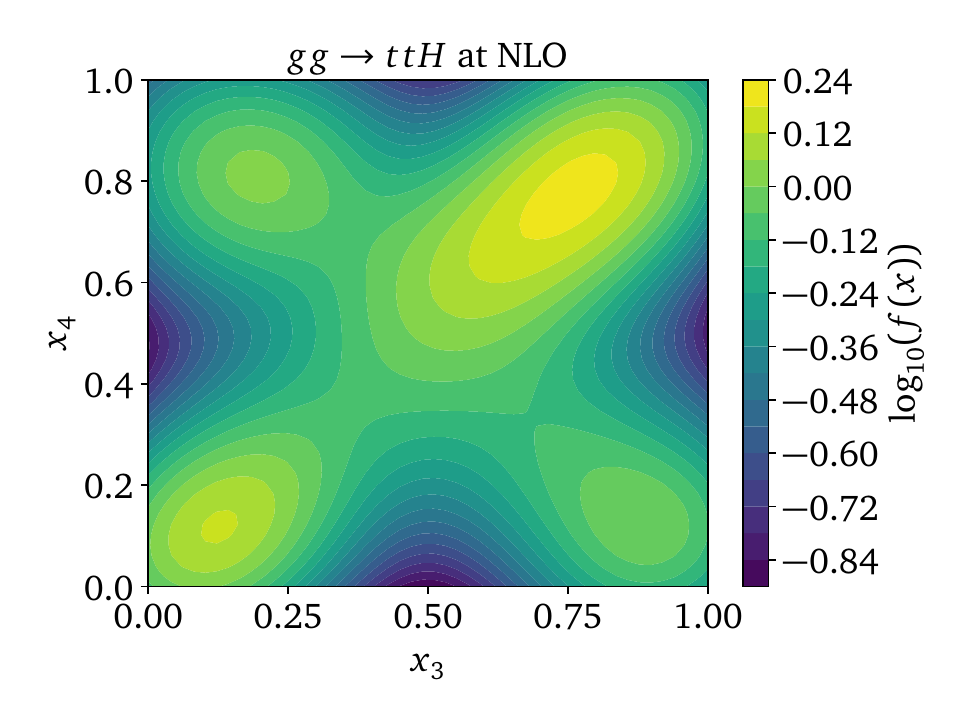}
  \caption{Density plot of the marginalized KDE-PDF for $gg\rightarrow ttH$ in the $x_3$-$x_4$ plane at LO (left) and at NLO (right).}
  \label{fig:ttH_NLO_densities}
\end{figure}

Fig.~\ref{fig:ttH_NLO_densities} shows the marginalized KDE-PDF for the $gg\rightarrow ttH$ channel in the $x_3$-$x_4$ plane at LO and NLO. The KDE-PDF can clearly be observed to reproduce the diagonal structure formed by the two clusters identified in the top right panel of Figs.~\ref{fig:ggttH_LO} and \ref{fig:ggttH_NLO}, leading to a KDE sample focused in the forward/backwards scattering regions.

\clearpage
\section{\texorpdfstring{$q\bar{q} \xrightarrow{} t \bar{t} H $}{qq->ttH} results}
\label{app:qqttH}

In this Appendix, we collect the results for the $q\bar q \to t\bar t H$ amplitude, which we investigated in the same way as the $gg\to t\bar t H$ amplitude.

\subsubsection*{Leading order}

\begin{figure}[H]
  \centering
  \includegraphics[width=0.495\textwidth]{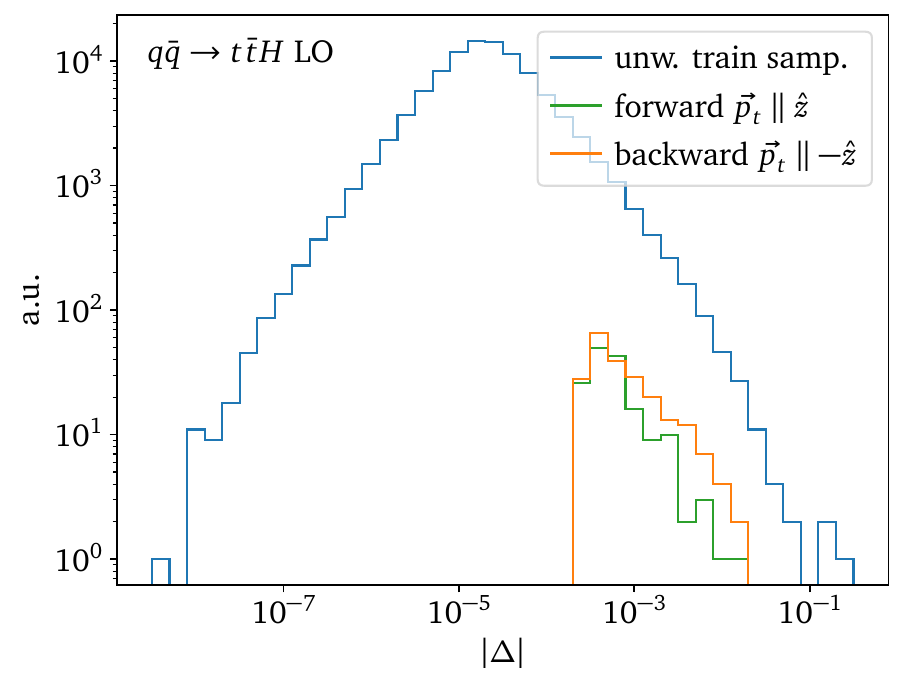}
  \includegraphics[width=0.495\textwidth]{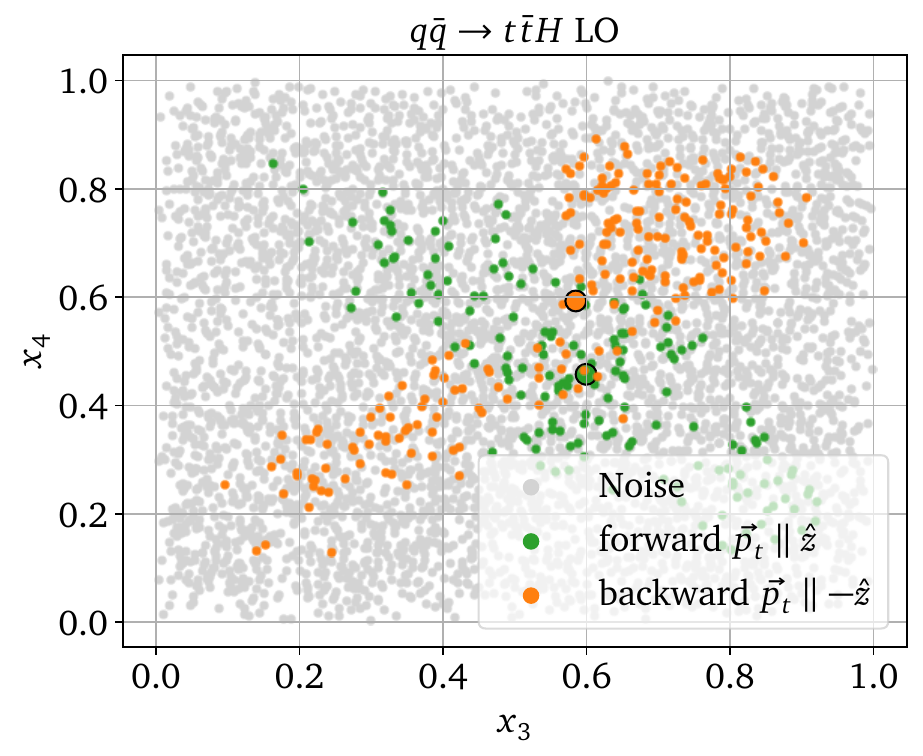}
  \includegraphics[width=0.495\textwidth]{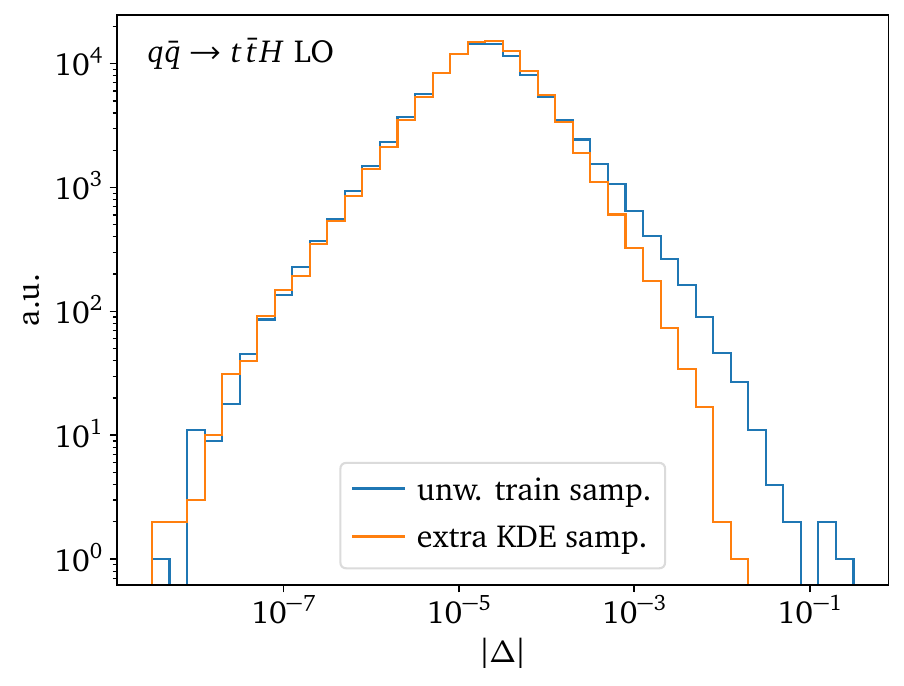}
  \includegraphics[width=0.495\textwidth]{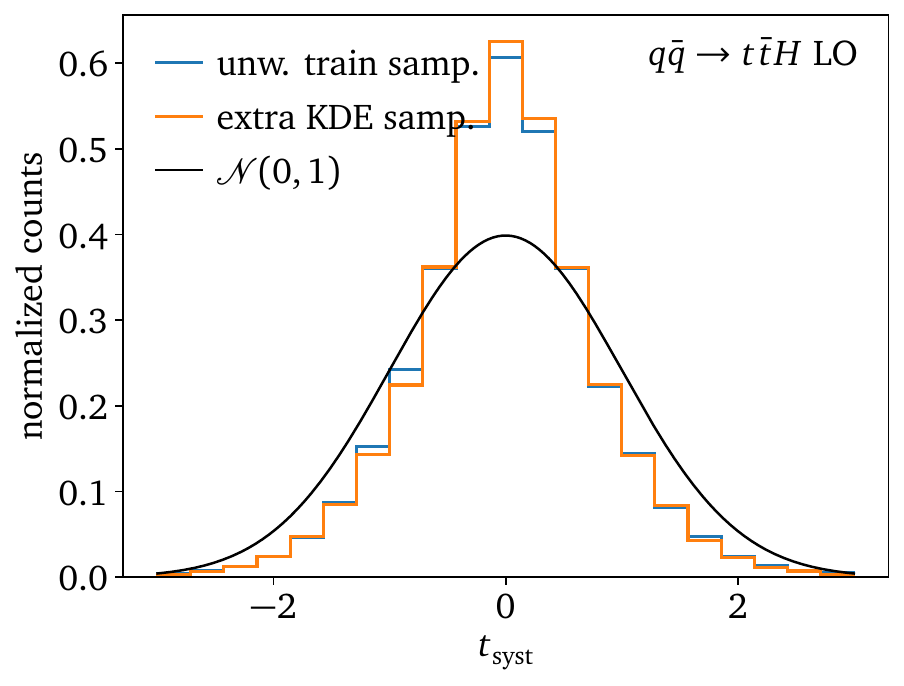}
  \caption{Upper left: Relative accuracy of $q \bar{q}\to t \bar{t}H$ LO surrogates highlighting the identified low-accuracy clusters. Upper right: identified clusters in the $(x_3, x_4)$ plane. Lower left: Relative accuracy comparing the surrogates trained with and without additional KDE training samples. Lower right: Systematic pull distributions.}
  \label{fig:qqttH_LO}
\end{figure}

First, we discuss the results for the LO amplitude shown in Fig.~\ref{fig:qqttH_LO}. As for the $gg\to t\bar t H$ amplitude, the clustering algorithm identifies the forward/backward regions. But due to the lack of a $t$-channel diagram, the forward/backward scattering singularities are significantly weaker than for the $gg\to t\bar t H$ amplitude and the tail of the $|\Delta|$ distribution is dominated by noise. Including an additional KDE sampling significantly improves the performance. We find the uncertainty estimate to be slightly underconfident.

\subsubsection*{Next-to-leading order }

\begin{figure}[H]
  \centering
  \includegraphics[width=0.495\textwidth]{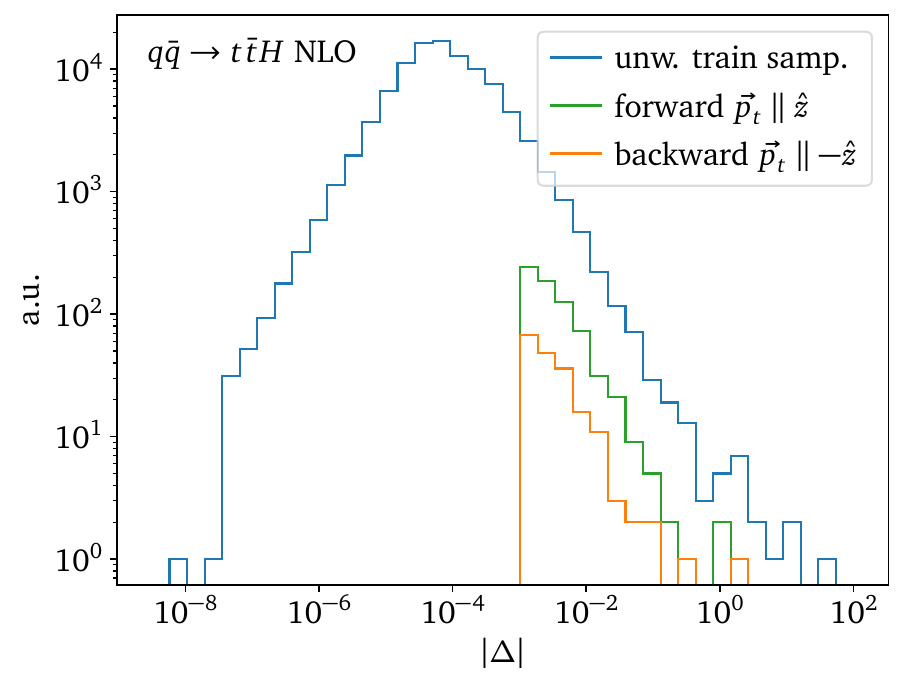}
  \includegraphics[width=0.495\textwidth]{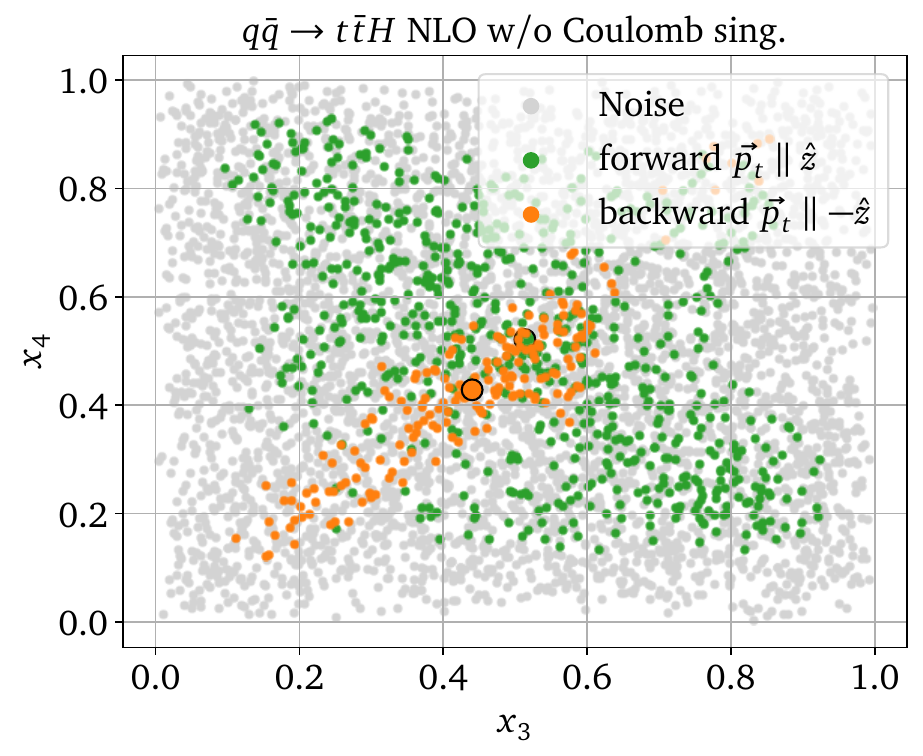}
  \includegraphics[width=0.495\textwidth]{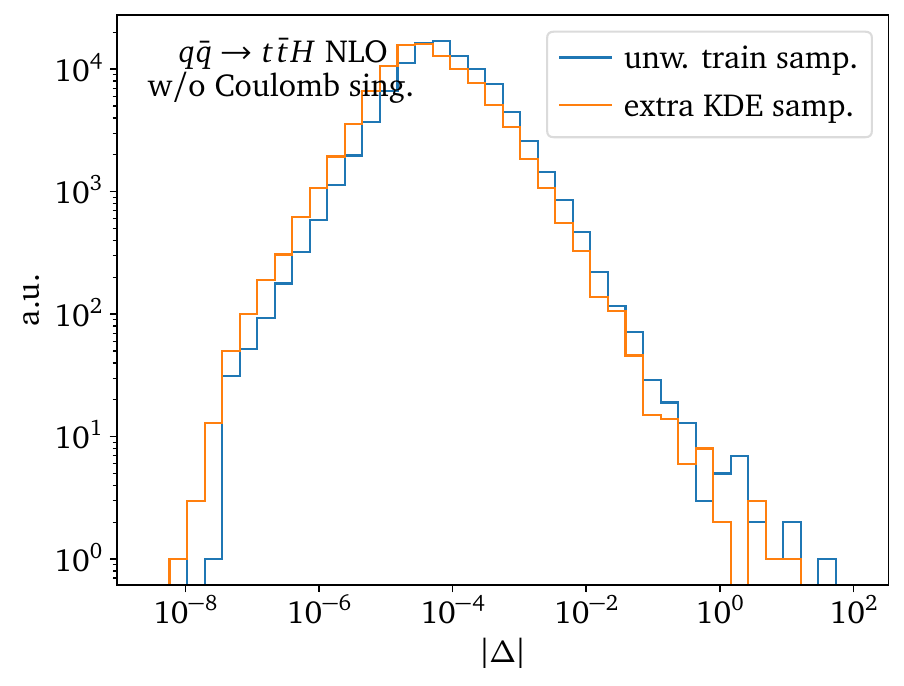}
  \includegraphics[width=0.495\textwidth]{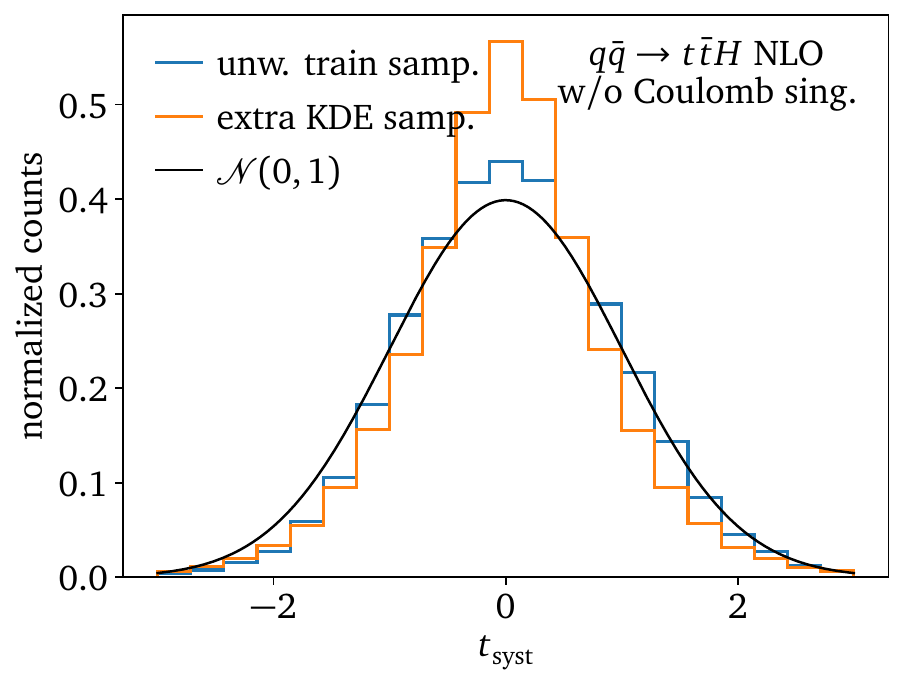}
  \caption{Upper left: Relative accuracy of $q \bar{q}\to t \bar{t}H$ NLO surrogates highlighting the identified low-accuracy clusters. Upper right: identified clusters in the $(x_3, x_4)$ plane. Lower left: Relative accuracy comparing the surrogates trained with and without additional KDE training samples. Lower right: Systematic pull distributions.}
  \label{fig:qqttH_NLO}
\end{figure}

The results for the NLO amplitude are shown in Fig.~\ref{fig:qqttH_NLO}.  The results are very similar to the LO amplitude. The forward/backward regions again only slightly contribute to the upper $|\Delta|$ tail, the KDE sampling significantly improves performance, and the uncertainty estimates are slightly underconfident.

\subsubsection*{NLO $q\bar{q} \xrightarrow{} t \bar{t} H $ with Coulomb singularity}

\begin{figure}[H]
  \centering
  \includegraphics[width=0.495\textwidth]{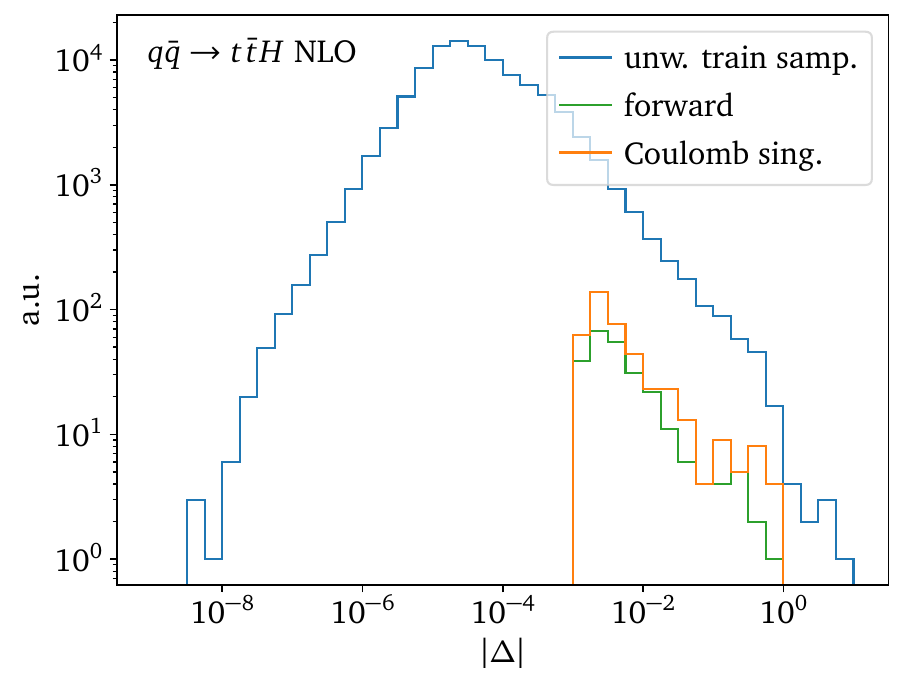}
  \includegraphics[width=0.495\textwidth]{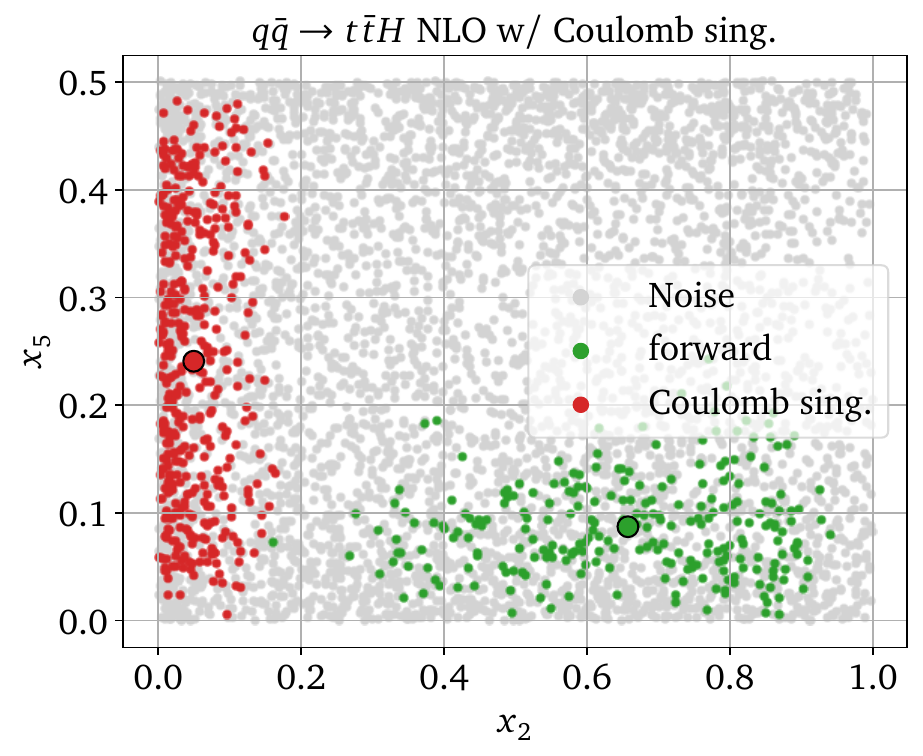}
  \includegraphics[width=0.495\textwidth]{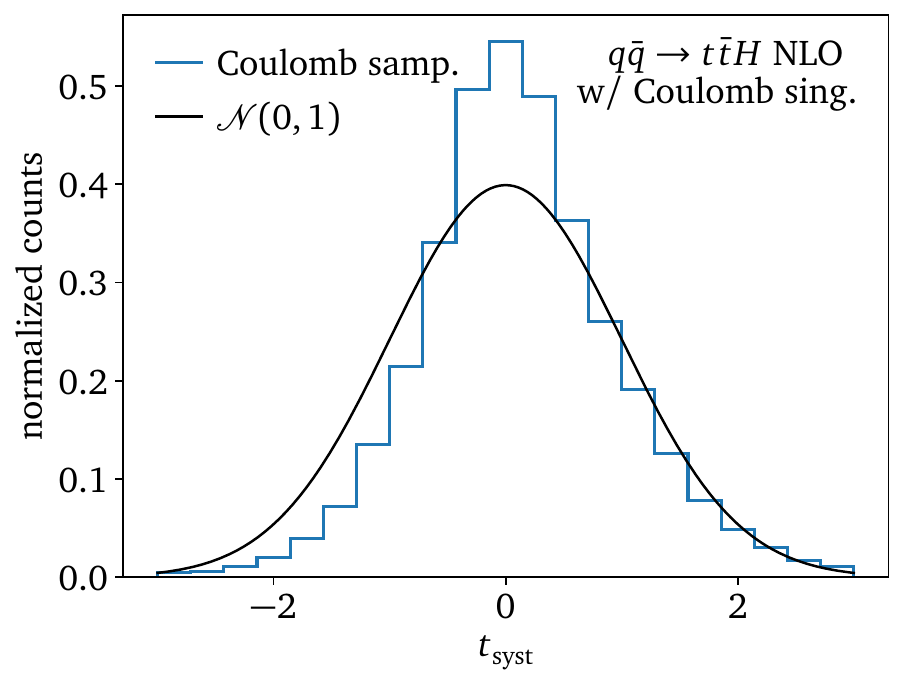}
  \caption{Upper left: Relative accuracy of $q \bar{q}\to t \bar{t}H$ NLO surrogates without subtraction of Coulomb singularity highlighting the identified low-accuracy clusters. Upper right: identified clusters in the $(x_2, x_5)$ plane. Lower left: Relative accuracy comparing the surrogates trained with and without additional KDE training samples. Bottom: Systematic pull distributions.}
  \label{fig:qqttH_NLO_D_extra}
\end{figure}

Finally, we show in \ref{fig:qqttH_NLO_D_extra} the results for the NLO $q\bar{q} \xrightarrow{} t \bar{t} H $ amplitude without subtraction of the Coulomb singularity. The accuracy is comparable to the case with subtraction. The clustering algorithm picks up the Coulomb singularity but fails to identify the backward scattering region due to the large amount of noise. We find the uncertainty estimate to be again slightly underconfident.

\clearpage
\bibliography{tilman,refs}

\end{document}